\documentclass[11pt]{article}

\usepackage{amsmath,amssymb,amsfonts,amsthm}
\usepackage{graphicx}
\usepackage[utf8]{inputenc}
\usepackage[margin=1in]{geometry} % Standard 1-inch margins
\usepackage{natbib}
\usepackage{subcaption}
\usepackage{multirow}
\usepackage{pdfpages}

\usepackage{authblk}

\usepackage[colorlinks=true, linkcolor=blue, citecolor=blue, urlcolor=blue]{hyperref}

\title{Shapes, forces, and torques of compressed elastic fluid interfaces: beyond axisymmetric configurations}

\author[1]{Xinyi Liu\thanks{XinyiLiu2025@u.northwestern.edu}}
\author[2,1]{Neelesh A. Patankar\thanks{n-patankar@northwestern.edu}}
\author[3]{Leroy L. Jia\thanks{royjia117@gmail.com}}

\affil[1]{\small Department of Engineering Sciences and Applied Mathematics, 
Northwestern University,\authorcr
2145 Sheridan Road, Evanston, Illinois 60208, USA}
\affil[2]{Department of Mechanical Engineering, 
Northwestern University, \authorcr
2145 Sheridan Road, Evanston, Illinois 60208, USA}
\affil[3]{Independent researcher, North Bethesda, MD 20852, USA}

\date{\today} % Or empty \date{} to hide the date

\begin{document}

\maketitle

\begin{abstract}
Inspired by the classical Plateau--Douglas problem for soap films bounded by two closed curves, we solve an analogous problem for fluid interfaces with fixed surface area and resistance to out-of-plane bending. The boundaries are planar but need not be symmetric or concentric. The equilibrium surfaces minimize the Willmore bending energy and generalize minimal surfaces such as the catenoid while also exhibiting characteristic features of confined elastic interfaces such as buckling. We systematically classify all possible buckling modes and solution branches by performing a weakly nonlinear analysis and developing a fully nonlinear spectral solver. Our mathematical framework provides insight into the forces and torques required to stabilize cellular membranes and other soft materials and shows 
that physically relevant asymmetric states can arise even in symmetric systems.
\end{abstract}

\section{Introduction}
\label{sec:intro}
%Membranes---thin fluid layers with a resistance to out-of-plane bending---are a simple physical model that see countless applications in biology, chemistry, and materials science. Their fluid nature allows them to reconfigure shape and topology with ease, while their elasticity bestows them with a type of permanence that classical fluids lack. The ubiquity of membranes throughout nature is no doubt thanks to their ability to merge the most useful properties of these two classes of materials. Understanding the mechanical laws that govern membranes is and will continue to be crucial to delivering innovations in fields such as regenerative medicine, bioinspired engineering, and metamaterials design. 

We study the mechanics of topologically cylindrical fluid interfaces with a resistance to out-of-plane bending that are attached to two closed boundaries. Such configurations occur naturally during cellular processes such as scission and endo- and exocytosis~\citep{tagliatti_imaging_2022,kozlovsky_membrane_2003}. They also appear \emph{in vitro} in  mechanical tests, in which instruments such as cantilevers, optical tweezers, or pipettes are used to deform a microscopic object, and the resulting shape is measured with the goal of inferring physical parameters such as elastic moduli~\citep{wubshet_methods_2023}. Because these systems involve small length and time scales, measurements can be subject to substantial uncertainty, highlighting the need for accurate models that connect observed shapes with underlying mechanics.

A celebrated mathematical framework for predicting the shapes of elastic fluid interfaces is the Willmore energy and its extensions, which continues to find use as a model for a wide range of biological and soft-matter systems, including lipid bilayer membranes~\citep{nitschke_hydrodynamic_2026}, giant unilamellar vesicles~\citep{reboucas_stationary_2024}, colloidal membranes composed of rod-like viruses~\citep{adkins_topology_2025}, and amphiphilic polymer layers~\citep{luo_periodic_2024}. Despite the inherently three-dimensional nature of these systems, many theoretical studies focus on highly symmetric configurations, often because symmetry reduces the governing nonlinear partial differential equation (PDE) to more tractable forms. This simplification, however, can exclude physically important nonaxisymmetric configurations; indeed, recent studies have demonstrated that nonaxisymmetric effects can play a crucial role in shape transitions and mechanical responses that are more representative of realistic biological systems~\citep{omar_nonaxisymmetric_2020,vasan_mechanical_2020}. Developing a tractable framework for nonaxisymmetric interfaces is therefore important both for understanding the broader range of physically accessible configurations and for connecting experimentally observed shapes to their underlying mechanical properties.

In this study, we develop such a framework to theoretically and computationally explore the mechanics of fluid layers with  bending stiffness under compression and tension. In so doing, we dramatically generalize previous work such as~\citep{jia_axisymmetric_2021} by discarding the restrictive assumption of axisymmetry, offering a way to analyze more complex shapes. Our article is structured as follows: first, we outline our problem and its relation to the classical minimal surface and buckling elastica problems, which provide useful intuition.
We then thoroughly catalog the different shapes that can arise and compute the external axial force and torque, using both a  perturbative approach and a pseudospectral solver for the full nonlinear problem. Our analysis is focused around two representative boundary configurations: the simplest case of parallel and concentric circles, which yields the catenoid in the Plateau--Douglas problem, and a pair of perturbed circles, which demonstrates the effects of more general, nonaxisymmetric boundaries. Both scenarios are found to yield physically relevant nonaxismmetric shapes that potentially affect stability. %Throughout our analysis, we pay special attention to nonaxisymmetric effects. %The possible shapes that can form are connected to the solutions of an eigenvalue problem, and perturbing the ring from its initial circular shape lifts the degeneracy certain eigenvalues. 
%The effects of more general shapes may be deduced from this analysis in a straightforward manner. %For these configurations, we consider axial pulling and torsional rotation modes of deformation. %We then discuss the stability of these surfaces and finally conclude in Section V. Geometric formulas that are not immediately essential to the presentation are relegated to the Appendix.

\section{\label{sec:background} Background}

%\subsection{Shape equation,  boundary conditions, and geometry}

\begin{figure}
(a)
    \includegraphics[trim={12.9cm 2cm 11.8cm 2cm},clip,width=0.33\linewidth]{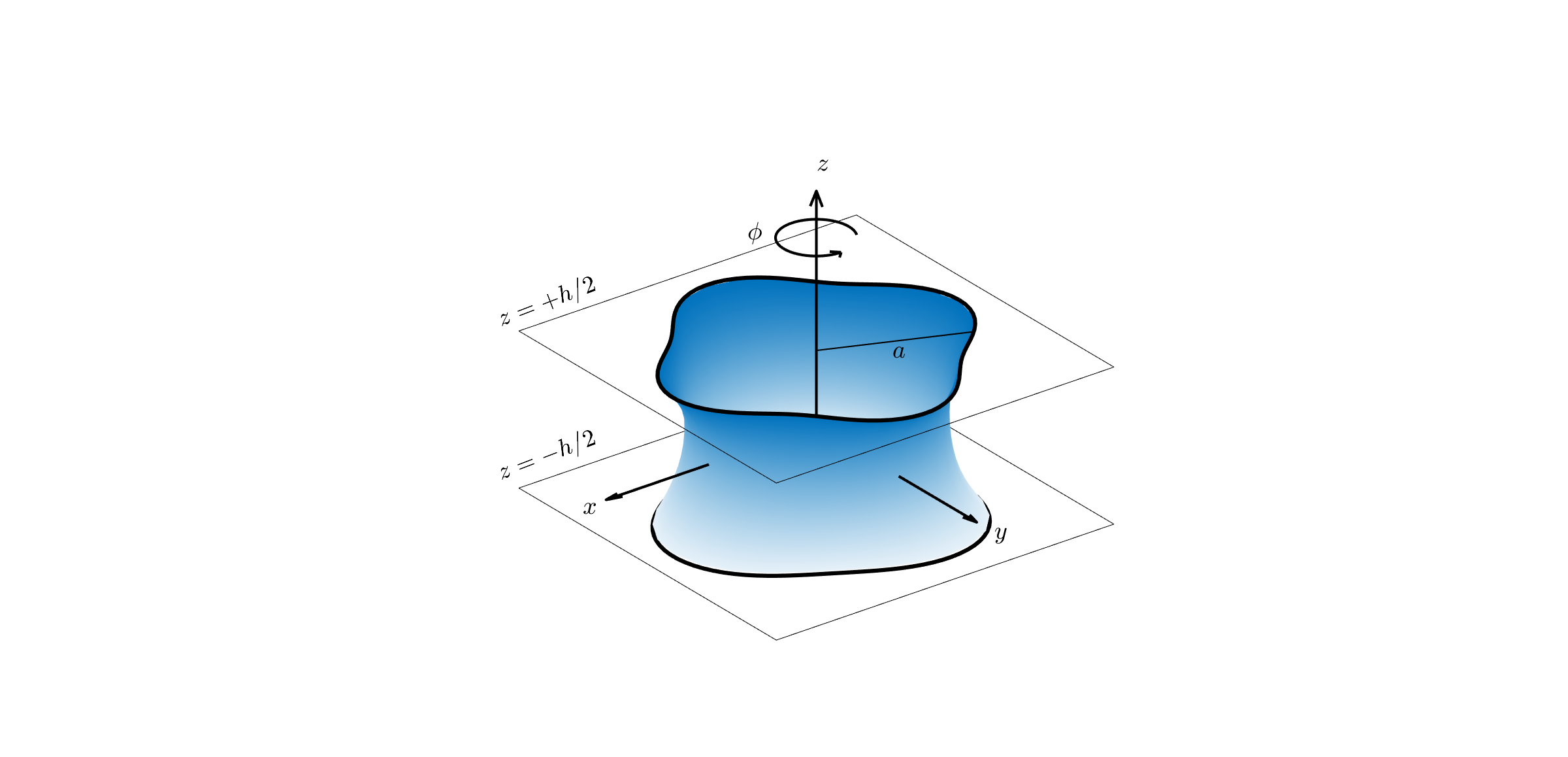}
(b)\includegraphics[trim={0.05cm 0.5cm 1.5cm 0.5cm},width=0.42\linewidth]{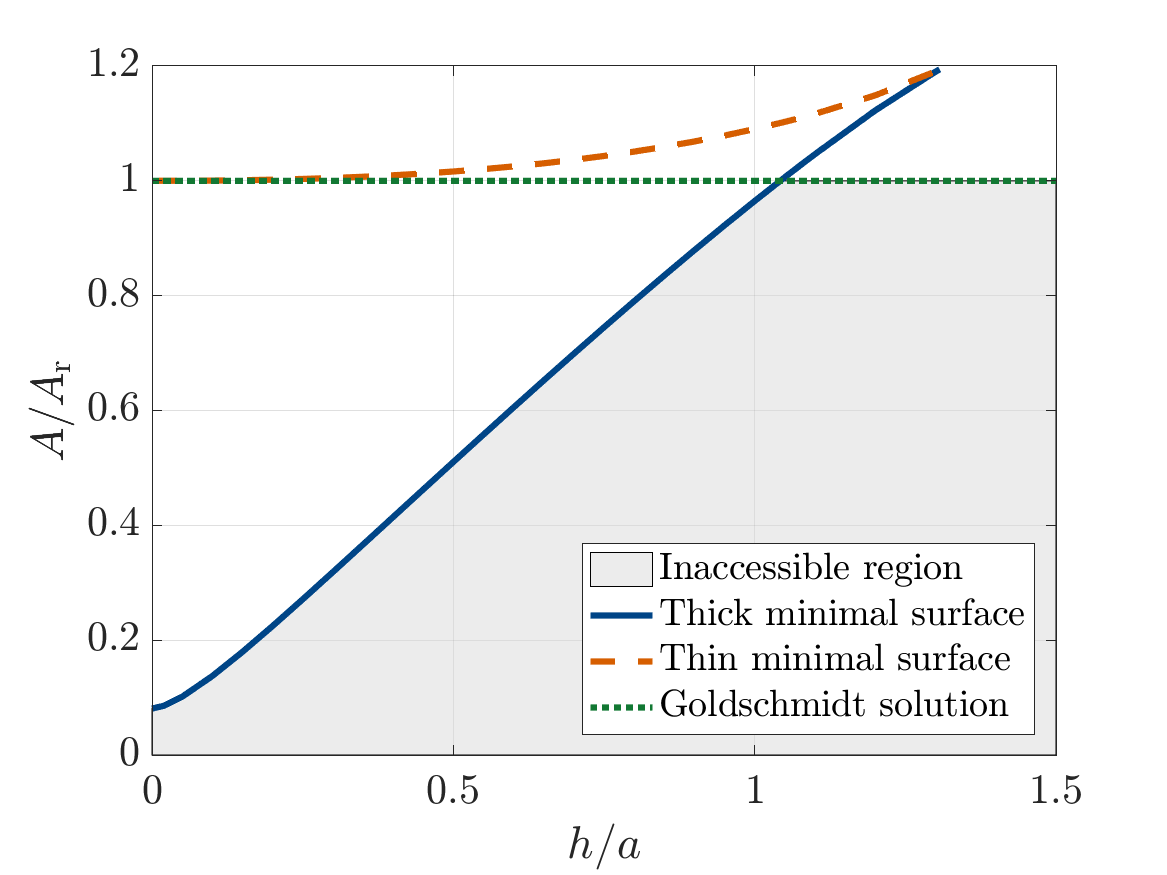}
(c)\begin{subfigure}[b]{0.14\textwidth}
\includegraphics[trim={2.5cm 2.5cm 2.5cm 2cm},clip,width=\linewidth]{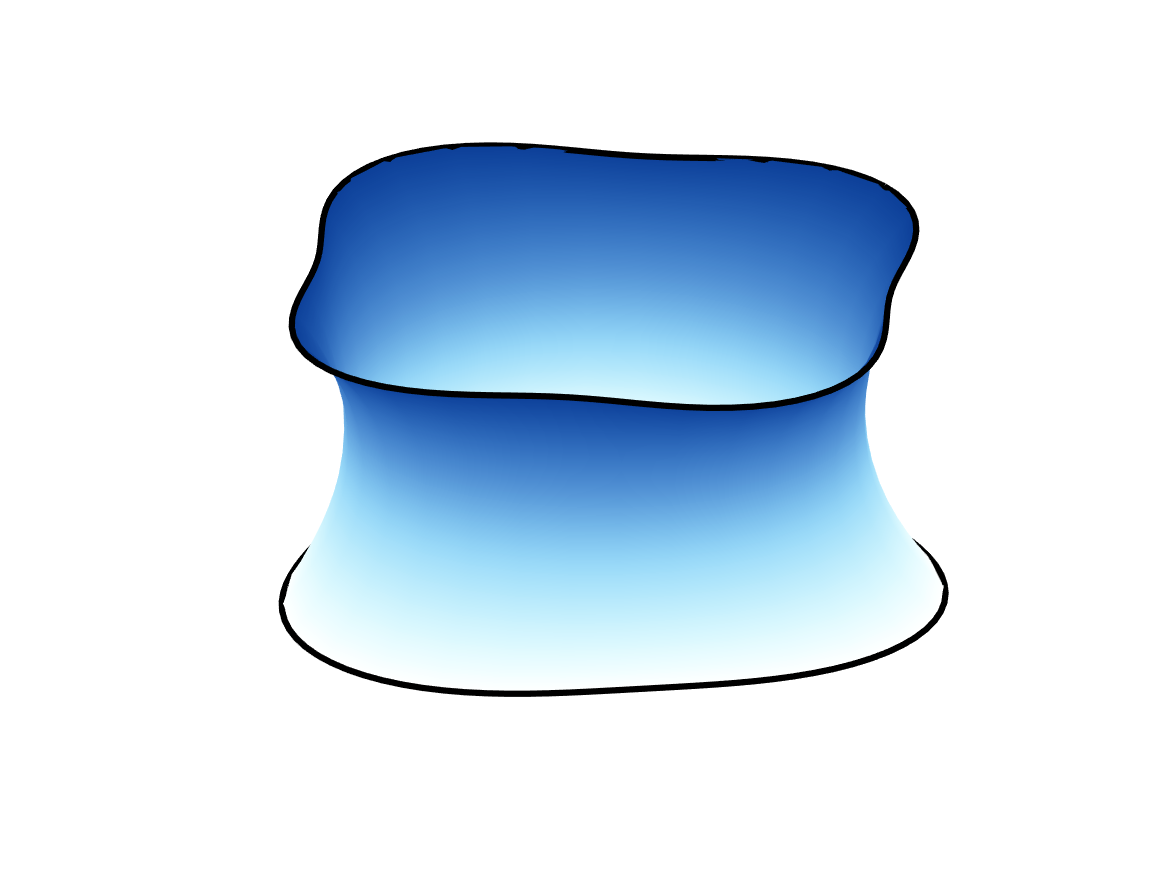}
\includegraphics[trim={2.5cm 2.5cm 2.5cm 2cm},clip,width=\linewidth]{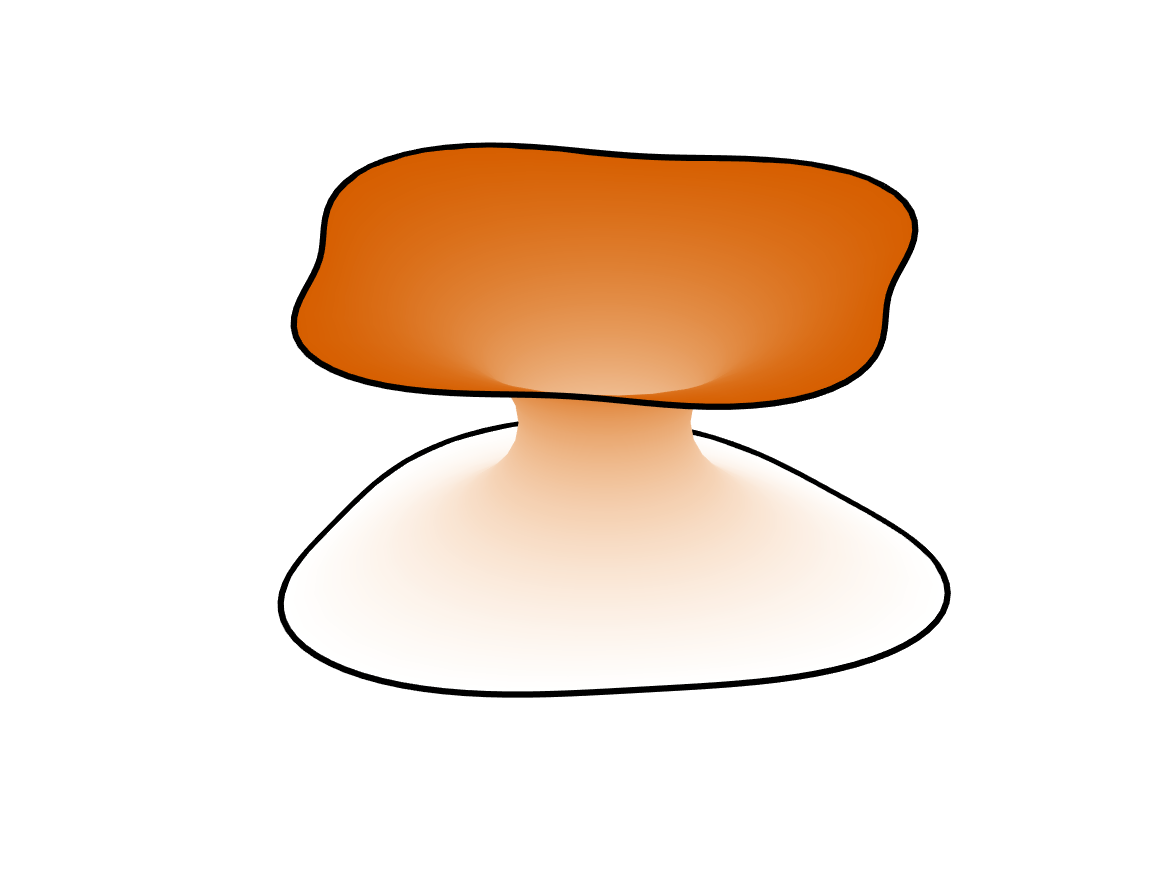}
\includegraphics[trim={2.5cm 2.5cm 2.5cm 2cm},clip,width=\linewidth]{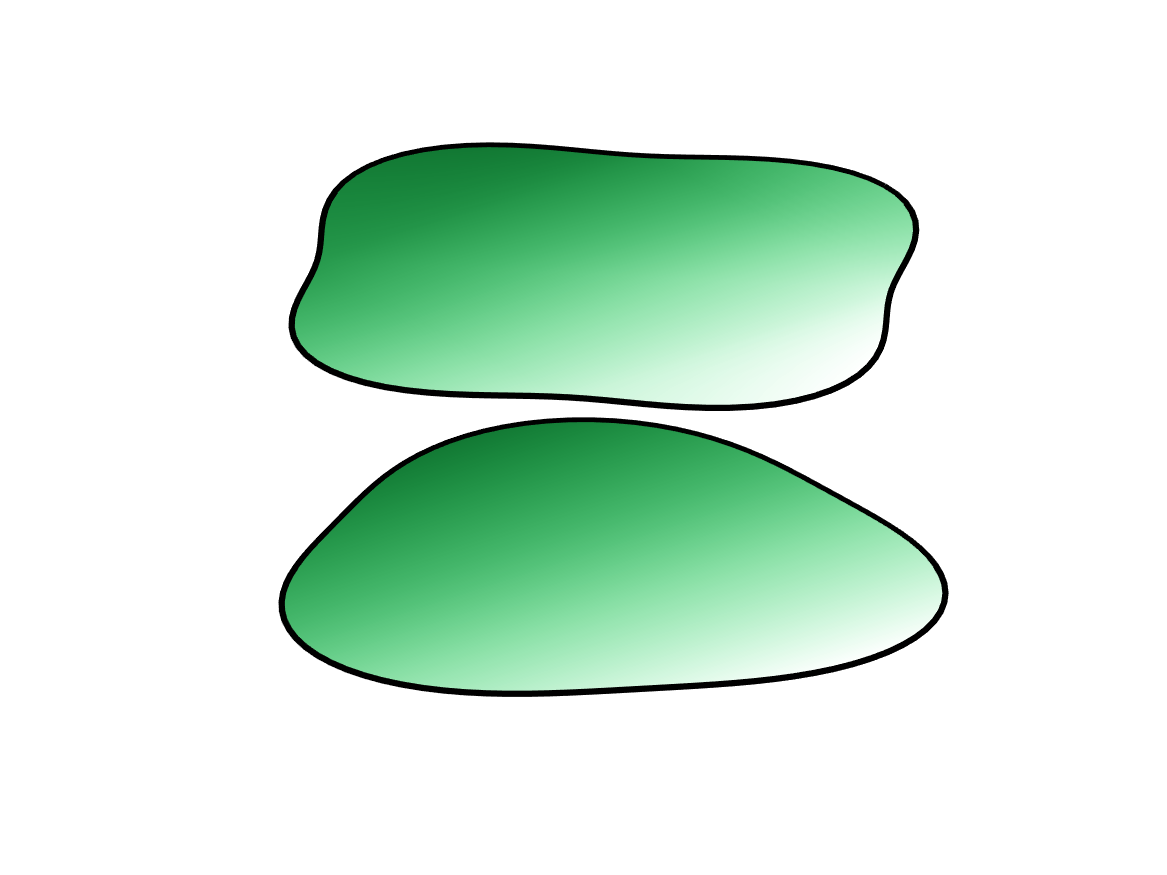}
\end{subfigure}
    \caption{(a) A minimal surface connecting two parallel planar rings with separation $h=a$. The surface is parameterized by an axial coordinate $z$ and azimuthal coordinate $\phi$. The boundaries are $r_+(\phi)=a(1+0.1\cos 4\phi)$, and $r_-(\phi)=a(1+0.1\cos 3\phi)$. (b) Area versus extension for minimal surfaces with these  boundary rings. No surfaces exist in the grey portion of the diagram. (c) The thick, thin, and Goldschmidt minimal surfaces for $h^*=a$.}
    \label{fig:geometry}
\end{figure}

We consider a surface $\mathcal{M}$ of prescribed area $A$ whose energy $E$ is given by the Willmore bending energy~\citep{willmore_riemannian_2023} plus an area term:
\begin{equation}
E = \frac{\kappa}{2}\int_\mathcal{M} \mathrm{d}A~(2H)^2 + \mu \int_{\mathcal{M}} \mathrm{d}A,
\label{helfrichE}
\end{equation}
where $H$ is the mean curvature, $\kappa$ is a bending modulus,  $\mathrm{d}A$ is the area element, and $\mu$ is the tension, a Lagrange multiplier enforcing the constraint of fixed surface area. %The integrals are surface integrals taken over  the surface $\mathcal{M}$ with area element $\mathrm{d}A$. 
%Variants of this energy also appear in other contexts under the name Helfrich energy~\citep{helfrich_elastic_1973}.
The Euler--Lagrange equation obtained by demanding the criticality of $E$ is
\begin{equation}
    \kappa (\Delta H -2HK + 2H^3) = \mu H,
    \label{ELeqn}
\end{equation}
where $K$ is the Gaussian curvature and $\Delta$ is the surface Laplace--Beltrami operator~\citep{tu_geometric_2004}. %Note that this equation is independent of $\bar\kappa$; this is a consequence of the Gauss--Bonnet theorem, which asserts that $\int \mathrm{d}A~K$ is equal to a boundary integral, up to a topological constant. % \fix{While the effects of nonzero $\bar\kappa$ are interesting~\cite{jia_axisymmetric_2021}, we nonetheless restrict ourselves to the nonsingular $\bar\kappa = 0$ case in this paper.} 
%For this article, we restrict ourselves to the case of $\bar\kappa = 0$. 
The associated boundary conditions are
\begin{equation}
    \boldsymbol{X}|_{\partial  \mathcal{M}}=\boldsymbol{Y} \quad \text{ and } \quad
    \kappa  (2H) |_{\partial \mathcal{M}} = 0,
    %\kappa  (2H) +  \bar\kappa K  |_{\partial \mathcal{M}} = 0,
    \label{notorqueBC}
\end{equation}
where $\boldsymbol{X}$ is the surface position (to be determined) and  $\boldsymbol{Y}$ is  the position of the boundary (assumed to be given), which may consist of multiple components. 
Physically, Eqn.~(\ref{notorqueBC}) represents the conditions of fluid--ring contact and  vanishing of the bending moment at the boundary, respectively. 
%We will consider surfaces embedded in three-dimensional Euclidean space whose boundaries are a distance $h>0$ apart. 
Using standard cylindrical coordinates $\{\rho,\phi,z\}$, %%with corresponding basis vectors $\{\boldsymbol{E}_\rho,\boldsymbol{E}_\phi,\boldsymbol{E}_z\}$, 
we consider surfaces of the form $\boldsymbol{X}(\phi,z) = (r(\phi,z)\cos \phi, r(\phi,z)\sin \phi, z)$,  $|z|<h/2$, with boundaries $\boldsymbol{Y}(\phi)=(r^\pm(\phi)\cos\phi, r^\pm(\phi) \sin\phi, \pm h/2)$, as illustrated in Fig.~\ref{fig:geometry}a. While this parameterization is unsuitable for modeling highly buckled shapes for which $r$ is not a function of $z$, it nonetheless suffices for shapes under small and intermediate compression. In this parameterization, 
\begin{equation}
    2H = \frac{1}{g^{3/2}} \left[{r(r_z^2+1)(r_{{\phi\phi}}-r)-2r_\phi^2-2rr_zr_\phi r_{z\phi} + rr_{zz}(r^2+r_\phi^2)}\right],% \approx -\frac{(r_{\phi\phi }- r)(r_z^2 +1 ) + r^2r_{zz}}{r^3 (r_z^2+1)^{3/2}} 
    \label{Heqn}
\end{equation}
\begin{equation}
    K = \frac{1}{g^2}\left[{rr_{zz}[r(r_{\phi\phi}-r)-2r_\phi^2] - (rr_{z\phi} - r_zr_\phi)^2}\right],
    \label{Keqn}
    %\approx \frac{r_{zz} (r_{\phi\phi}-r)}{r^2 (r_z^2 + 1)^2}
\end{equation}
\begin{equation}
    \Delta = \frac{1}{\sqrt{g}} \left[\frac{\partial }{\partial\phi} \left(\frac{1+r_z^2}{\sqrt{g}}\frac{\partial}{\partial \phi}- \frac{r_\phi r_z}{\sqrt{g}} \frac{\partial}{\partial z}\right) + \frac{\partial}{\partial z} \left(- \frac{r_\phi r_z}{\sqrt{g}} \frac{\partial}{\partial \phi} + \frac{r^2+r_\phi^2}{\sqrt{g}}\frac{\partial}{\partial z}\right) \right],
    \label{lapbel}
\end{equation}
where $r_z=\partial r/\partial z$, $r_\phi = \partial r/\partial \phi$, etc., 
and $g = r^2(1+r_z^2)+r_\phi^2$.   Substituting Eqns.~(\ref{Heqn})--(\ref{lapbel}) into Eqn.~(\ref{ELeqn}) then yields a fourth-order PDE for the unknown function $r$. %While the general solution to this PDE is complicated, we recognize that the two-boundary configuration studied here is very reminiscent of two classic problems in the calculus of variations, whose solutions offer valuable guiding intuition. We briefly review these problems in the next section.
External forces and torques at the boundary rings are generally needed in order to impose Eqn.~(\ref{notorqueBC}a).  These quantities are necessarily constant, as follows from Noether's theorem: since the integrand of the energy functional Eqn.~(\ref{helfrichE}), $\mathcal{E} = [(\kappa/2)(2H)^2 + \mu] \sqrt{g}$, does not depend on $z$ or $\phi$ explicitly, conserved quantities associated with these variables exist. We find that
\begin{equation}
    F=\oint \mathrm{d}\phi ~ \left[\mathcal{E} - r_z\frac{\partial\mathcal{E}}{\partial r_z}-r_{zz}\frac{\partial\mathcal{E}}{\partial r_{zz}} + r_z \frac{\partial}{\partial z}\left(\frac{\partial\mathcal{E}}{\partial r_{zz}}\right) - r_{z\phi} \frac{\partial \mathcal{E}}{\partial r_{z\phi}}\right]
    \label{Feqn}
\end{equation}
is the conserved quantity associated with $z$ and
\begin{equation}
    T=-\oint \mathrm{d}\phi ~ \left[r_\phi\frac{\partial\mathcal{E}}{\partial r_z}+r_{z\phi}\frac{\partial\mathcal{E}}{\partial r_{zz}} - r_\phi \left(\frac{\partial\mathcal{E}}{\partial r_{zz}}\right)_z - r_{\phi} \frac{\partial}{\partial \phi}\left(\frac{\partial \mathcal{E}}{\partial r_{z\phi}}\right) \right]
    \label{Teqn}
\end{equation}
is the conserved quantity associated with $\phi$~\citep{gelfand_calculus_2000}. Based on dimensions %($\mathcal{E}$ has dimensions of energy per length, i.e. force) 
and the expressions in the axisymmetric limit, we might expect $F$ to represent the axial force and $T$ to represent the torque about the $z$-axis; the principle of virtual work can be used to confirm this (see Supplementary Material and also~\citep{walzel_perturbing_2022} for the $\kappa=0$ case). %We remark that the above expressions are connected to more general principles of momentum and pseudomomentum conservation in mechanical systems that exhibit translation symmetry within a medium, as discussed in references such as~\citet{dharmavaram_shear_2025}. 

%The main takeaway of this article is to show that, broadly speaking, the critical surfaces of the functional Eqn.~(\ref{helfrichE}) can be understood as amalgamations of the solutions to two classic problems in the calculus of variations. In the case of $\kappa=0$, one can see that unconstrained shapes will tend to be minimal surfaces: $H=0$, like for a soap film. However, the presence of external confining forces, in conjunction with the area constraint, introduces the possibility of buckled solutions, analogous to how compressing the ends of a slender inextensible elastic rod gives rise to the Euler buckling instability. We briefly review these two problems before proceeding to the general case. %Consequently, our shapes can have both a maximal extension and an infinitely large solution space of buckled configurations. There is also the possibility of fluid tethers when the surface area exceeds the area spanned by the two boundary rings.

% \subsection{Relation to the Plateau--Douglas and elastica problems}
% \label{sec:plateau}
While the general solution to Eqn.~(\ref{ELeqn}) is complicated, we recognize that the two-boundary configuration studied here is very reminiscent of two classic problems in the calculus of variations, whose solutions offer valuable guiding intuition.
Noting that the trivial solution $H=0$ satisfies both the shape equation Eqn.~(\ref{ELeqn}) and the no-bending-moment boundary condition Eqn.~(\ref{notorqueBC}b), and assuming that this shape is compatible with the fixed area constraint, we see that our problem becomes the problem of Plateau (or Douglas)~\citep{nitsche_plateaus_1974}: what is the surface of minimal area that has a given boundary? 
%Historically, this question kicked off the mathematical study of  minimal surfaces, which have vanishing mean curvature and are realized in everyday life by fixed-tension soap films and other simple fluid interfaces~\citep{nitsche_plateaus_1974}. %One anticipates that the similarity between our problem and the Plateau--Douglas problem implies that minimal surfaces are related to the set of possible solutions.
%to find a minimal surface given a fixed boundary. 
% Our question can be cast as an extension of this problem whose
% solutions are realized by membranes. %: %\emph{what is the shape assumed by a fluid surface with a resistance to bending that has a given boundary?} %One anticipates that the similarity between our problem and the Plateau problem implies that minimal surfaces are related to the set of possible solutions.
%Thus, we anticipate that the solutions of Eqn.~(\ref{ELeqn}) are related to minimal surfaces.
The answer to this question is well known when the boundary consists of parallel concentric circles of radius $a$: in this case, the only connected surfaces with $H=0$ (i.e., minimal surfaces) are catenoids and the Goldschmidt solution. The former are surfaces of revolution with meridians of the form $r=b\cosh(z/b)$, $0<b<1$, while the latter is simply two discs connected center to center by a line. Catenoids cease to be a solution when $h/a>1.3255...$, as the transcendental equation $b\cosh(h/(2b))=a$, which encodes the boundary condition of fluid--boundary contact, has no solution for $b$. For $h/a$ less than this value, there are actually two possible catenoids, a thick one (larger $b$) and a thin one (smaller $b$). Substituting $r=b\cosh (z/b)$ into Eqns.~(\ref{Feqn}) and (\ref{Teqn}) shows that $F=2\pi \mu b$ and $T=0$ for catenoids. %The area of this catenoid of maximal extension is $2\pi(1.1997...)a^2$.

Generalizations of the Plateau--Dogulas problem to noncircular boundaries have been studied~\citep{walzel_perturbing_2022,alimov_soap_2021}, and many properties of the axisymmetric case carry over.
%, and many of the properties of catenoids carry over, particularly the existence of a maximal extension that depends on the exact boundary shape and the existence of two non-Goldschmidt minimal surfaces when $h$ is less than this maximal extension. 
A plot of the solution area as a function of extension for a particular choice of noncircular boundary is shown in Fig.~\ref{fig:geometry}b, and visualizations of the three  minimal surfaces for $h=a$, analogous to the two types of catenoid and Goldschmidt solution, are shown in Fig.~\ref{fig:geometry}c. In Fig.~\ref{fig:geometry}b, the solid blue line represents a family of relatively ``thick'' surfaces that are global minimizers of area. There is also a branch of ``thin''  solutions (dashed red line) that are theoretically feasible with the same $h$. However, these surfaces are not area-minimizing: such films are unstable to perturbations and thus do not occur in typical experiments. The last possible branch is the nonaxisymmetric Goldschmidt solution (dotted green line), whose area is  $A_\mathrm{r}$, the total area spanned by the two rings. Beyond the maximal extension for which the thick and thin branches intersect, this Golschmidt solution is the only possible one.

%A recent series of papers by Alimov et al.~\citep{alimov_soap_2021,alimov_skew_2022,alimov_minimal_2020} exploits an analogy with Chaplygin flow to analytically calculate these shapes when the boundaries have special symmetries. \citet{walzel_perturbing_2022} used experiments and perturbation theory to study minimal surfaces formed from noncircular boundaries.

Our problem can additionally be viewed as a higher-dimensional analogue of the classical elastica problem, in which an inextensible, one-dimensional rod buckles when compressed at its ends. This straight-to-buckled transition is the quintessential example of an elastic  instability~\citep{holmes_elasticity_2019}. If the rod remains planar, its energy can be expressed as $(B/2 )\int\mathrm{d}s~k^2  +  \lambda  \int\mathrm{d}s$, where $k$ is the curvature of the rod, $B$ is its bending stiffness, $\mathrm{d}s$ is the arc length element, and $\lambda$ is the tension that preserves total length. Note the resemblance between this energy and Eqn.~(\ref{helfrichE}): the first term is an integral of extrinsic curvature squared, and the second imposes a constant geometric measure. 
%The deformation of an elastica from its initially straight configuration into a bent shape when a critical load is attained is the quintessential example of a buckling instability~\citep{holmes_elasticity_2019}. 
Analysis of the linearized Euler--Lagrange equation of the elastica bending energy reveals an eigenvalue problem with infinitely many sinusoidal eigenfunctions, each requiring a compressive force $F_n=B(n\pi/\varsigma)^2$, where $n=1,2,\hdots$ is the mode number and $\varsigma$ is the rod length. In  the same way, an elastic fluid interface sandwiched between two rings can be expected to exhibit infinitely many oscillatory modes, each requiring a compressive axial force.

\section{Results: Circular Boundaries}
\label{sec:results}

%\subsection{Minimal surface solutions and maximal extension}

%The first fundamental observation regarding solutions to the membrane equation Eqn.~(\ref{ELeqn}) is that minimal surfaces, i.e., surfaces for which $H=0$, trivially satisfy this equation and the no-torque boundary condition Eqn.~(\ref{notorqueBC}). Thus, the Helfrich problem reduces to the Plateau problem, provided that a minimal surface with the given area exists for the given boundary ring shape.

%As discussed in the previous section, in the Plateau--Douglas problem, rings with a separation of $h$ will lead to one of three minimal surface solutions. 

%As discussed in Section~\ref{sec:plateau}, the Plateau problem has a critical separation $h^*$, dependent on the shape of the boundary, beyond which no smooth nondegenerate minimal surfaces exist. This maximal separation $h^*$ is dependent on the ring shape and the area $A_0$, as shown in Fig.~\ref{fig:geometry}b. This maximal extension has consequences for our problem as well.
We now consider the shape of a fixed-area fluid interface with out-of-plane bending resistance as the rings are pulled apart. In the phase diagram of Fig.~\ref{fig:geometry}b, this action corresponds to moving horizontally to the right. %There are two cases to  consider, depending on the value of $\bar{A}=A/A_\mathrm{r}$, where $A_{\mathrm{r}}$ is the area spanned by the two boundary rings.
If $\bar{A}=A/A_\mathrm{r}$ is less than or equal to 1, then moving to the right will eventually take one to the blue branch of thick minimal surfaces. The rings cannot be pulled any farther since this would entail entering the grey region of impossible surfaces with area less than that of a thick minimal surface. Therefore, our surfaces must have a maximum extension of $h^*(A)$, the extension corresponding to a thick minimal surface of area $A$. Furthermore, because $H=0$ is a solution to the shape equation, the interface shape for $h=h^*$ will be precisely the minimal surface solution to the Plateau--Douglas problem. %This simple solution serves as a natural base case for a perturbation theory, which we examine next.

%In the case where $\bar\kappa \neq 0$, $H=0$ continues to satisfy the bulk equation but no longer satisfies the no-torque boundary condition. Incompatibility with boundary conditions yields a singularity. The resolution of this paradox is that as the rings are pulled toward the maximal extension, the mean curvature will either develop a boundary layer near the rings or will oscillate rapidly to satisfy $H=0$ in an averaged sense. Some numerical results for this case are given in the Supplementary Material. In principle, this case can be analyzed with singular perturbation theory, as~\citet{jia_axisymmetric_2021} did for the axisymmetric case.

%\subsection{Linearized eigenvalue problem}
\label{sec:linevp}

\begin{figure}
    \centering
    (a)\includegraphics[width=0.44\linewidth]{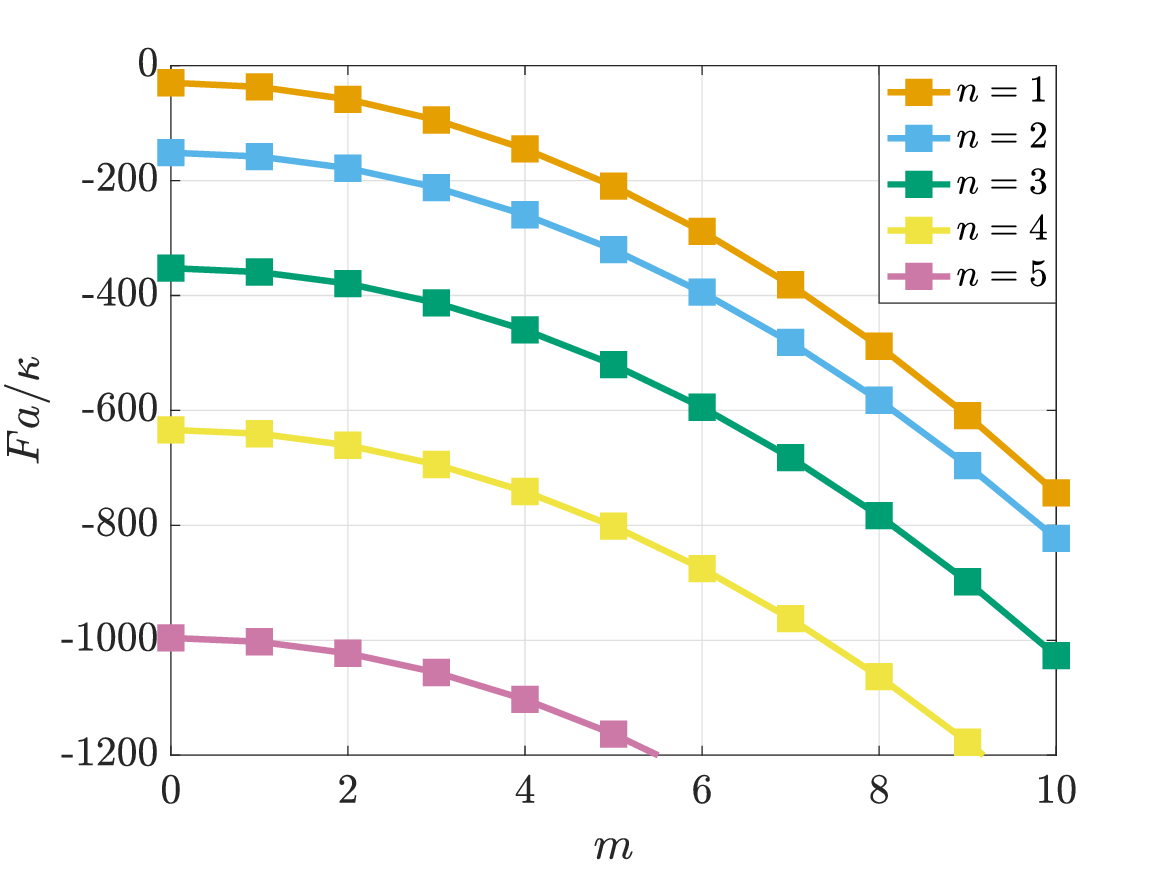}
    (b)\includegraphics[trim = 35pt 10pt 55pt 10pt, clip,width=0.46\linewidth]{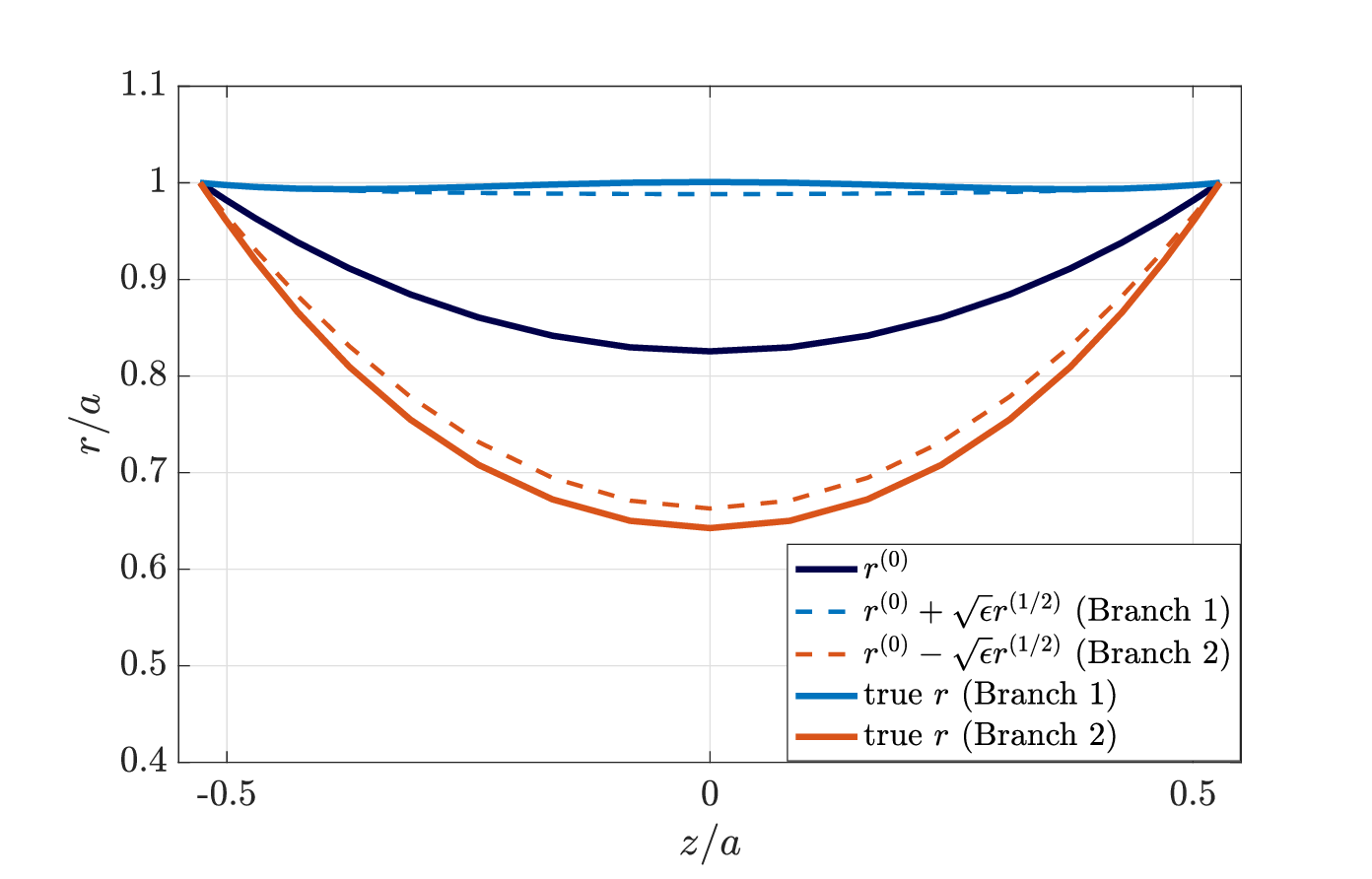}
    \caption{(a) Axial force for the catenoid ($\bar{A}=1$) as a function of azimuthal mode number $m$ for various axial mode numbers $n$. (b) Base (catenoid) and two area-preserving $O(\sqrt{\epsilon})$ perturbation solutions for $n=1,m=0$ ($h^*/a=1.0554...$, $\epsilon = 0.05 h^*/a$).}
    \label{fig:f_vs_m}
\end{figure}

%Minimal surfaces of appropriate area, if they exist, are solutions to the shape equation Eqn.~(\ref{ELeqn}) for $h=h^*$, but  
It remains to calculate the force needed to hold these minimal surfaces in equilibrium. %However, in this equation, the tension is a Lagrange multiplier and the area is fixed.
%What is the surface tension that allows a minimal surface to be held? 
To do this, we expand upon an idea  from~\citep{jia_axisymmetric_2021} and %perturb about a base minimal surface $r^{(0)}$. %We take a small perturbation $r^{(1)}$ about the minimal surface base state $r^{(0)}$. 
define a dimensionless parameter $\epsilon = |h-h^*|/a$; this parameter is small when the fluid interface is stretched to nearly its maximal extension. We expand $r=r^{(0)} + \sqrt{\epsilon} r^{(1/2)} + \epsilon r^{(1)}+\hdots$ and $H$, $K$, $\Delta$ and $\mu$ similarly. If the base quantity $r^{(0)}$ is a minimal surface,  using these representations and $H^{(0)} =0 $ in Eqns.~(\ref{ELeqn}) and (\ref{notorqueBC}b) yields the  $O(\sqrt{\epsilon})$ eigenvalue problem
\begin{equation}
    -\kappa \mathcal{J}^{(0)} H^{(1/2)} = \mu^{(0)}H^{(1/2)}, \quad H^{(1/2)}(\phi,z=\pm h/2) =0
    \label{evproblem}
\end{equation}
for $H^{(1/2)}$. Here, $\mathcal{J}^{(0)}=-\Delta^{(0)}+2K^{(0)}$ is the Jacobi operator of the base minimal surface; this eigenvalue problem is entirely analogous to the ones obtained when analyzing buckling instabilities of elastic rods, plates, and shells~\citep{love_treatise_1990}. %The Jacobi operator characterizes the stability of critical surfaces of the area functional. 
In the special case of a catenoid base state $r^{(0)} = b\cosh (z/b)$,  $K^{(0)}=-2/\cosh^4(z/b)$, and $   \Delta^{(0)} = \text{sech}^2(z/b) (\partial_z^2 + b^{-2} \partial_\phi^2)$.
Separation of variables can then be used: letting $H^{(1)} = F_m(z)G_m(\phi)$, we find that $    G_m(\phi) = A_m \cos m\phi + B_m \sin m\phi$, 
where $A_m$ and $B_m$ are constants, $m$ is an integer, and $F_m(z)$ solves
\begin{equation}
    \frac{1}{\cosh^2 \frac{z}{b}}
 \left(\partial_z^2 - \frac{m^2}{b^2}+ \frac{2}{b^2 \cosh^2 \frac{z}{b}}\right) F_m = \frac{\mu^{(0)}}{\kappa} F_m, \quad F_m(z=\pm h/2)=0.
\label{evp}
\end{equation}
% \begin{equation}
%     \left[\frac{1}{(L^{(0)})^2 r^{(0)}} \partial_t r^{(0)}\partial_t - 2K^{(0)}- \frac{m^2}{(r^{(0)})^2}\right]F_m = \frac{\mu^{(0)}}{\kappa} F_m
%     \label{evp}
% \end{equation}
The modes are thus characterized by ordered pairs of integers $(n,m)$,  $n\geq 1$, $m\geq 0$. The permissible values of the tension (scaled by $\kappa$) are the eigenvalues of $-\mathcal{J}^{(0)}$; recalling that $F=2\pi \mu b$ for a catenoid of neck $b$ gives the external force for each mode,  illustrated in Fig.~\ref{fig:f_vs_m}a. %We will see that these eigenfunctions are connected to the solution branches of the problem at arbitrary extension. 
When the boundary is circular, the cosine and sine eigenfunctions  are simply rotations of each other; thus, the eigenvalues corresponding to a given $n$ and $m$ are doubly degenerate (except for $m=0$, where the sine mode is spurious).
We remark that while Eqn.~(\ref{evp}) lacks a closed  solution in terms of common functions, it closely resembles the associated Legendre equation after the transformation $\eta =\tanh (z/b)$, admitting a simple Frobenius series solution in $\eta$. % so that the eigenvalues $\mu^{(0)}/\kappa$ can be approximated as roots of a truncated polynomial in $\eta$. 
However, we do not find this representation enlightening, so we will not pursue it here, relying instead on a numerical approach. %for finding $H^{(1/2)}$ and $r^{(1/2)}$. 

As an eigenfunction, $H^{(1/2)}$ is only determined up to a multiplicative constant, and thus so is $r^{(1/2)}$. To find this constant, we expand the area constraint:
\begin{equation}
   0= \sqrt{\epsilon}  \oint \mathrm{d}\phi \int_{-h/2}^{h/2}\mathrm{d}z~\sqrt{g}^{(1/2)} + \epsilon\left[- 2\left.\sqrt{g}^{(0)}\right|_{z=\frac{h}{2}}+ \oint \mathrm{d}\phi \int_{-h/2}^{h/2}\mathrm{d}z~\sqrt{g}^{(1)} \right].
   \label{areaconstraint}
\end{equation}
Expanding Eqn.~(\ref{notorqueBC}a) to $O(\epsilon)$ yields the boundary conditions $r^{(0)}=a$, $r^{(1/2)}=0$, and $r^{(1)}= \pm r_z^{(0)}$ at $z=\pm h/2$. Because the base solution is a minimal surface, the first variation of area vanishes in the bulk, so the $O(\sqrt{\epsilon})$ $z$-integral reduces to boundary terms proportional to $r^{(1/2)}$ at $z=\pm h/2$. Since $r^{(1/2)}=0$ at the boundary, the $O(\sqrt{\epsilon})$ term vanishes completely, and the area constraint is automatically satisfied at this order. We thus must examine the $O(\epsilon)$ terms to determine $H^{(1/2)}$.
To this end, we compute $\sqrt{g}^{(0)} |_{z=h/2}= b\cosh^2({h}/{2b}) = 1/b$ and
\begin{equation}
    %\sqrt{g}^{(1)} = \cosh \frac{z}{b} r^{(1)} + b\sinh \frac{z}{b}r_z^{(1)} + \frac{({r^{(1/2)}_\phi})^2}{2b\cosh^2 \frac{z}{b}}  + \tanh\frac{z}{b} r^{(1/2)}r_z^{(1/2)} + \frac{b}{2}\text{sech}^2 \frac{z}{b} (r_z^{(1/2)})^2
    \sqrt{g}^{(1)} = \cosh \frac{z}{b} r^{(1)} + b\sinh \frac{z}{b}r_z^{(1)} + \frac{({r^{(1/2)}_\phi})^2}{2b\cosh^2 \frac{z}{b}}  + \tanh\frac{z}{b} r^{(1/2)}r_z^{(1/2)} + \frac{b(r_z^{(1/2)})^2}{2 \cosh^2 \frac{z}{b}} .
\end{equation}
The $r^{(1)}$ terms again form a total derivative and their $z$-integral thus also reduces to boundary terms. As a result, Eqn.~(\ref{areaconstraint}) becomes
\begin{equation}
    4\pi b \sin^2 \frac{h}{2b} + \oint \mathrm{d}\phi\int_{-h}^h \mathrm{d}z\left[\frac{({r^{(1/2)}_\phi})^2}{2b\cosh^2 \frac{z}{b}}  + \tanh\frac{z}{b} r^{(1/2)}r_z^{(1/2)} + \frac{b(r_z^{(1/2)})^2}{2 \cosh^2 \frac{z}{b}} \right] = \frac{4\pi}{b},
\end{equation}
which determines the normalization constant. Notably, since $r^{(1/2)}$ and its derivatives appear in pairwise products, both negative and positive constants are valid. This suggests that there are two branches emanating from each eigensolution as $h$ is reduced from $h^*$. %, one with more positive $r$ values and one with more negative $r$ values. 
Figure~\ref{fig:f_vs_m}b shows these  branches for the $(n,m)=(1,0)$ catenoid of area $\bar{A}=1$ and their approximations. %The above argument holds for any $(n,m)$.% but by symmetry, the two branches are reflections of each other when $n$ is even. 

% \begin{figure*}
%     \centering  \includegraphics[width=0.8\linewidth]{figures/catenoid_modes.png}
%     \caption{Eigenmodes of a catenoid. Each row is one value of $n$, and each column is one value of $m$. Eigenmode amplitudes are scaled differently in each shape for ease of visualization.}
%     \label{fig:catenoidmodes}
% \end{figure*}

\begin{figure}
  \centering
  \begin{tabular}{c c c c c c c c}
  
    % --- COLUMN HEADERS ---
   & & \textbf{$m=0$} & \textbf{$m=1$} & \textbf{$m=2$} & \textbf{$m=3$} & \textbf{$m=4$}\\

    % --- ROW 1 ---
    & \rotatebox{90}{\textit{ 
    \quad Branch 1}} &
    \begin{subfigure}[b]{0.16\textwidth}
      \includegraphics[trim = 100pt 70pt 100pt 15pt, clip,width=\textwidth]{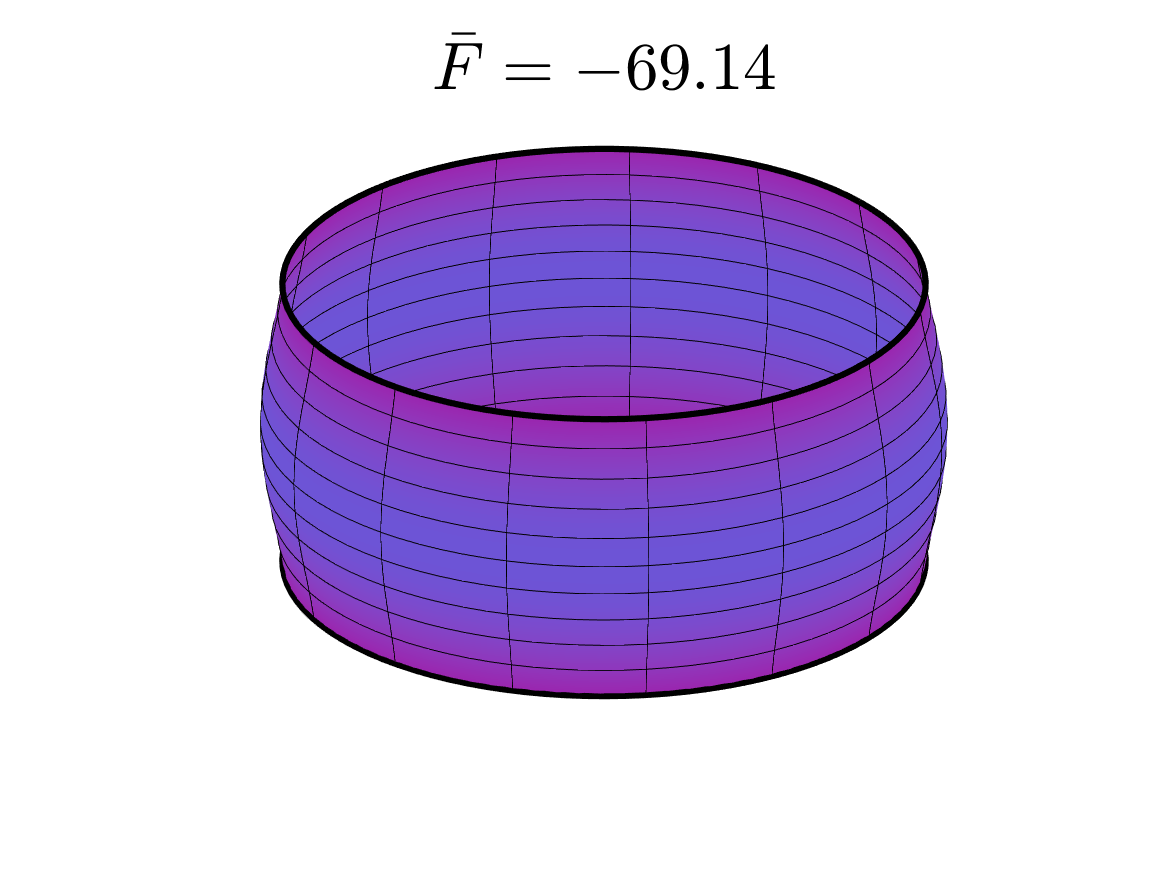}
    \end{subfigure} &
    \begin{subfigure}[b]{0.16\textwidth}
      \includegraphics[trim = 100pt 70pt 100pt 15pt, clip,width=\textwidth]{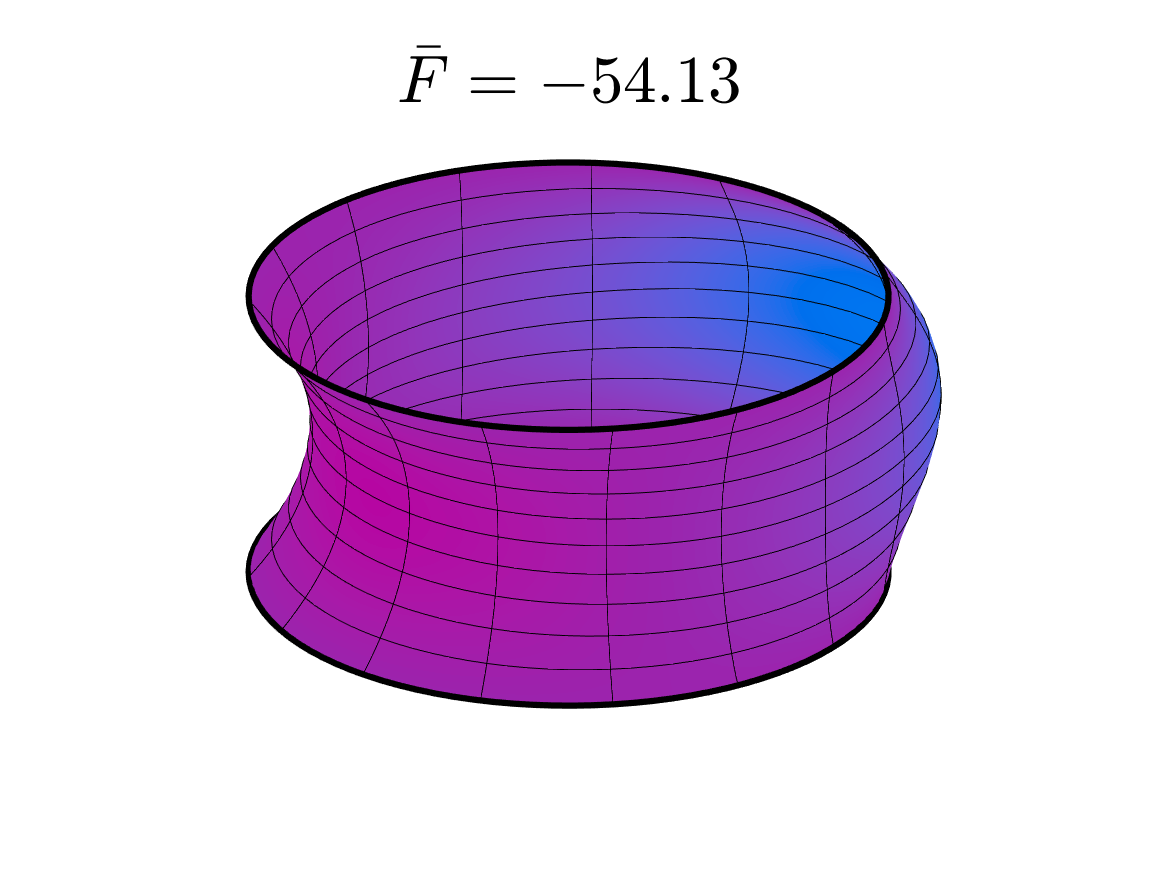}
    \end{subfigure} &
    \begin{subfigure}[b]{0.16\textwidth}
      \includegraphics[trim = 100pt 70pt 100pt 15pt, clip,width=\textwidth]{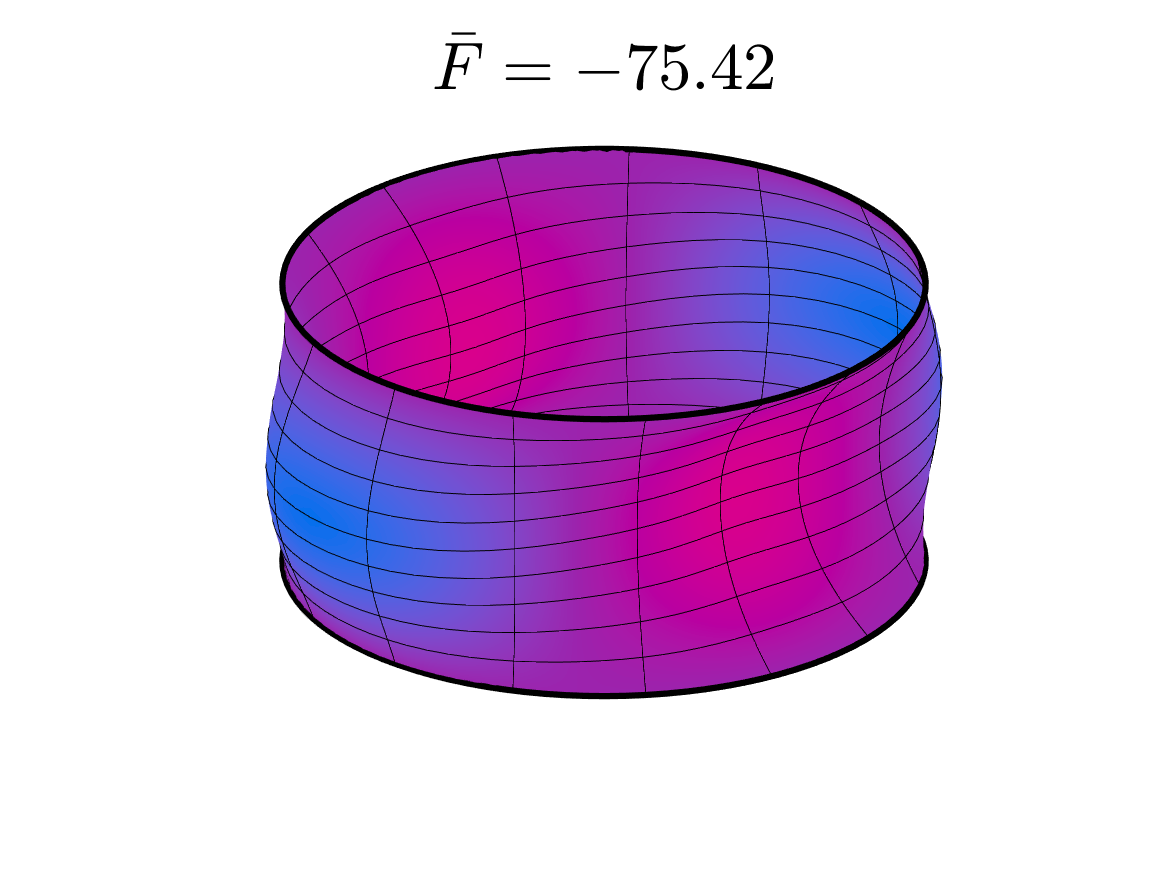}
    \end{subfigure} &
    \begin{subfigure}[b]{0.16\textwidth}
      \includegraphics[trim = 100pt 70pt 100pt 15pt, clip,width=\textwidth]{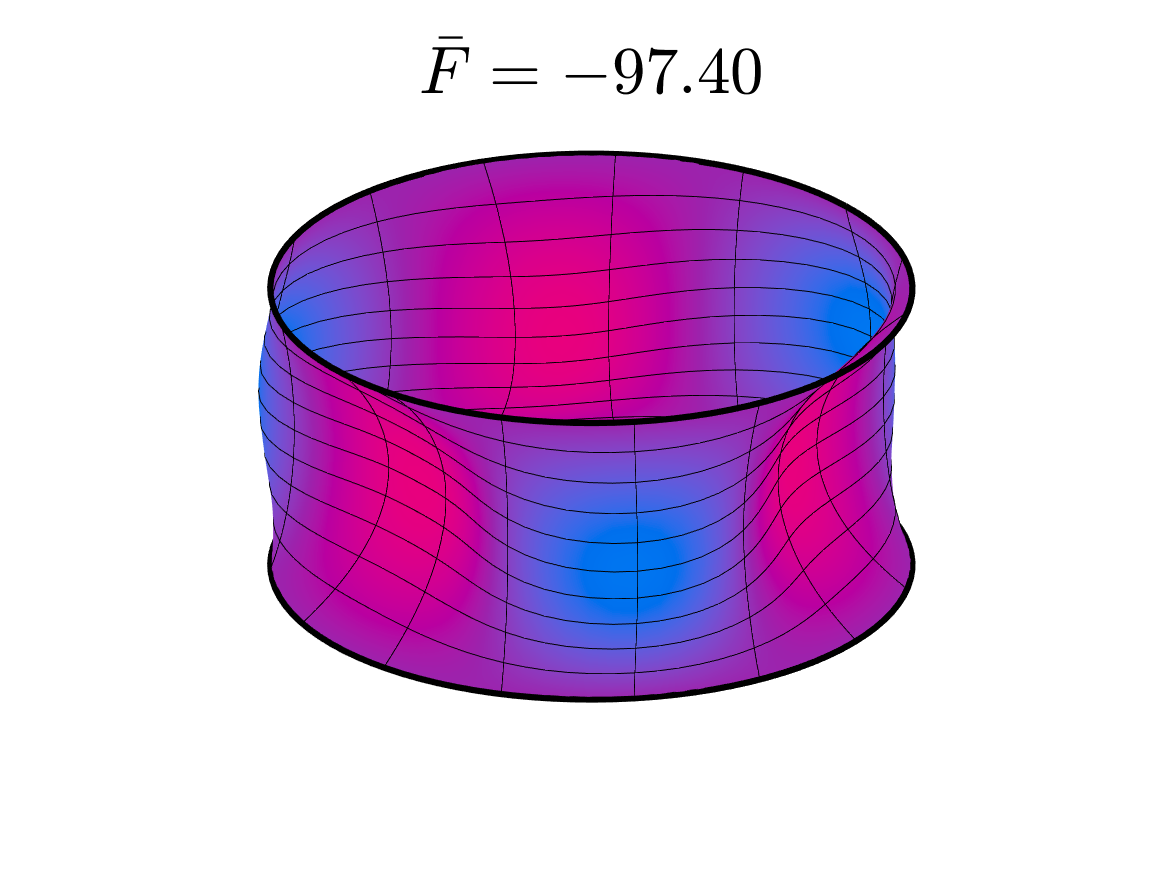}
    \end{subfigure} & \begin{subfigure}[b]{0.16\textwidth}
      \includegraphics[trim = 100pt 70pt 100pt 15pt, clip,width=\textwidth]{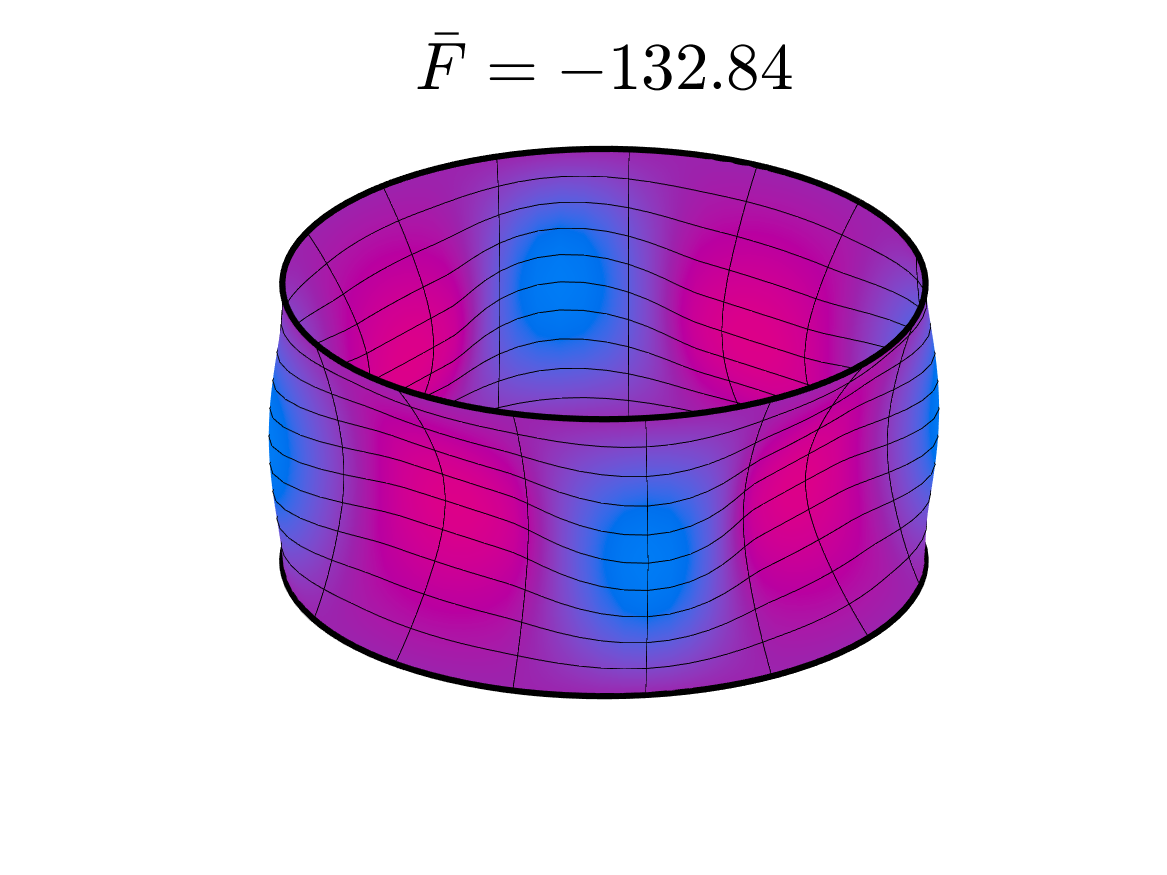}
    \end{subfigure}&
    \multirow{3}{*}{\begin{subfigure}[b]{0.04\textwidth}
        \includegraphics[trim = 475pt 10pt 35pt 30pt, clip,width=\textwidth]{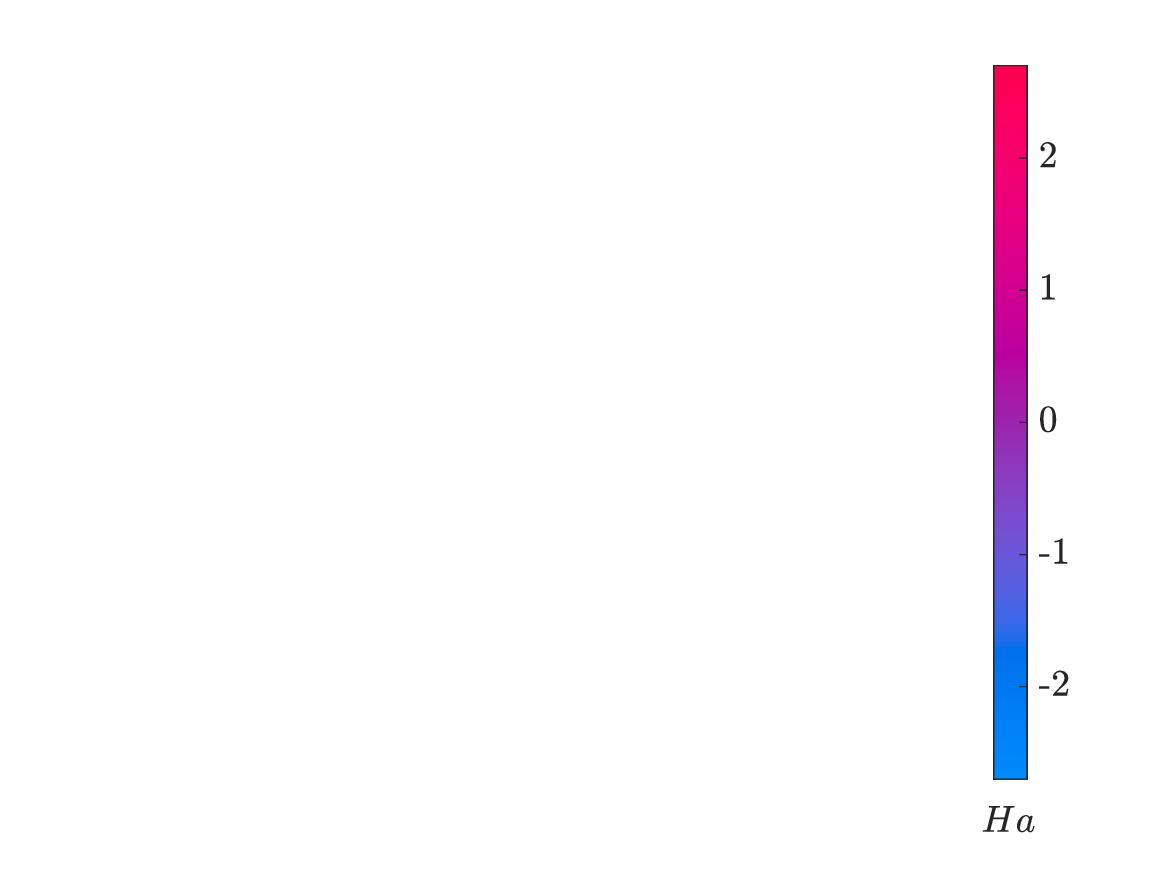}
    \end{subfigure}} \\

    % --- ROW 2 ---
    \rotatebox{90}{\quad \quad $n=1$}&\rotatebox{90}{\textit{ Eigensurfaces}} &
    \begin{subfigure}[b]{0.16\textwidth}
      \includegraphics[trim = 100pt 70pt 100pt 15pt, clip,width=\textwidth]{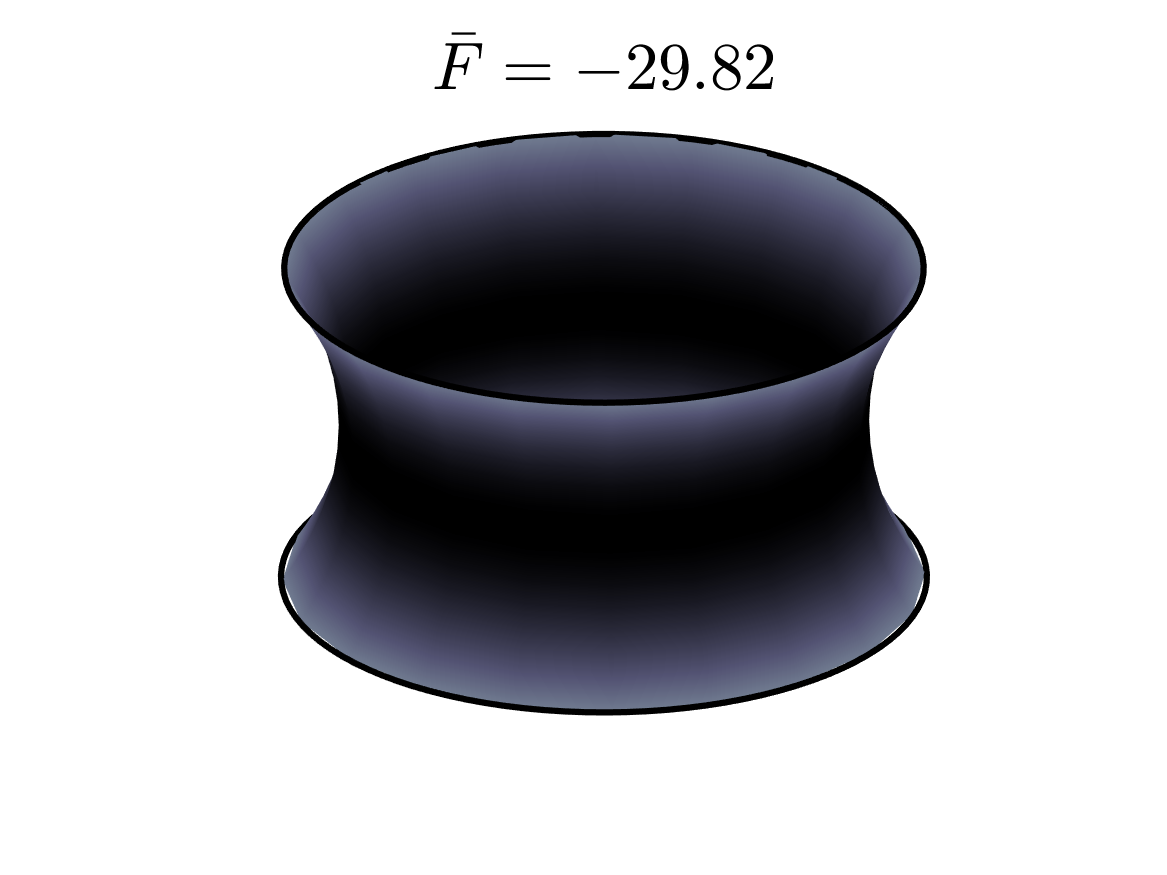}
    \end{subfigure} &
    \begin{subfigure}[b]{0.16\textwidth}
      \includegraphics[trim = 100pt 70pt 100pt 15pt, clip,width=\textwidth]{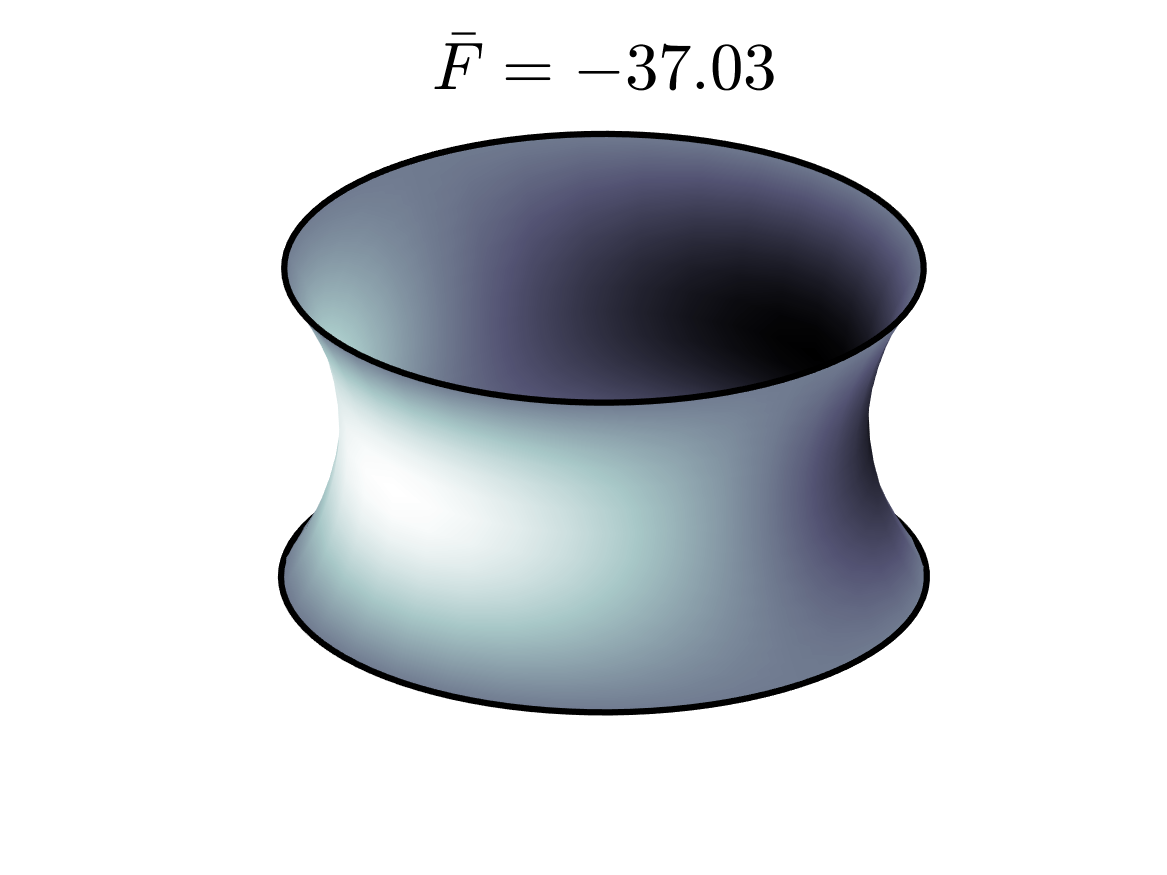}
    \end{subfigure} &
    \begin{subfigure}[b]{0.16\textwidth}
      \includegraphics[trim = 100pt 70pt 100pt 15pt, clip,width=\textwidth]{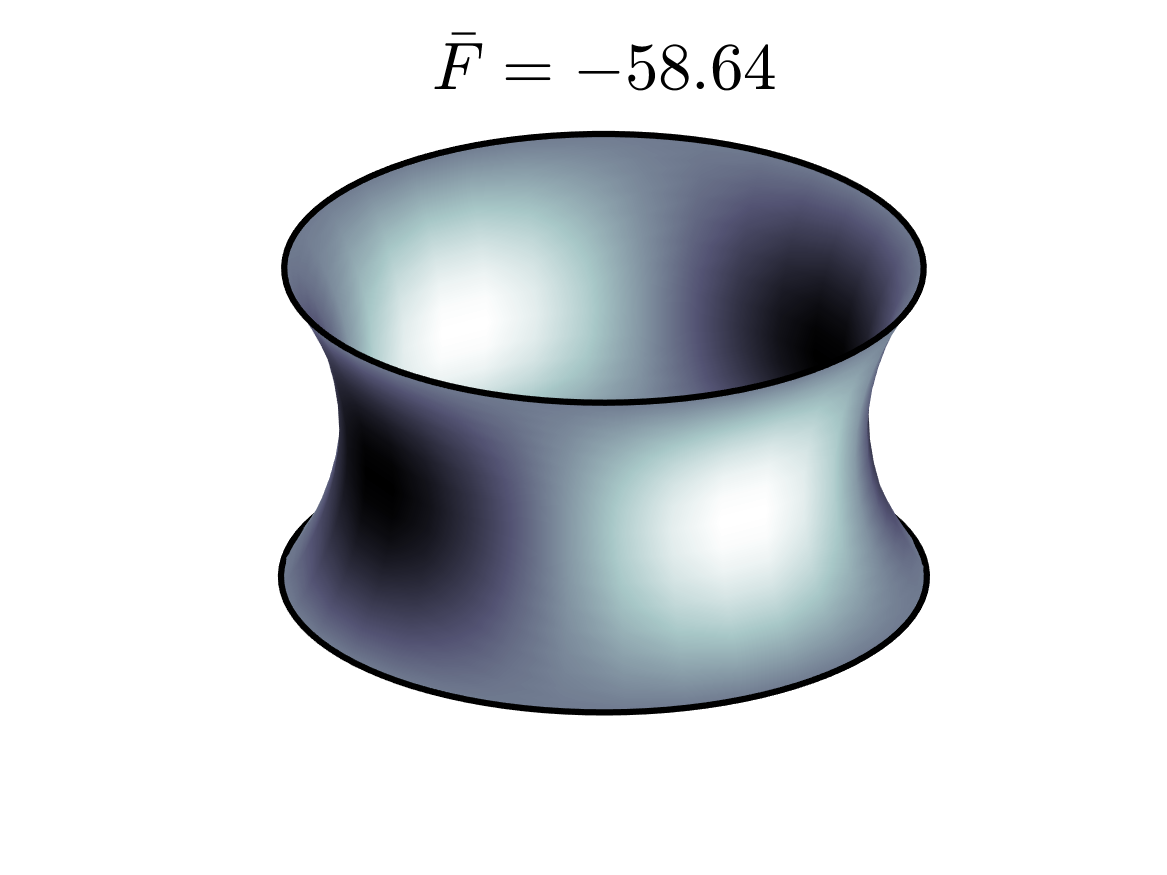}
    \end{subfigure} &
    \begin{subfigure}[b]{0.16\textwidth}
      \includegraphics[trim = 100pt 70pt 100pt 15pt, clip,width=\textwidth]{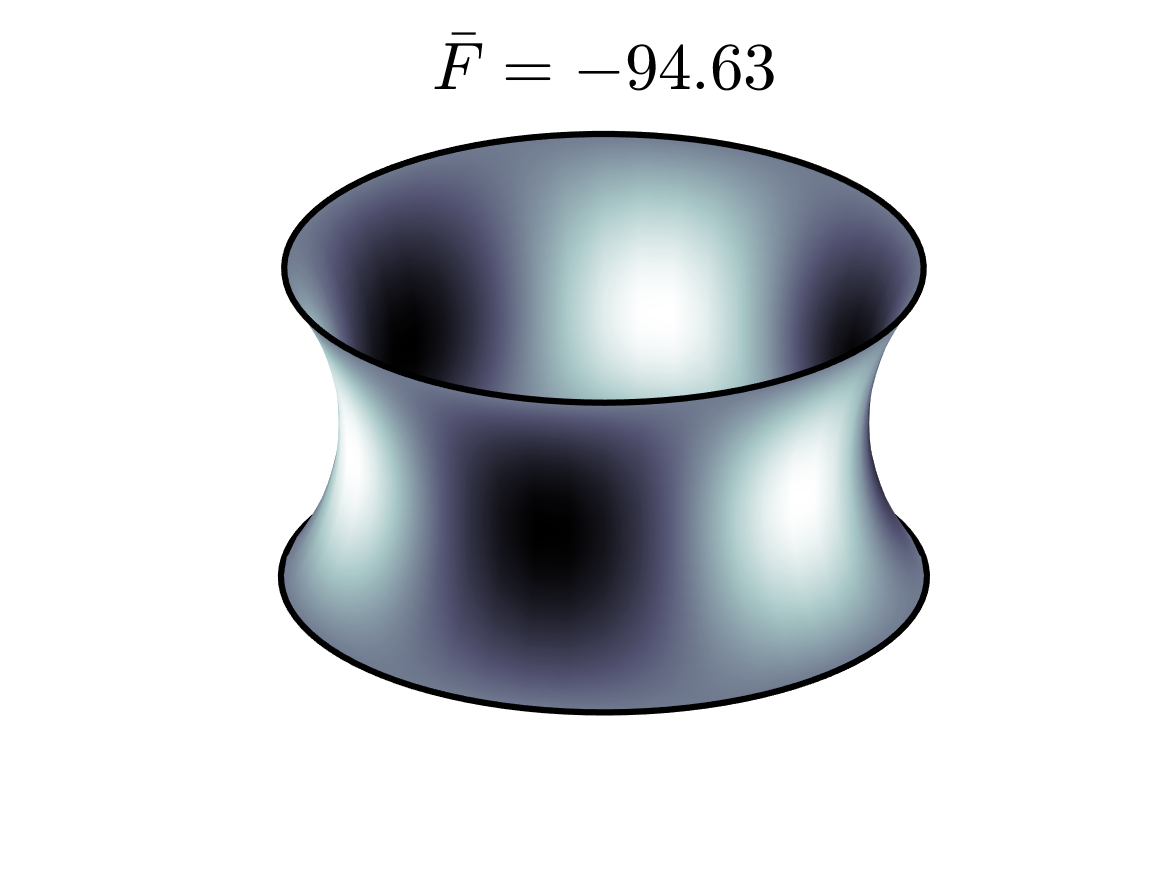}
    \end{subfigure} & \begin{subfigure}[b]{0.16\textwidth}
      \includegraphics[trim = 100pt 70pt 100pt 15pt, clip,width=\textwidth]{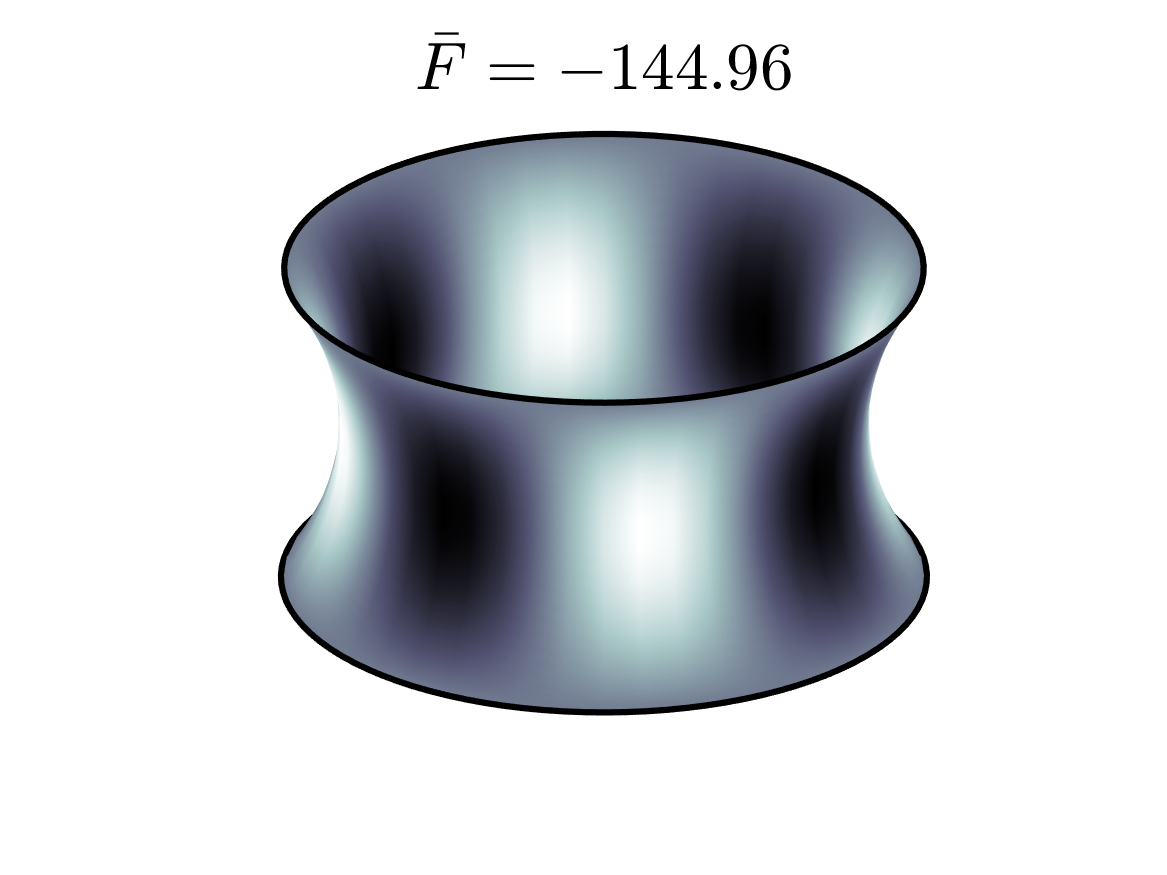}
    \end{subfigure}\\

    % --- ROW 3 ---
    &\rotatebox{90}{\textit{ \quad Branch 2}} &
    \begin{subfigure}[b]{0.16\textwidth}
      \includegraphics[trim = 100pt 85pt 100pt 15pt, clip,width=\textwidth]{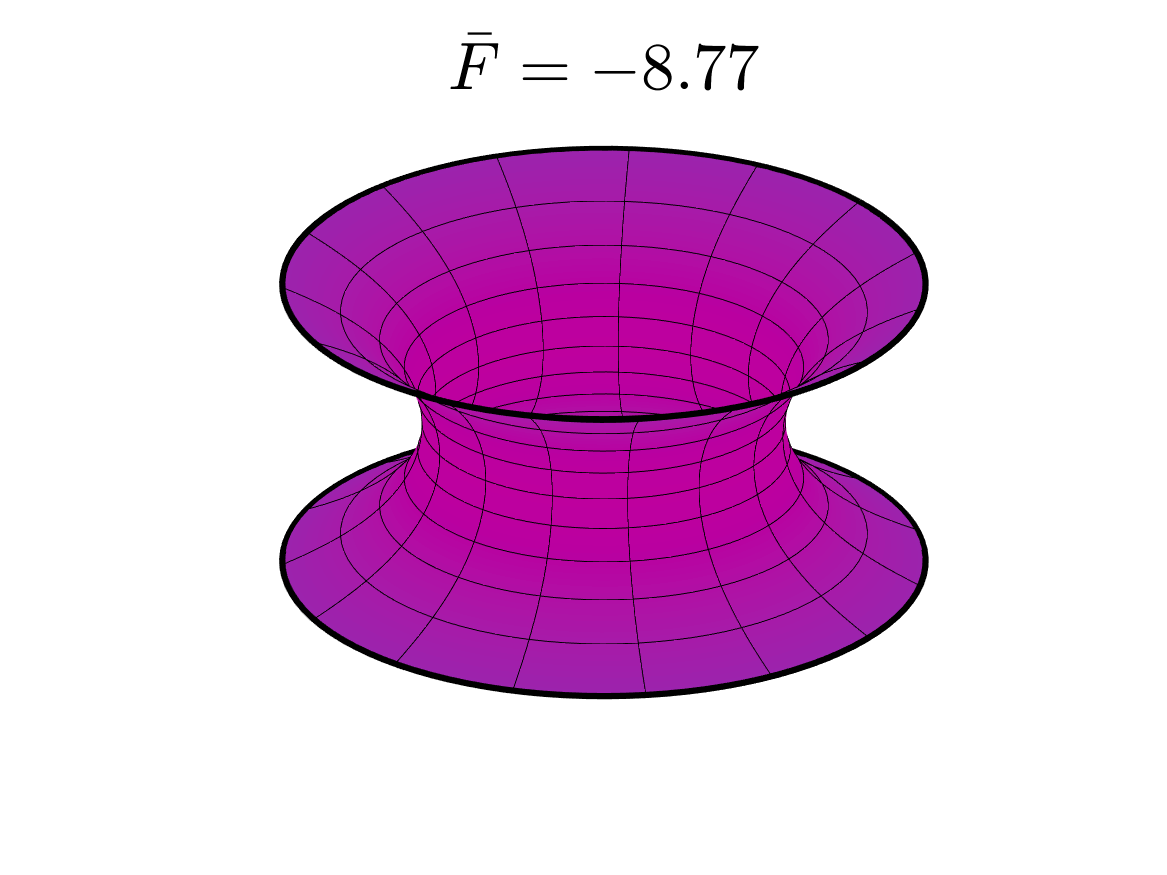}
    \end{subfigure} &
    \begin{subfigure}[b]{0.16\textwidth}
      \includegraphics[trim = 100pt 85pt 100pt 15pt, clip,width=\textwidth]{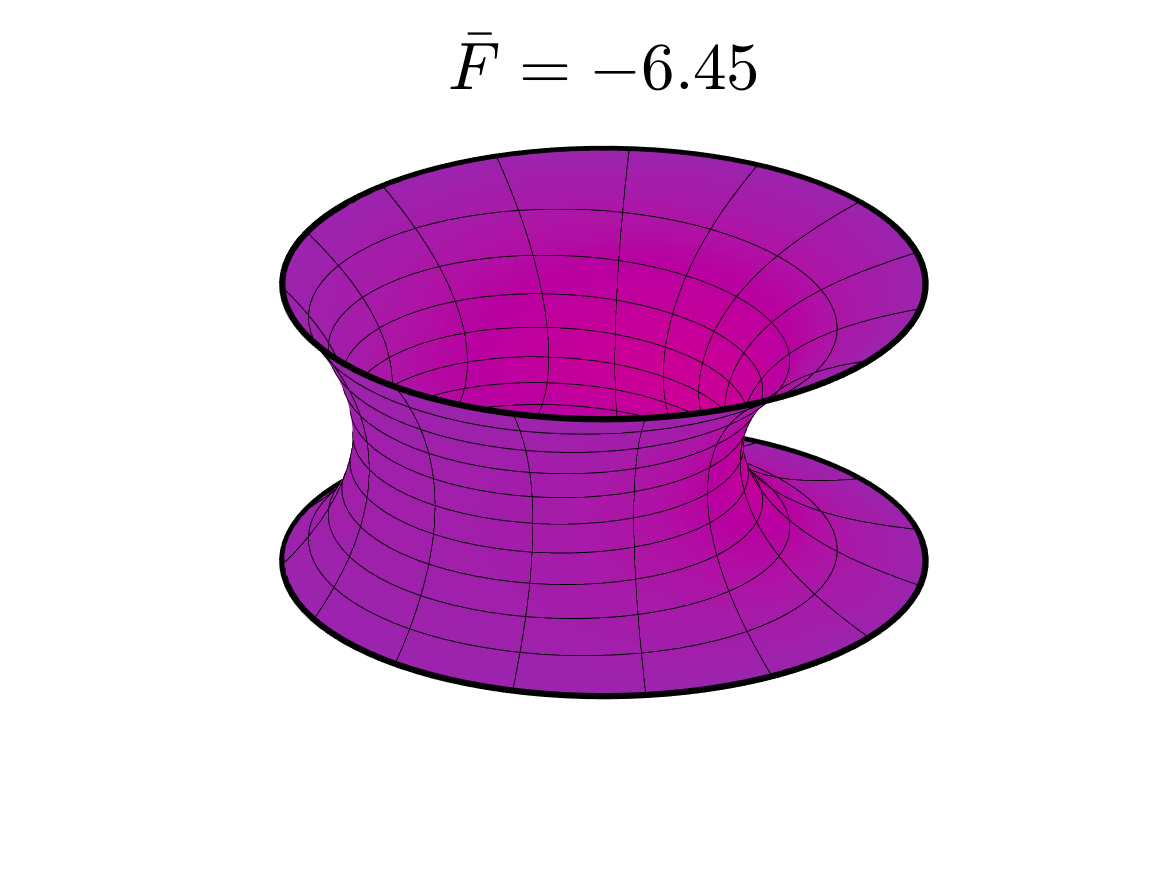}
    \end{subfigure} &
    \begin{subfigure}[b]{0.16\textwidth}
      \includegraphics[trim = 100pt 85pt 100pt 15pt, clip,width=\textwidth]{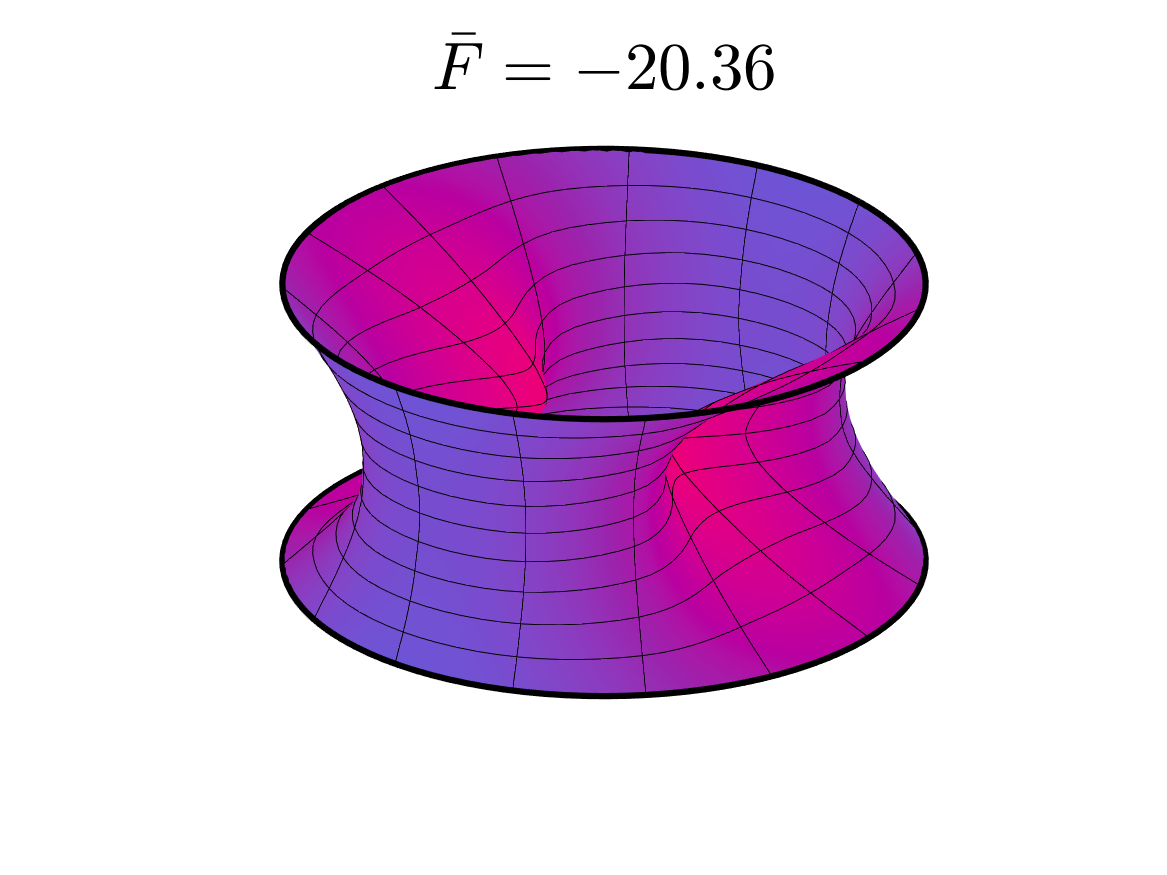}
    \end{subfigure} &
    \begin{subfigure}[b]{0.16\textwidth}
      \includegraphics[trim = 100pt 85pt 100pt 15pt, clip,width=\textwidth]{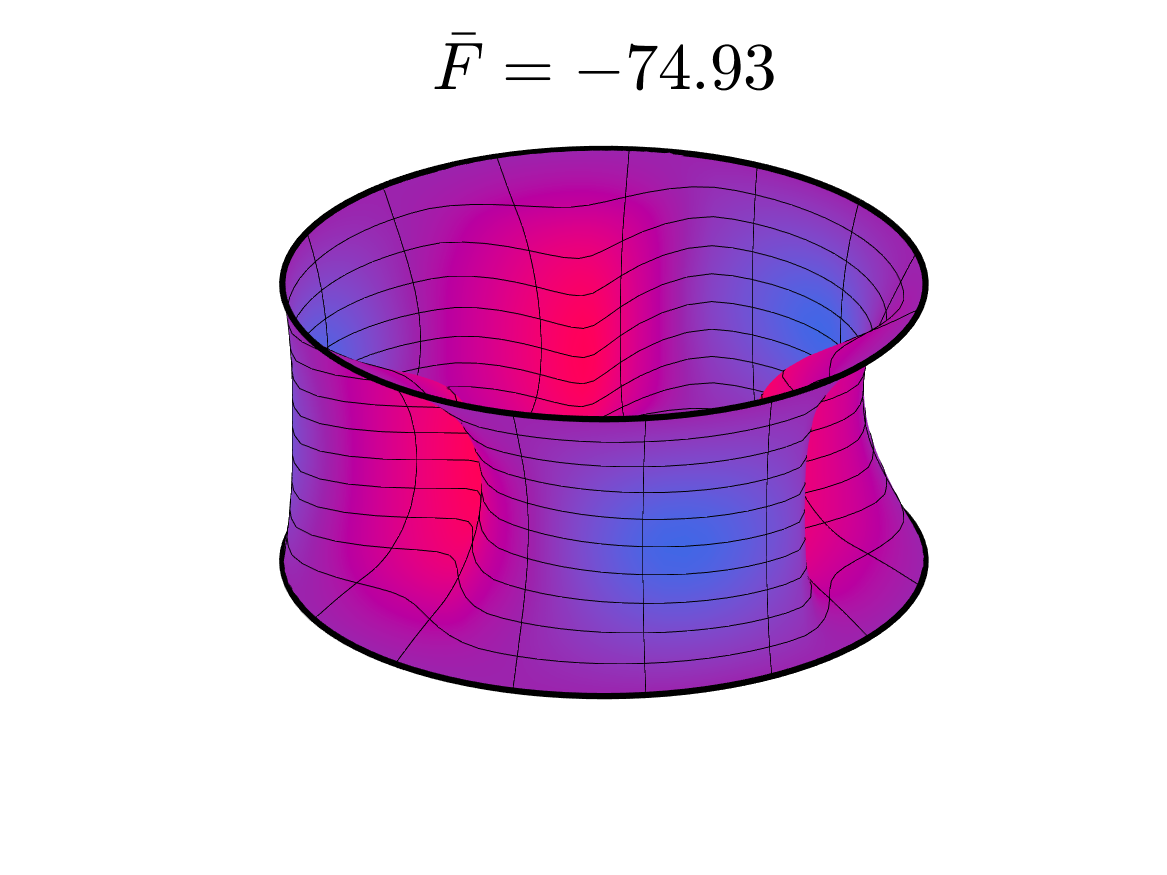}
    \end{subfigure} & \begin{subfigure}[b]{0.16\textwidth}
      \includegraphics[trim = 100pt 85pt 100pt 15pt, clip,width=\textwidth]{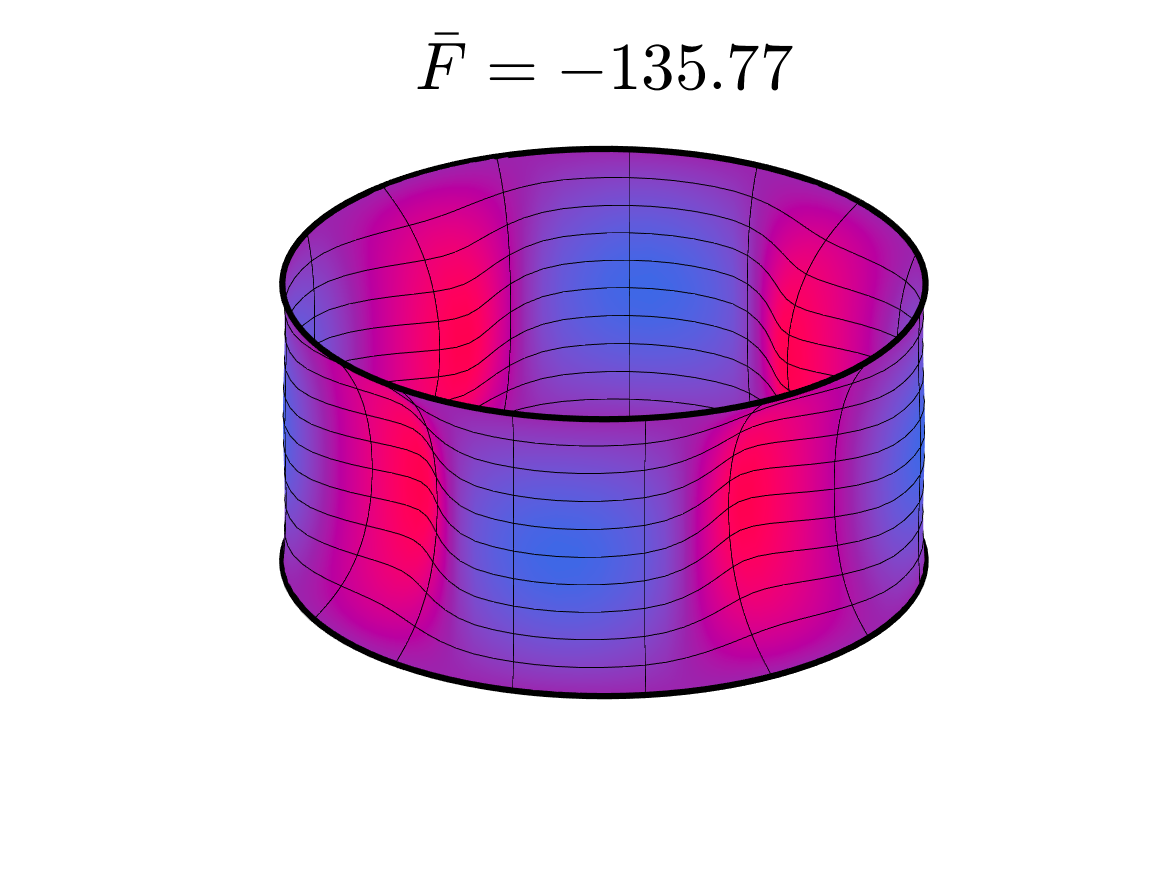}
    \end{subfigure} \\ \hline
    
        % --- ROW 4 ---
    &\rotatebox{90}{\textit{ 
    \quad Branch 1}} &
    \begin{subfigure}[b]{0.16\textwidth}
      \includegraphics[trim = 100pt 70pt 100pt 15pt, clip,width=\textwidth]{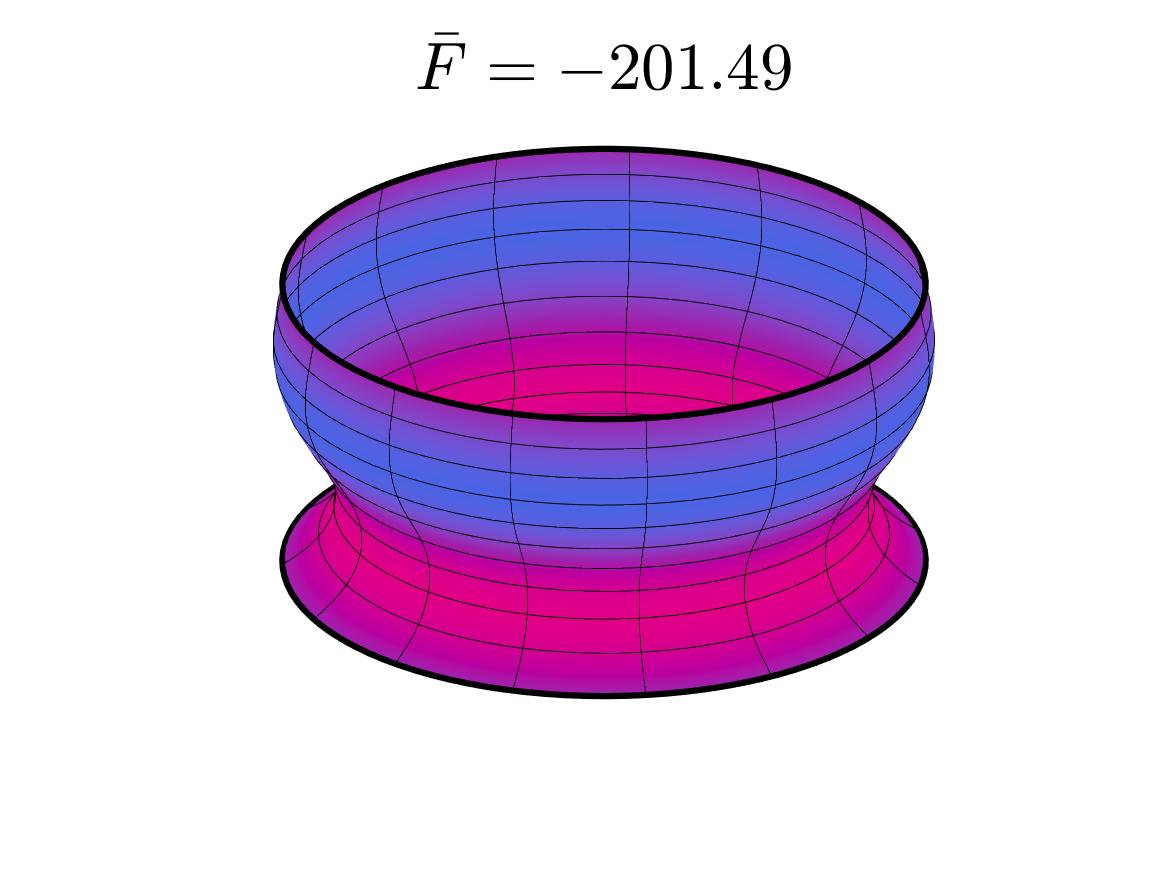}
    \end{subfigure} &
    \begin{subfigure}[b]{0.16\textwidth}
      \includegraphics[trim = 100pt 70pt 100pt 15pt, clip,width=\textwidth]{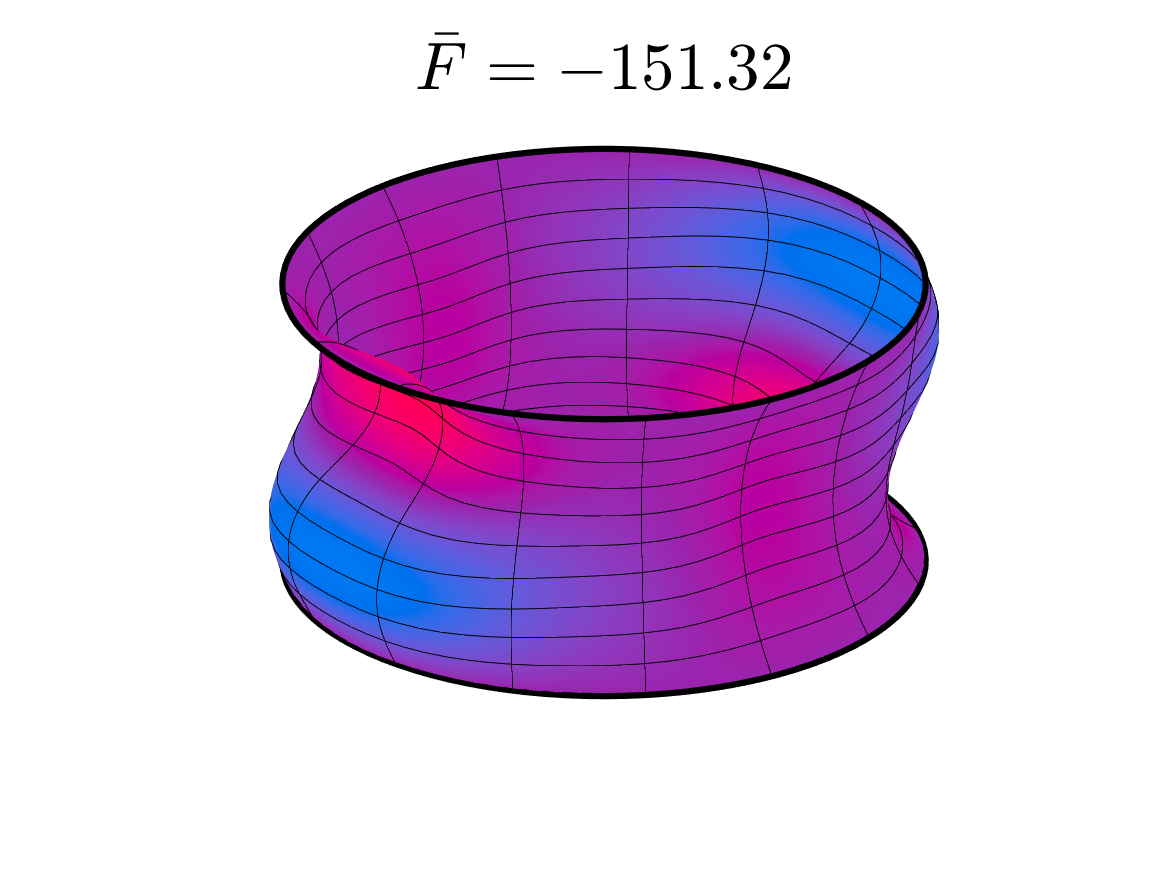}
    \end{subfigure} &
    \begin{subfigure}[b]{0.16\textwidth}
      \includegraphics[trim = 100pt 70pt 100pt 15pt, clip,width=\textwidth]{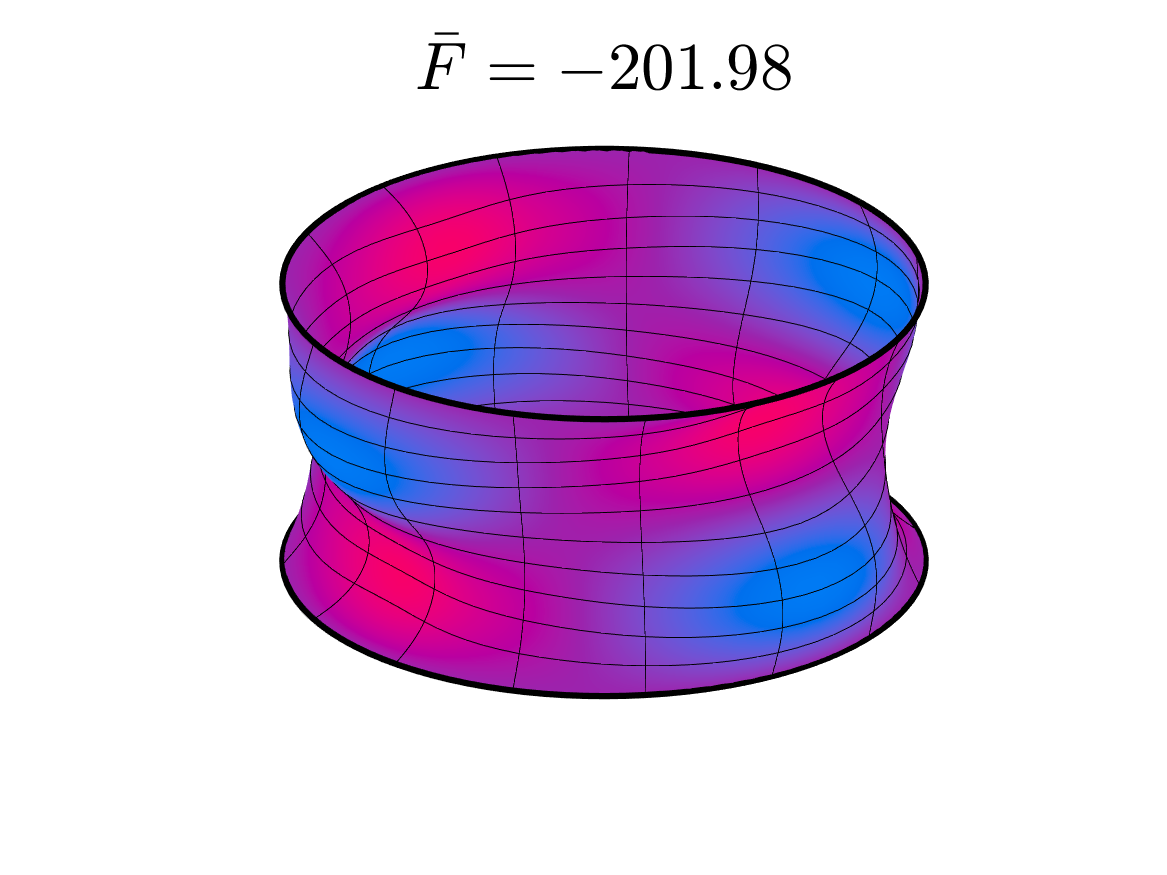}
    \end{subfigure} &
    \begin{subfigure}[b]{0.16\textwidth}
      \includegraphics[trim = 100pt 70pt 100pt 15pt, clip,width=\textwidth]{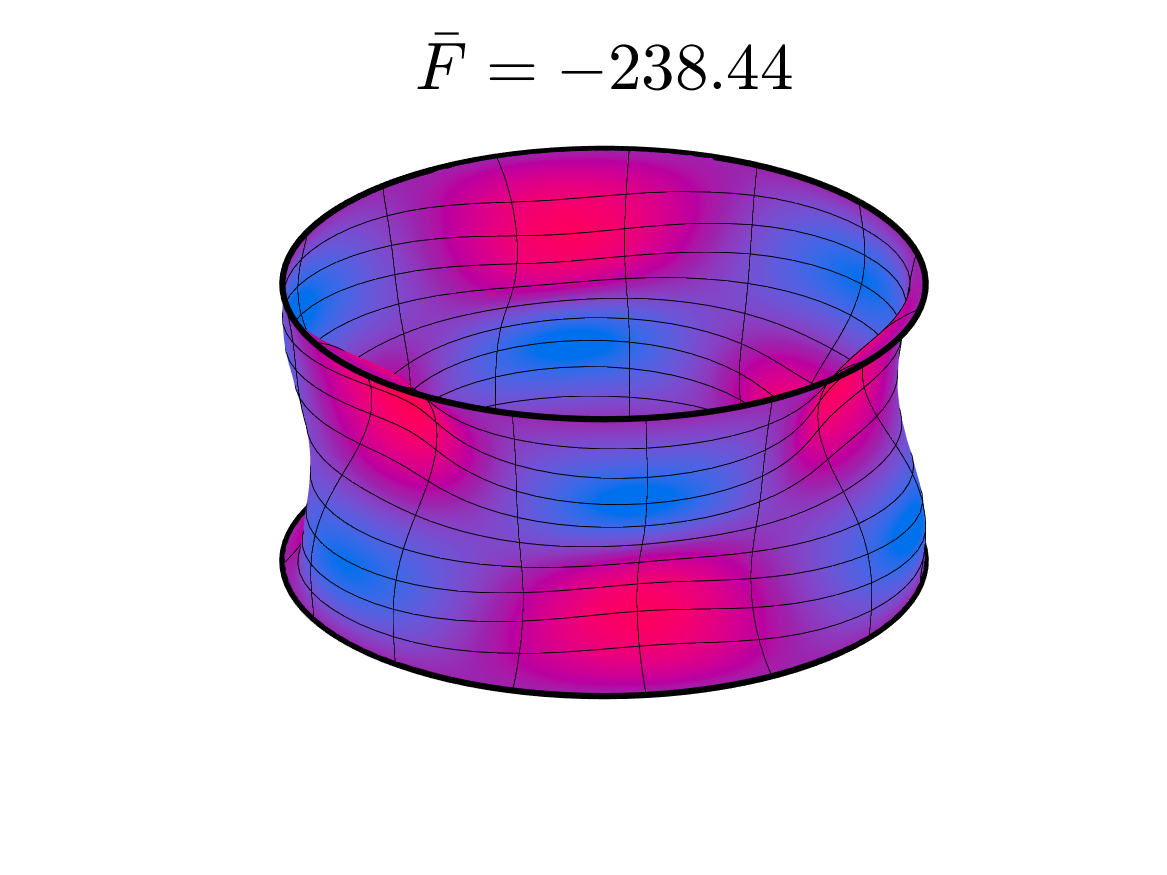}
    \end{subfigure} & \begin{subfigure}[b]{0.16\textwidth}
      \includegraphics[trim = 100pt 70pt 100pt 15pt, clip,width=\textwidth]{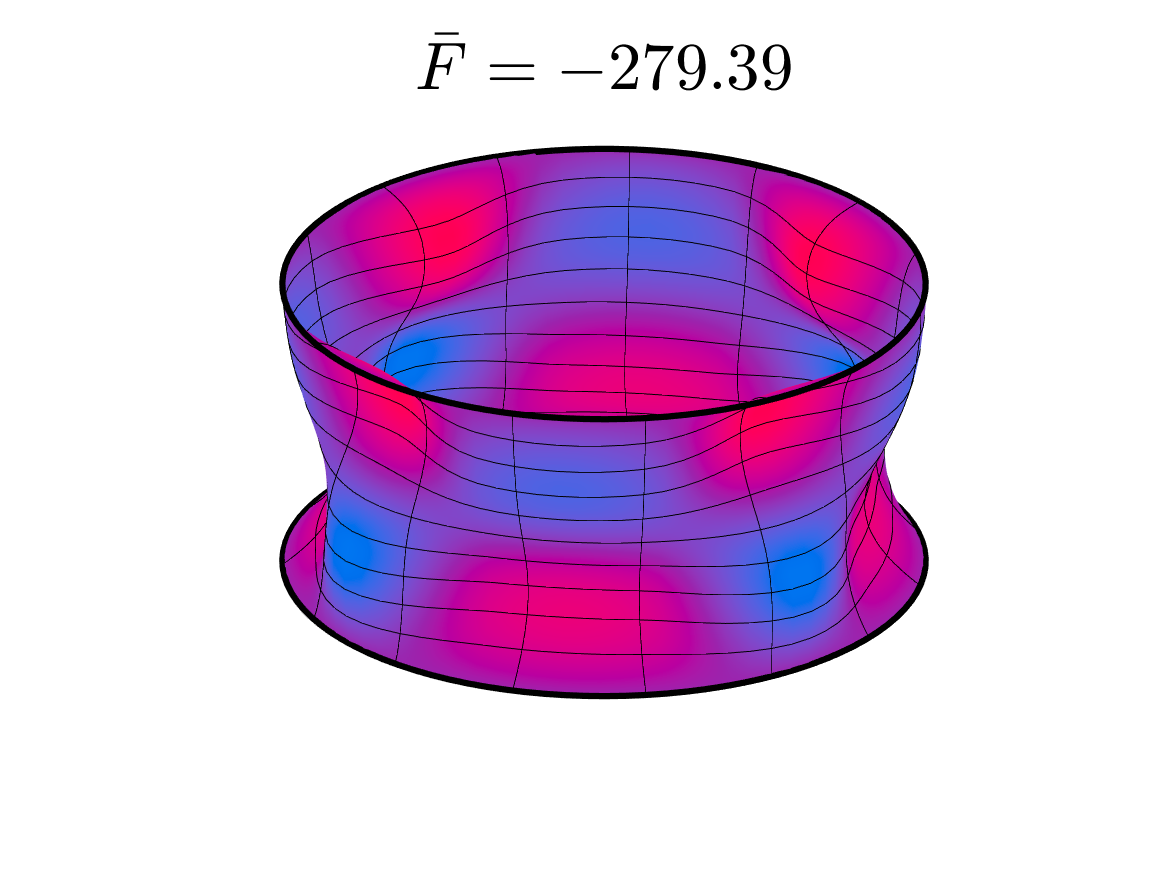}
    \end{subfigure}&
    \multirow{3}{*}{\begin{subfigure}[b]{0.04\textwidth}
        \includegraphics[trim = 475pt 10pt 35pt 30pt, clip,width=\textwidth]{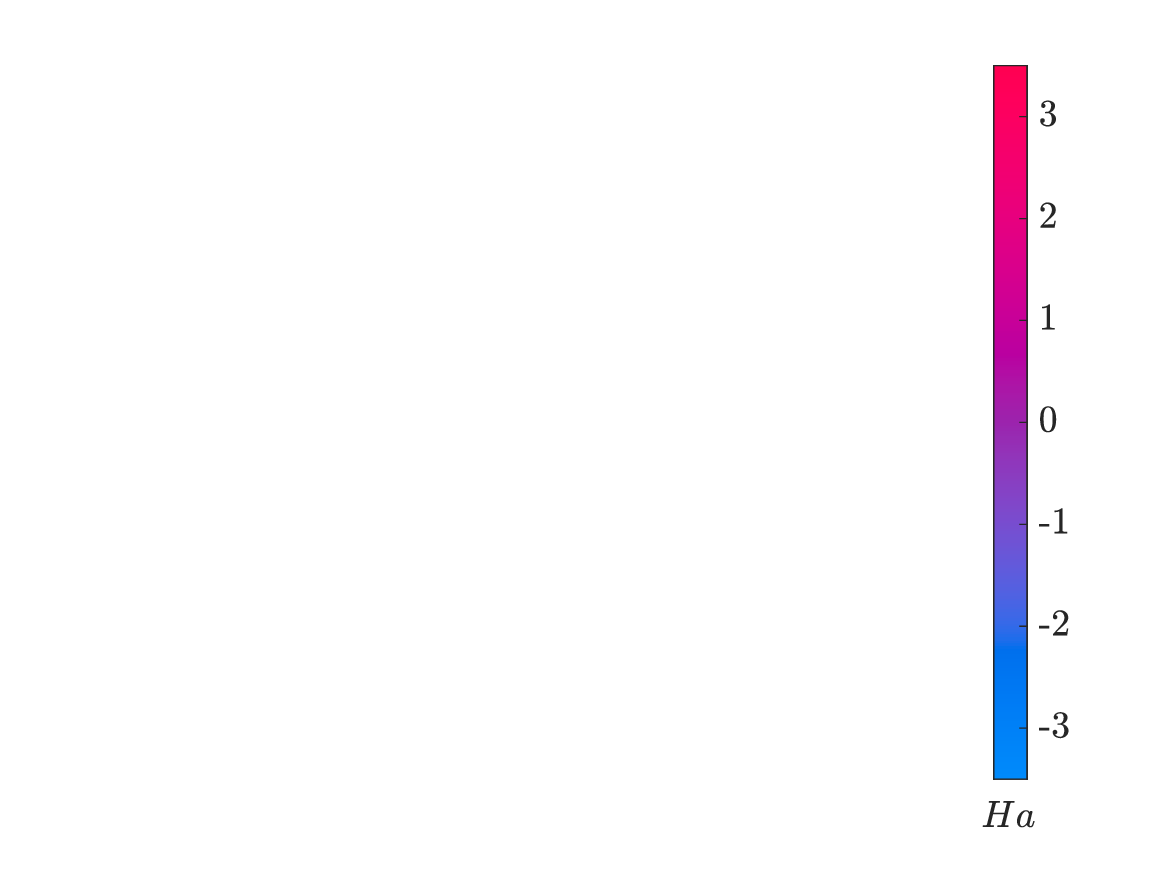}
    \end{subfigure}} \\

    % --- ROW 5 ---
   \rotatebox{90}{\quad \quad $n=2$} & \rotatebox{90}{\textit{ Eigensurfaces}} &
    \begin{subfigure}[b]{0.16\textwidth}
      \includegraphics[trim = 100pt 70pt 100pt 15pt, clip,width=\textwidth]{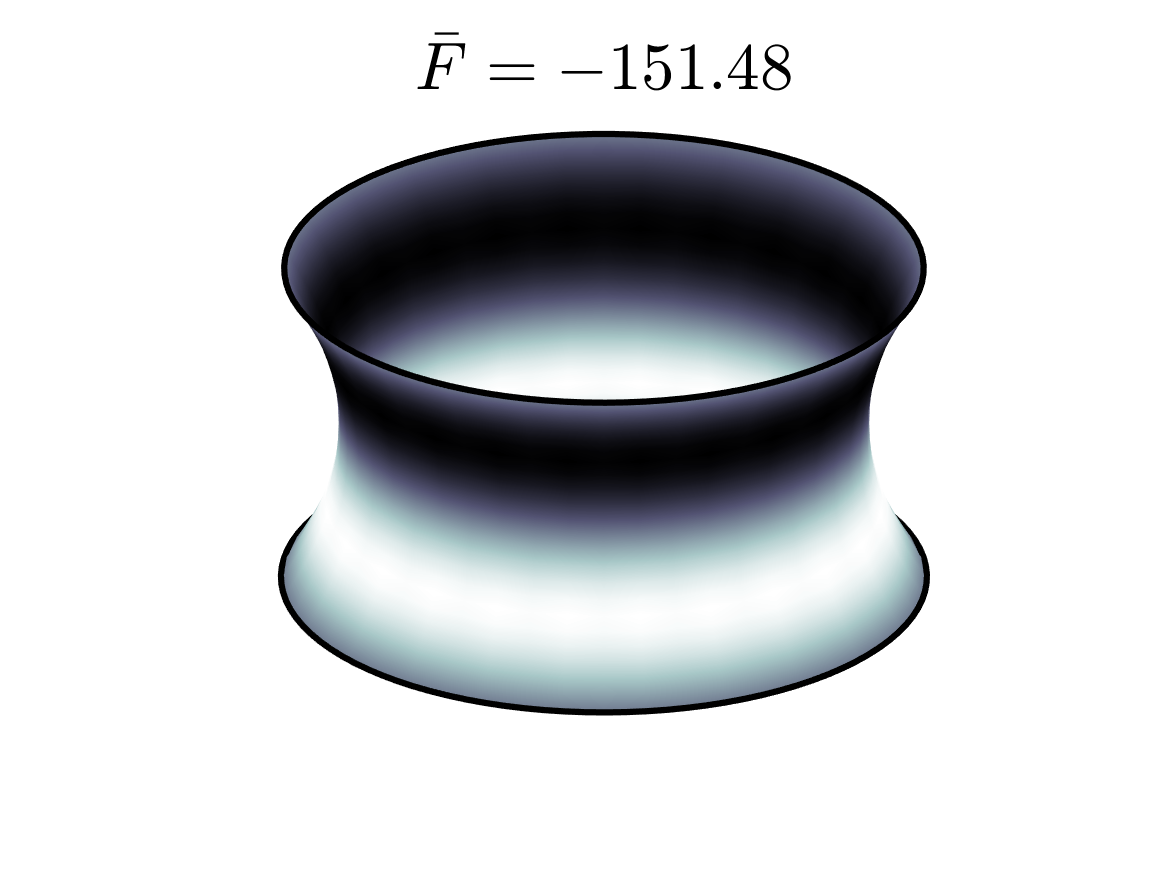}
    \end{subfigure} &
    \begin{subfigure}[b]{0.16\textwidth}
      \includegraphics[trim = 100pt 70pt 100pt 15pt, clip,width=\textwidth]{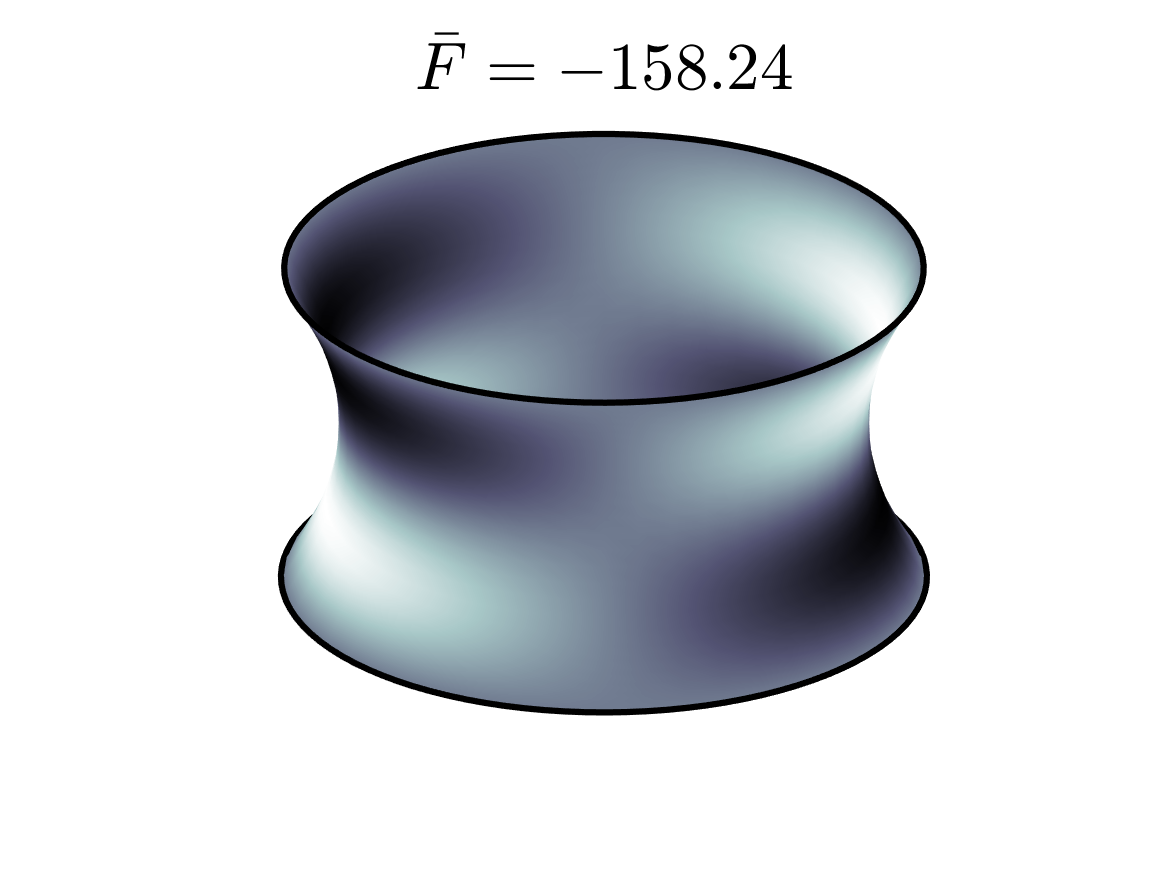}
    \end{subfigure} &
    \begin{subfigure}[b]{0.16\textwidth}
      \includegraphics[trim = 100pt 70pt 100pt 15pt, clip,width=\textwidth]{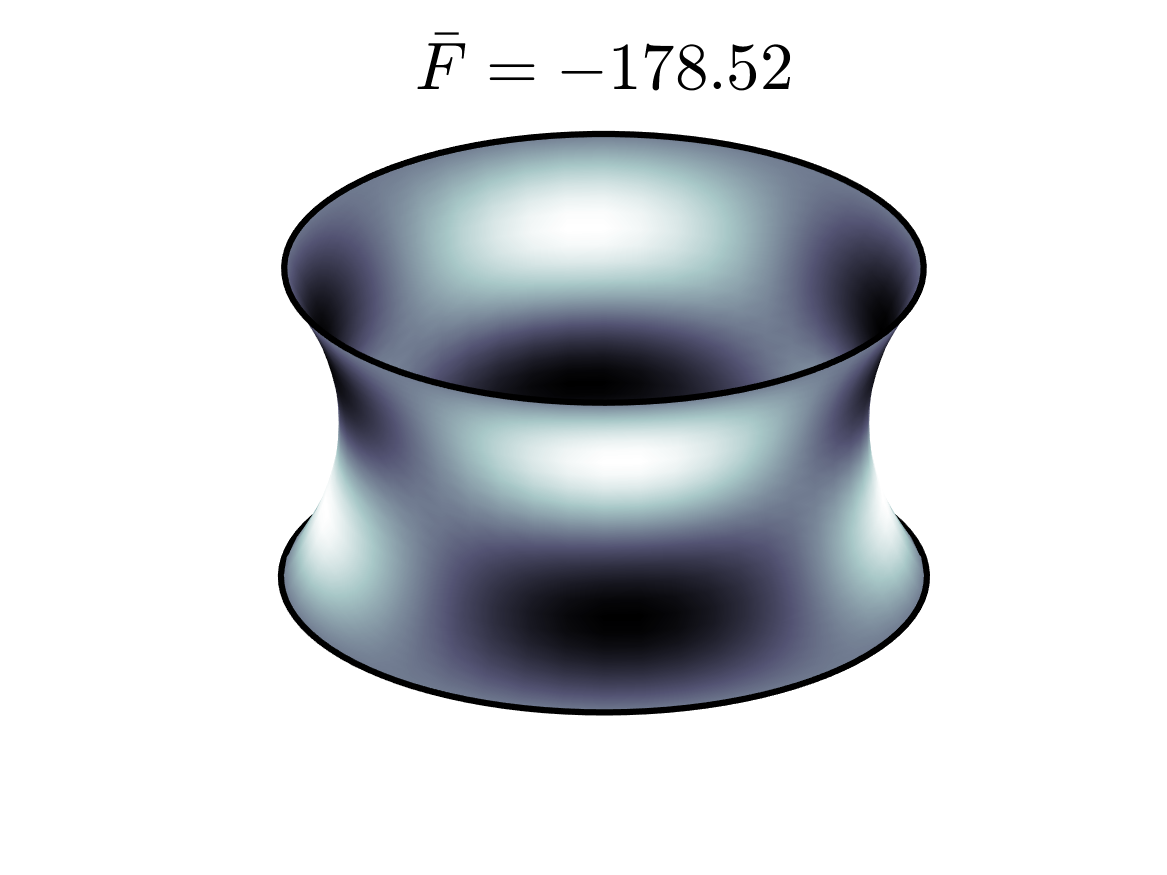}
    \end{subfigure} &
    \begin{subfigure}[b]{0.16\textwidth}
      \includegraphics[trim = 100pt 70pt 100pt 15pt, clip,width=\textwidth]{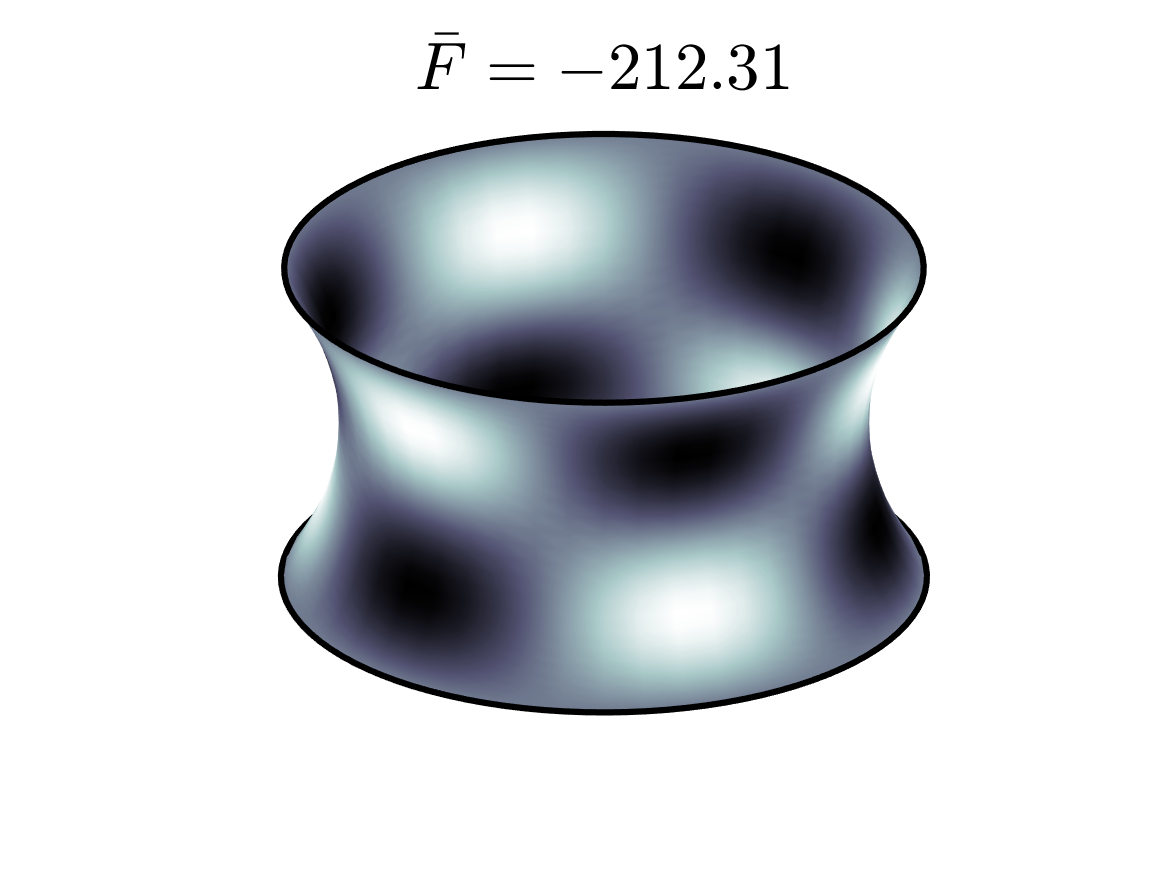}
    \end{subfigure} & \begin{subfigure}[b]{0.16\textwidth}
      \includegraphics[trim = 100pt 70pt 100pt 15pt, clip,width=\textwidth]{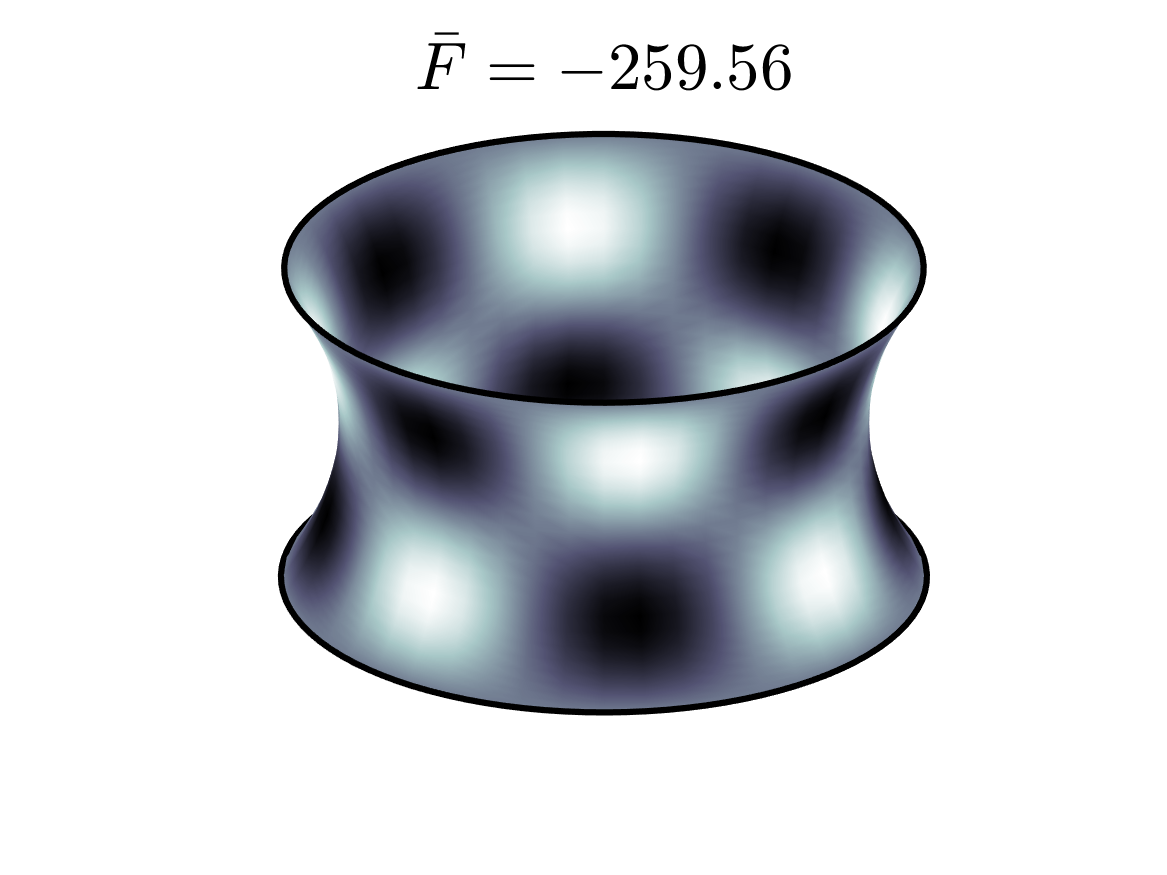}
    \end{subfigure}\\

    % --- ROW 6 ---
    &\rotatebox{90}{\textit{ \quad Branch 2}} &
    \begin{subfigure}[b]{0.16\textwidth}
      \includegraphics[trim = 100pt 85pt 100pt 15pt, clip,width=\textwidth]{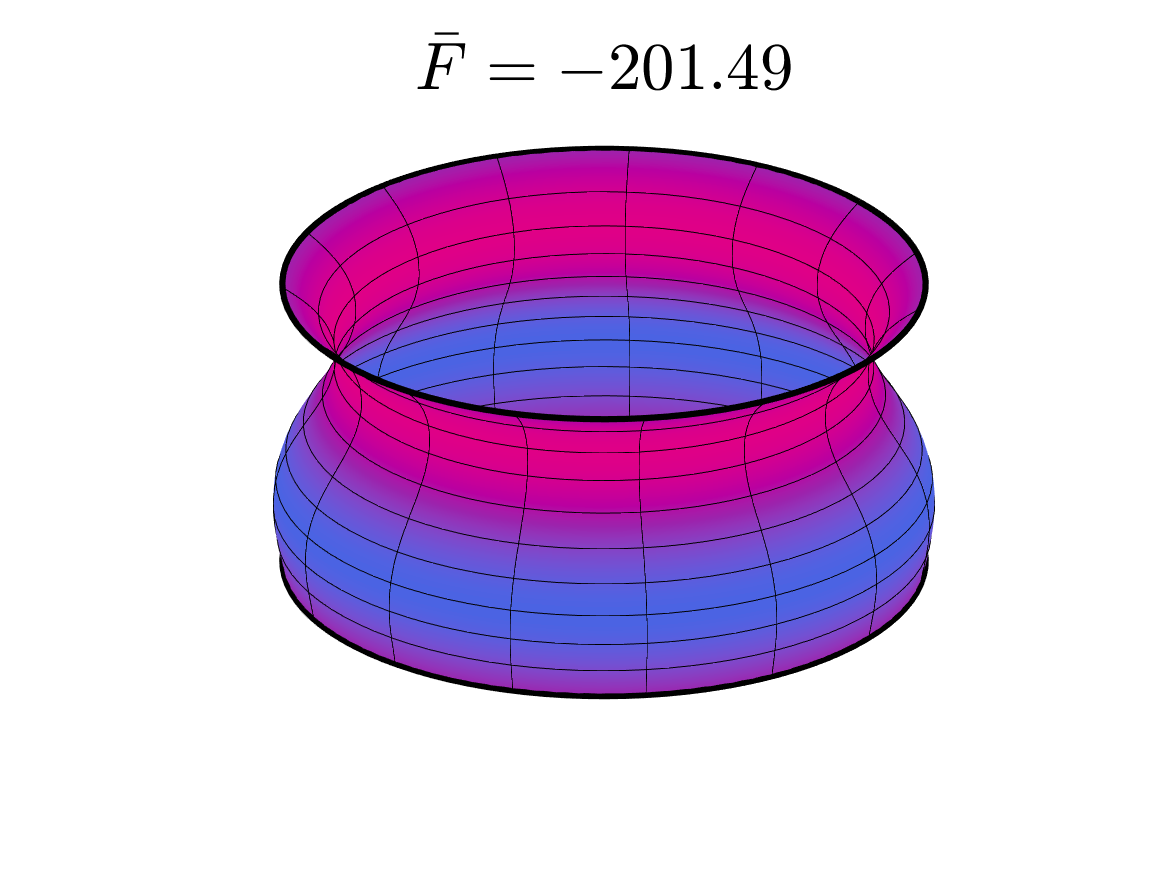}
    \end{subfigure} &
    \begin{subfigure}[b]{0.16\textwidth}
      \includegraphics[trim = 100pt 85pt 100pt 15pt, clip,width=\textwidth]{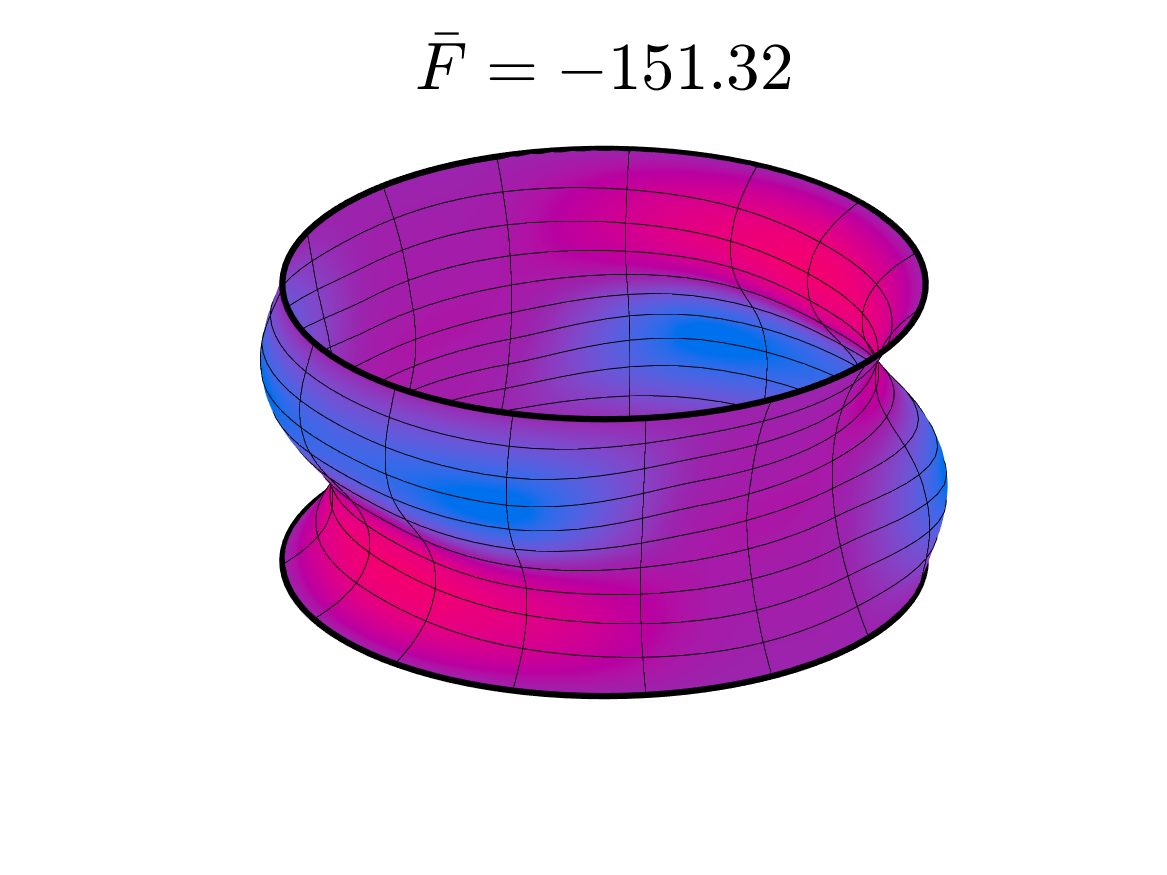}
    \end{subfigure} &
    \begin{subfigure}[b]{0.16\textwidth}
      \includegraphics[trim = 100pt 85pt 100pt 15pt, clip,width=\textwidth]{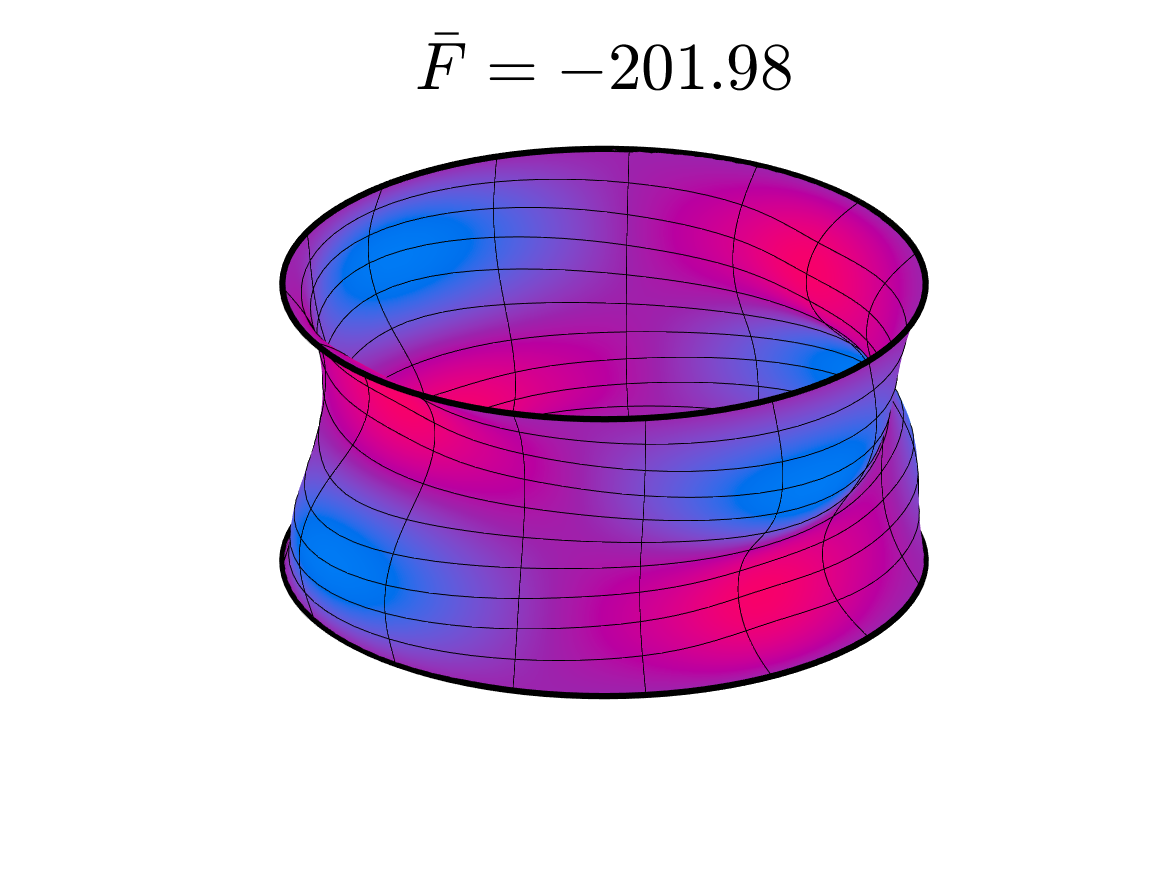}
    \end{subfigure} &
    \begin{subfigure}[b]{0.16\textwidth}
      \includegraphics[trim = 100pt 85pt 100pt 15pt, clip,width=\textwidth]{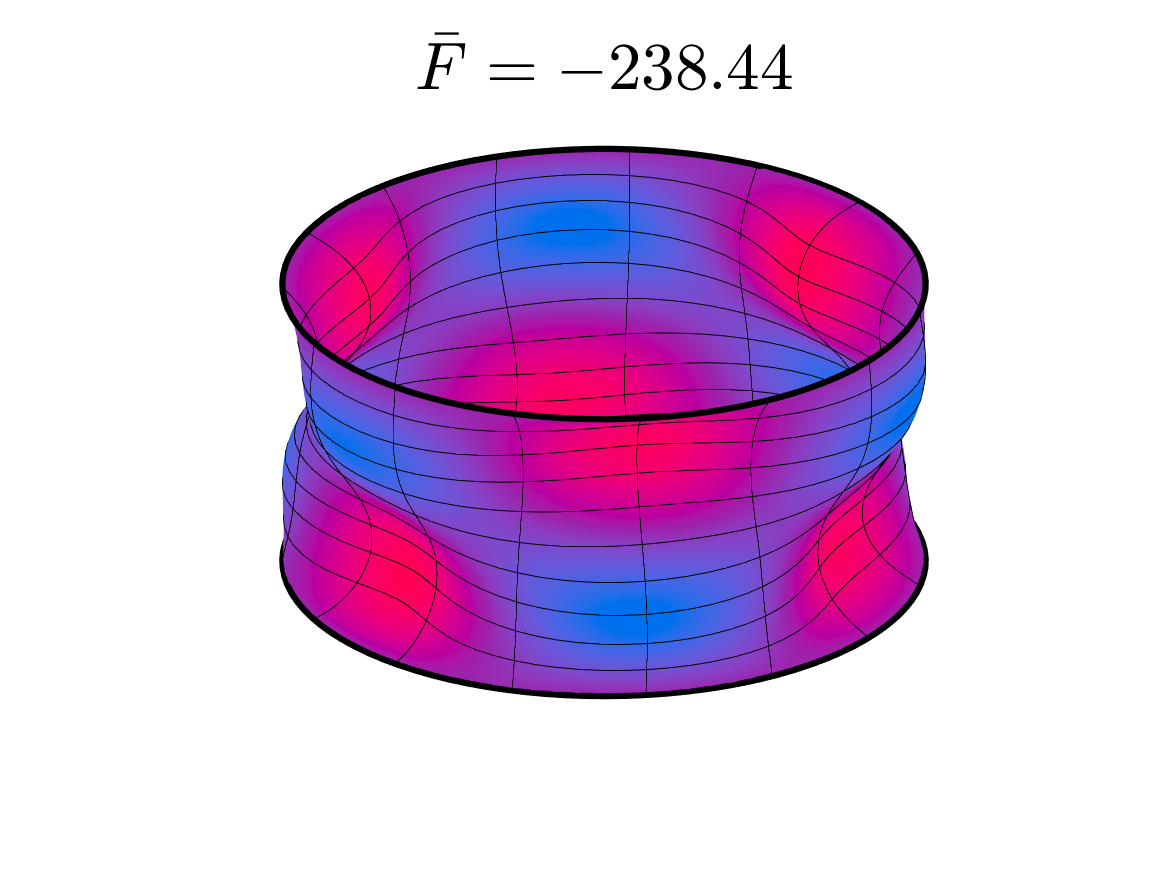}
    \end{subfigure} & \begin{subfigure}[b]{0.16\textwidth}
      \includegraphics[trim = 100pt 85pt 100pt 15pt, clip,width=\textwidth]{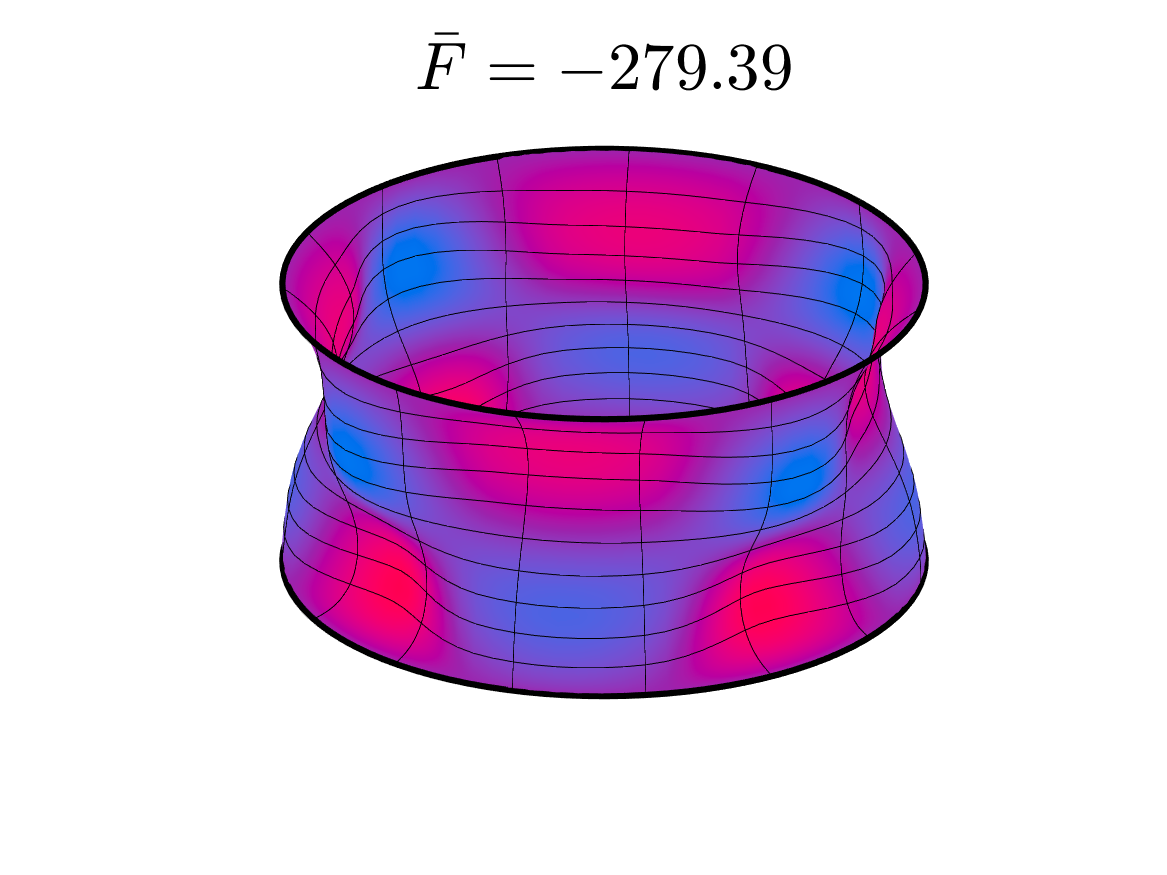}
    \end{subfigure} \\ \hline
    
    % --- ROW 7 ---
    &\rotatebox{90}{\textit{ 
    \quad Branch 1}} &
    \begin{subfigure}[b]{0.16\textwidth}
      \includegraphics[trim = 100pt 70pt 100pt 15pt, clip,width=\textwidth]{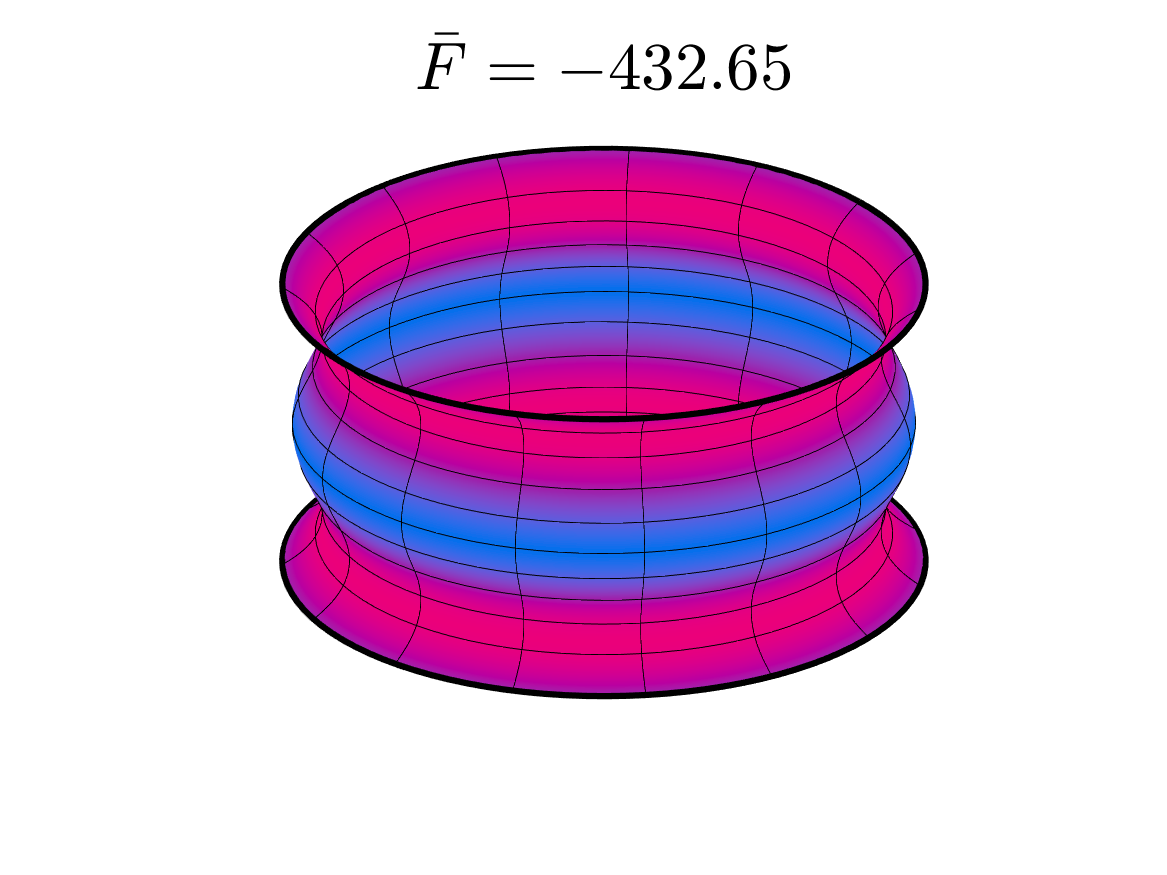}
    \end{subfigure} &
    \begin{subfigure}[b]{0.16\textwidth}
      \includegraphics[trim = 100pt 70pt 100pt 15pt, clip,width=\textwidth]{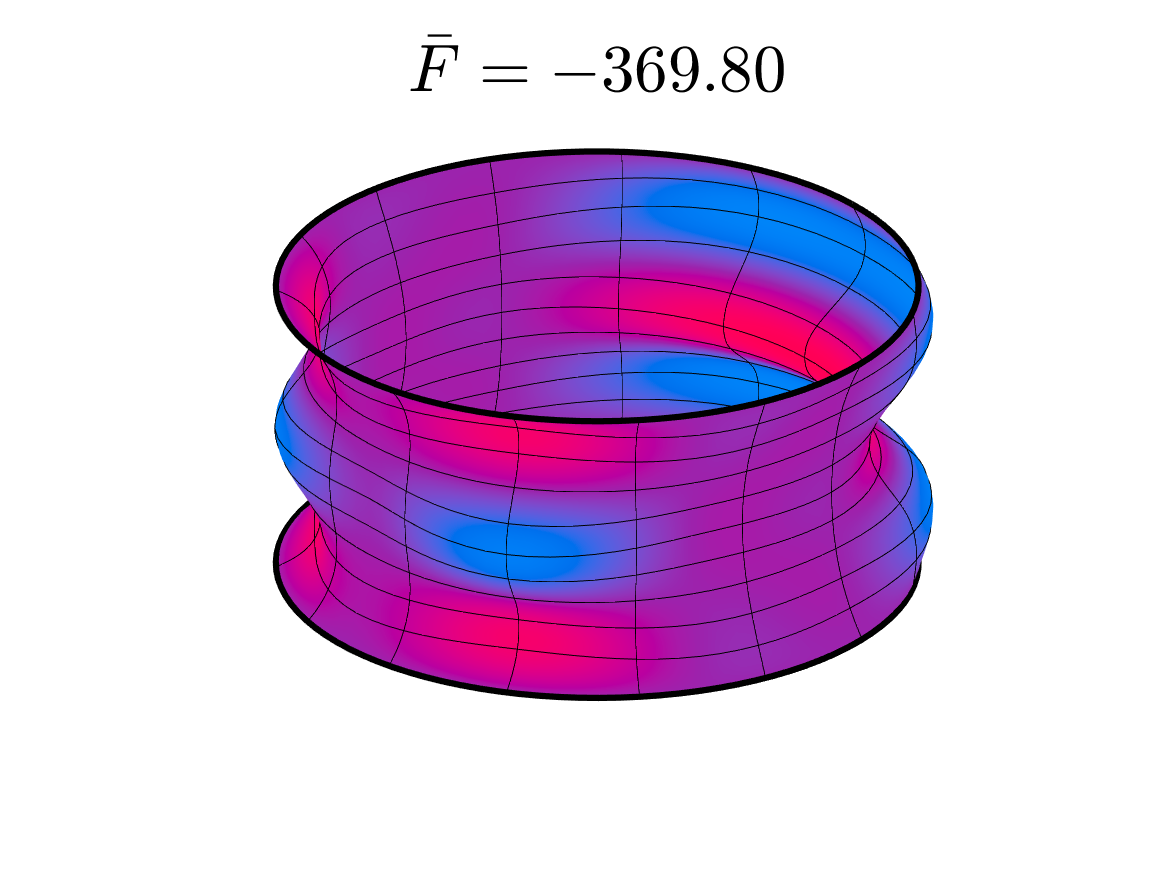}
    \end{subfigure} &
    \begin{subfigure}[b]{0.16\textwidth}
      \includegraphics[trim = 100pt 70pt 100pt 15pt, clip,width=\textwidth]{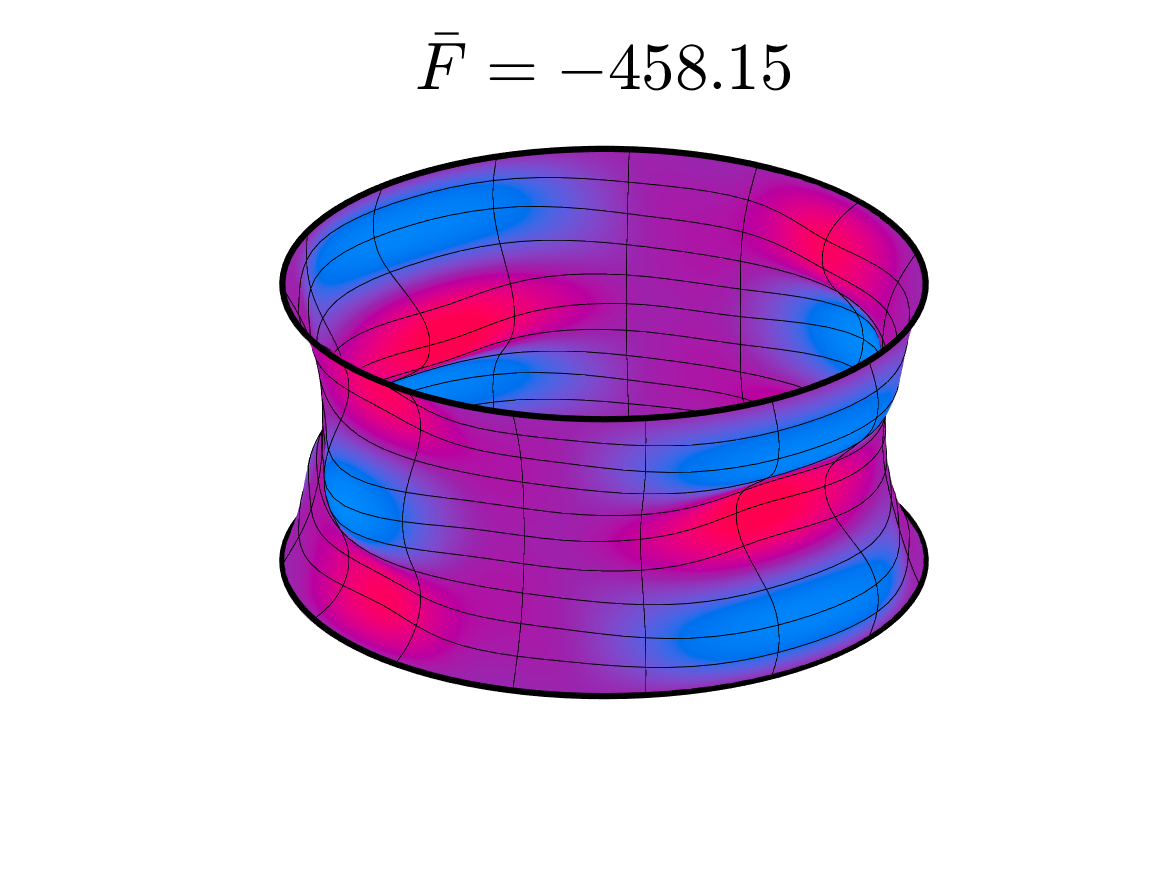}
    \end{subfigure} &
    \begin{subfigure}[b]{0.16\textwidth}
      \includegraphics[trim = 100pt 70pt 100pt 15pt, clip,width=\textwidth]{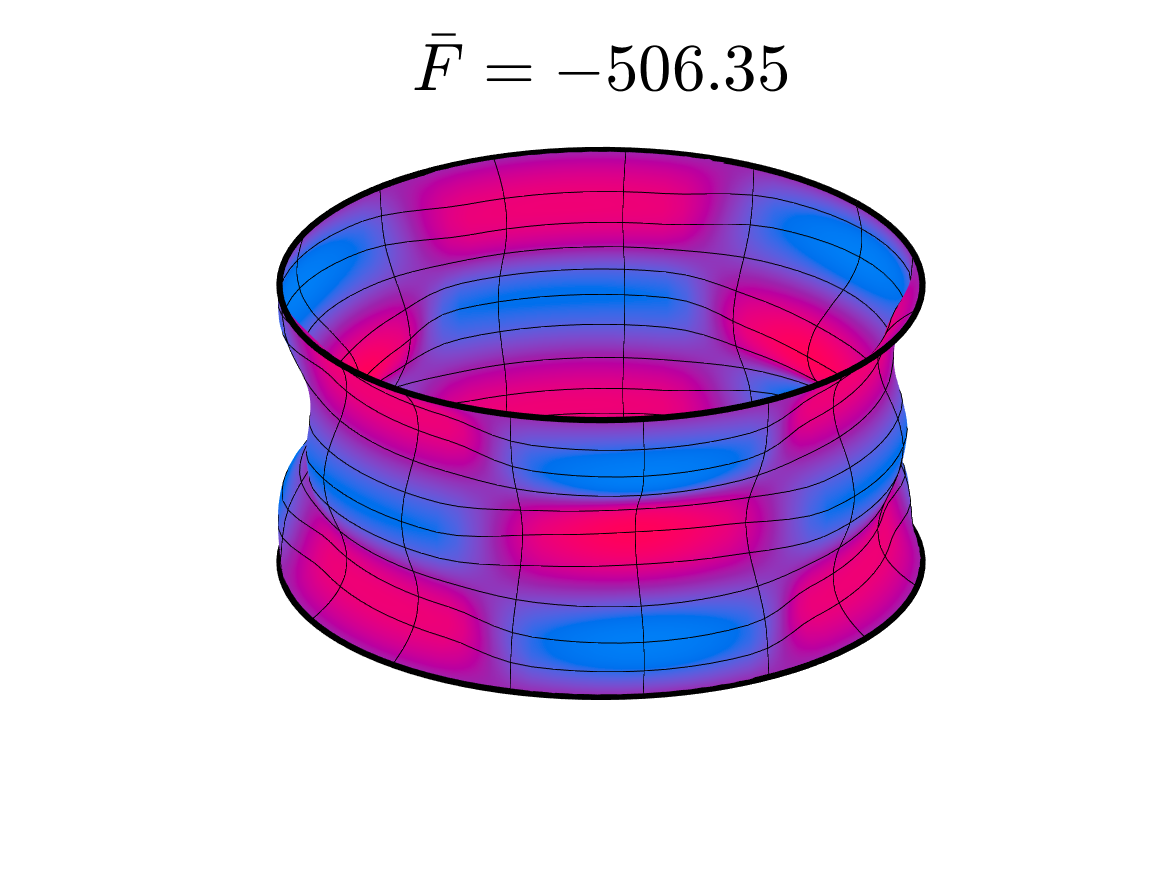}
    \end{subfigure} & \begin{subfigure}[b]{0.16\textwidth}
      \includegraphics[trim = 100pt 70pt 100pt 15pt, clip,width=\textwidth]{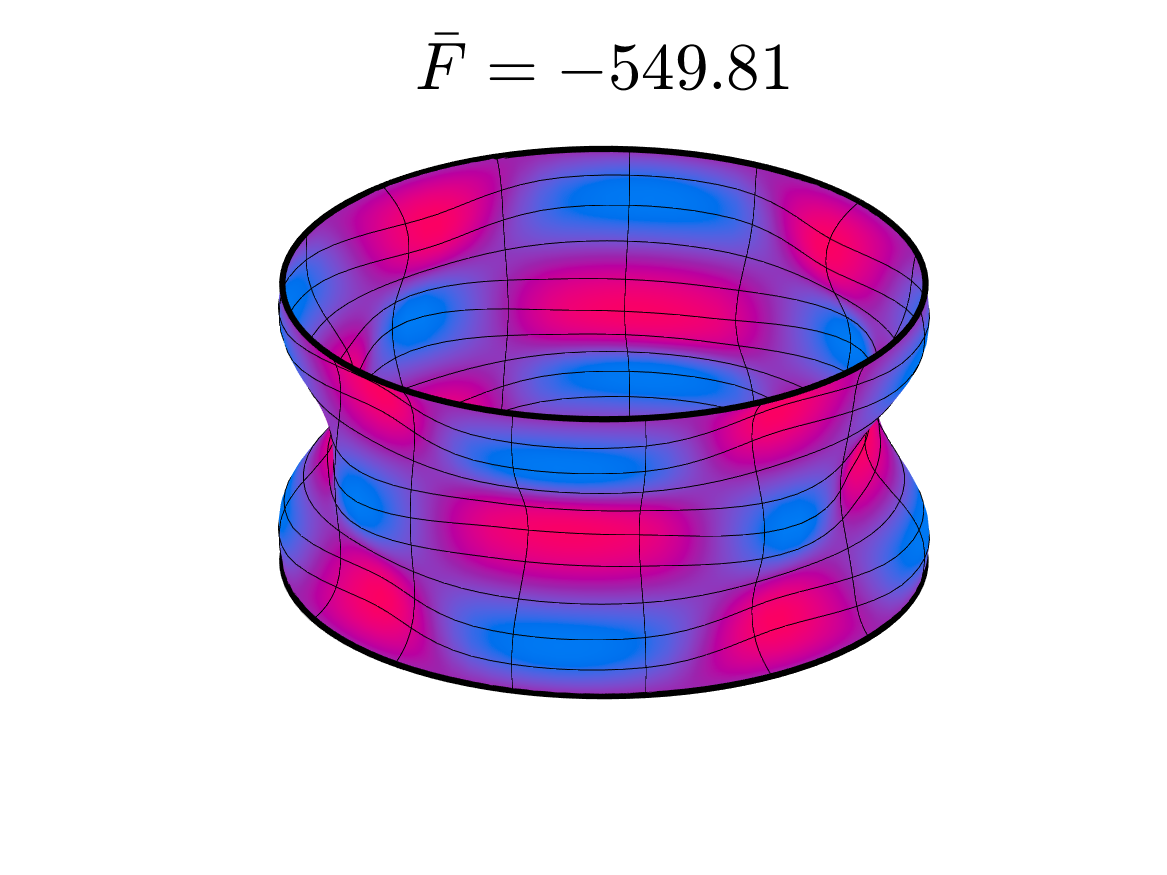}
    \end{subfigure}&
    \multirow{3}{*}{\begin{subfigure}[b]{0.04\textwidth}
        \includegraphics[trim = 475pt 10pt 35pt 30pt, clip,width=\textwidth]{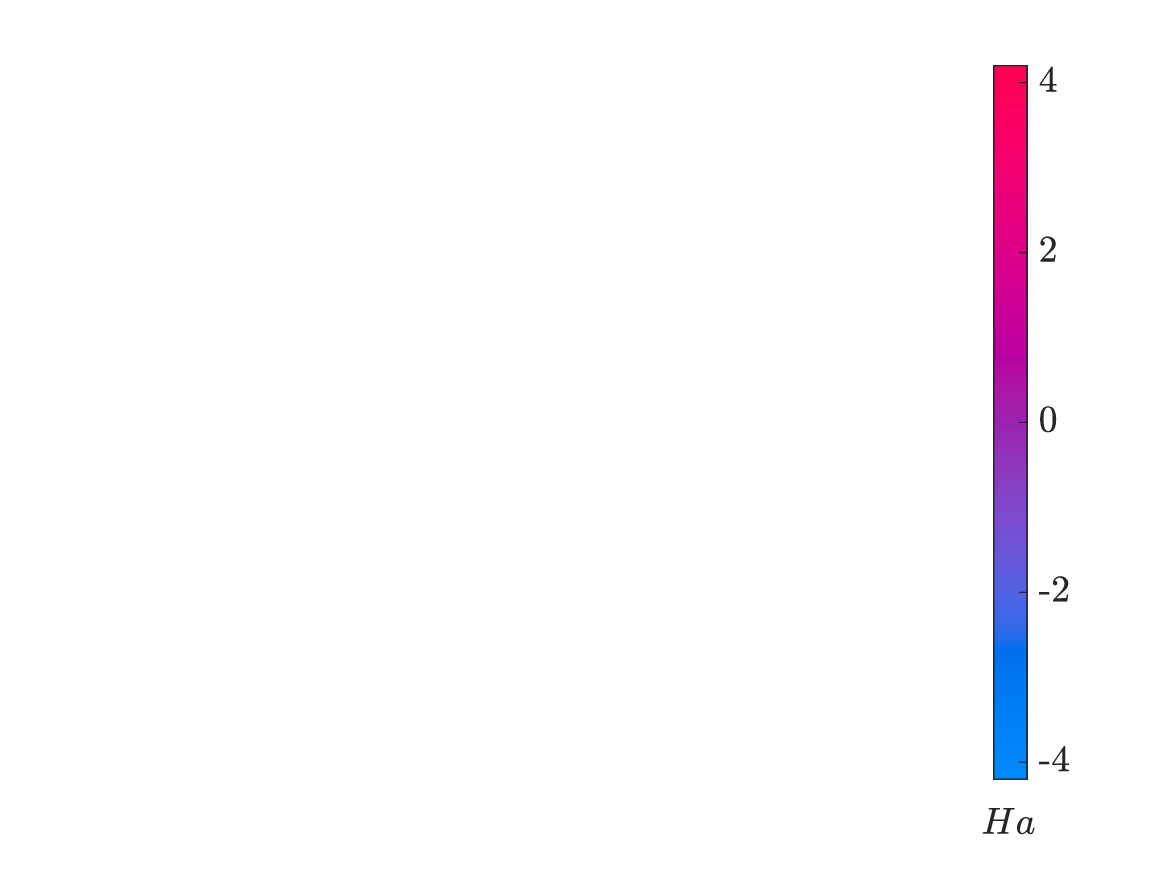}
    \end{subfigure}}\\

    % --- ROW 8 ---
    \rotatebox{90}{\quad \quad $n=3$}&\rotatebox{90}{\textit{ Eigensurfaces}} &
    \begin{subfigure}[b]{0.16\textwidth}
      \includegraphics[trim = 100pt 70pt 100pt 15pt, clip,width=\textwidth]{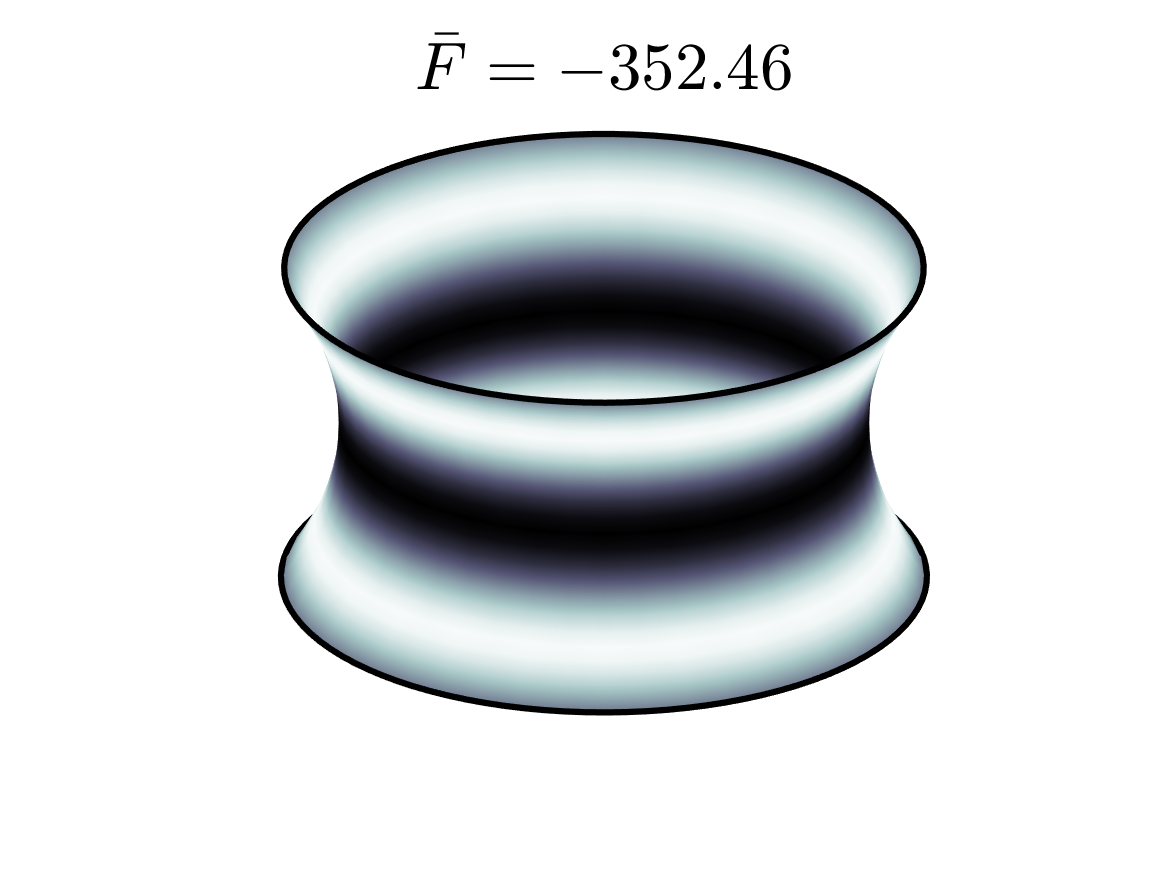}
    \end{subfigure} &
    \begin{subfigure}[b]{0.16\textwidth}
      \includegraphics[trim = 100pt 70pt 100pt 15pt, clip,width=\textwidth]{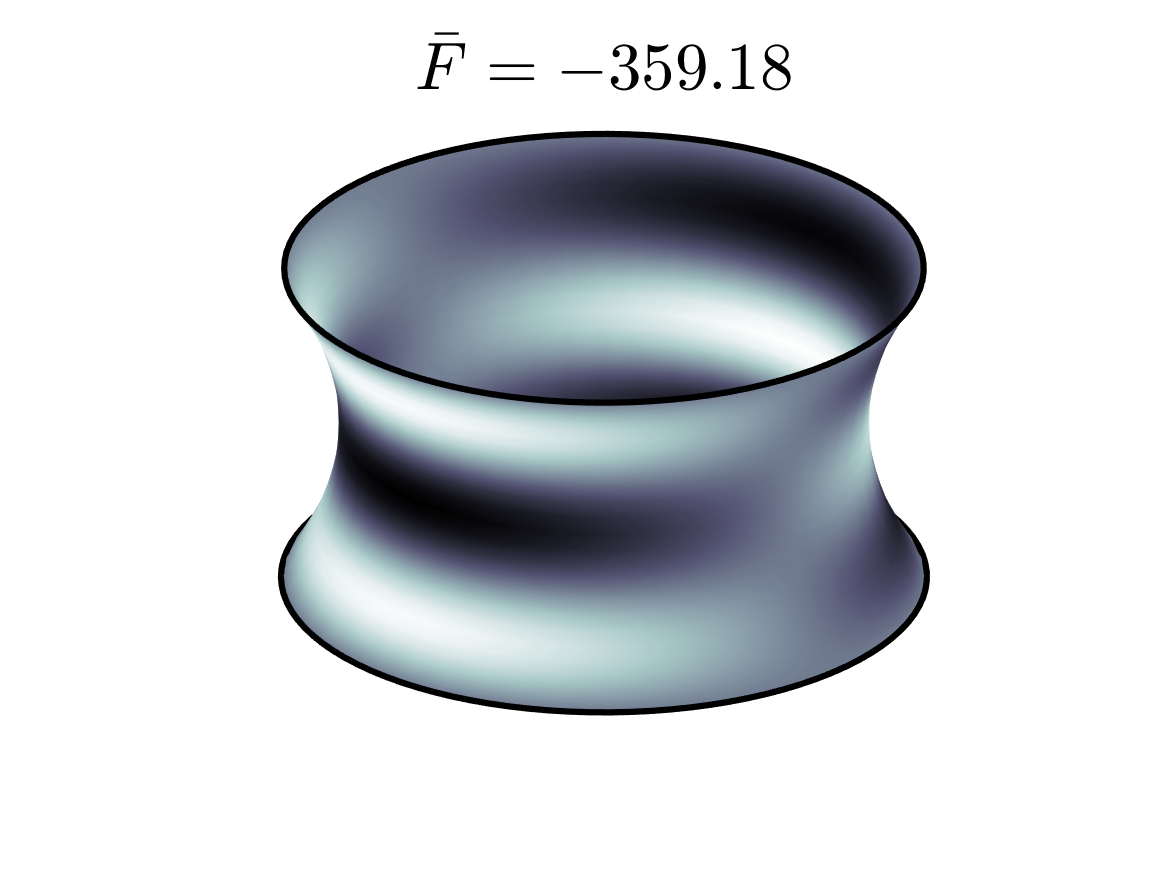}
    \end{subfigure} &
    \begin{subfigure}[b]{0.16\textwidth}
      \includegraphics[trim = 100pt 70pt 100pt 15pt, clip,width=\textwidth]{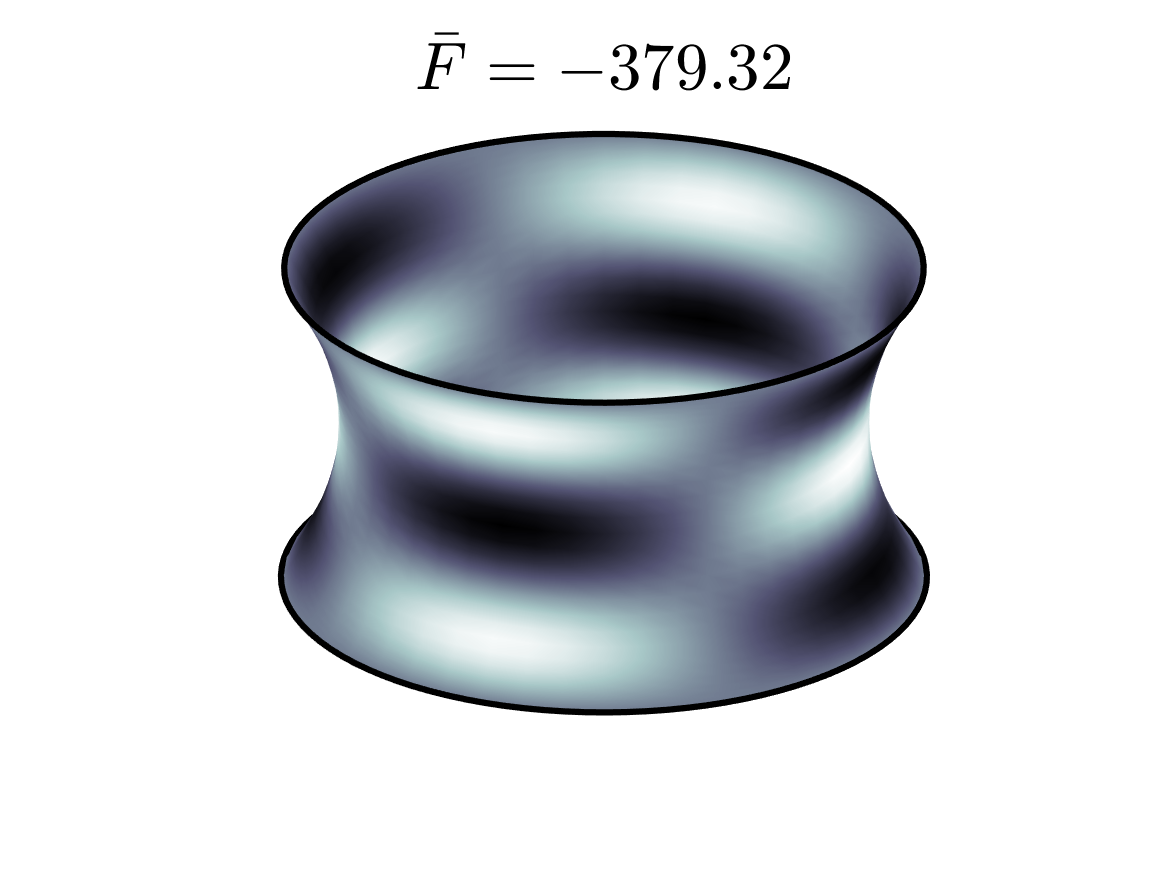}
    \end{subfigure} &
    \begin{subfigure}[b]{0.16\textwidth}
      \includegraphics[trim = 100pt 70pt 100pt 15pt, clip,width=\textwidth]{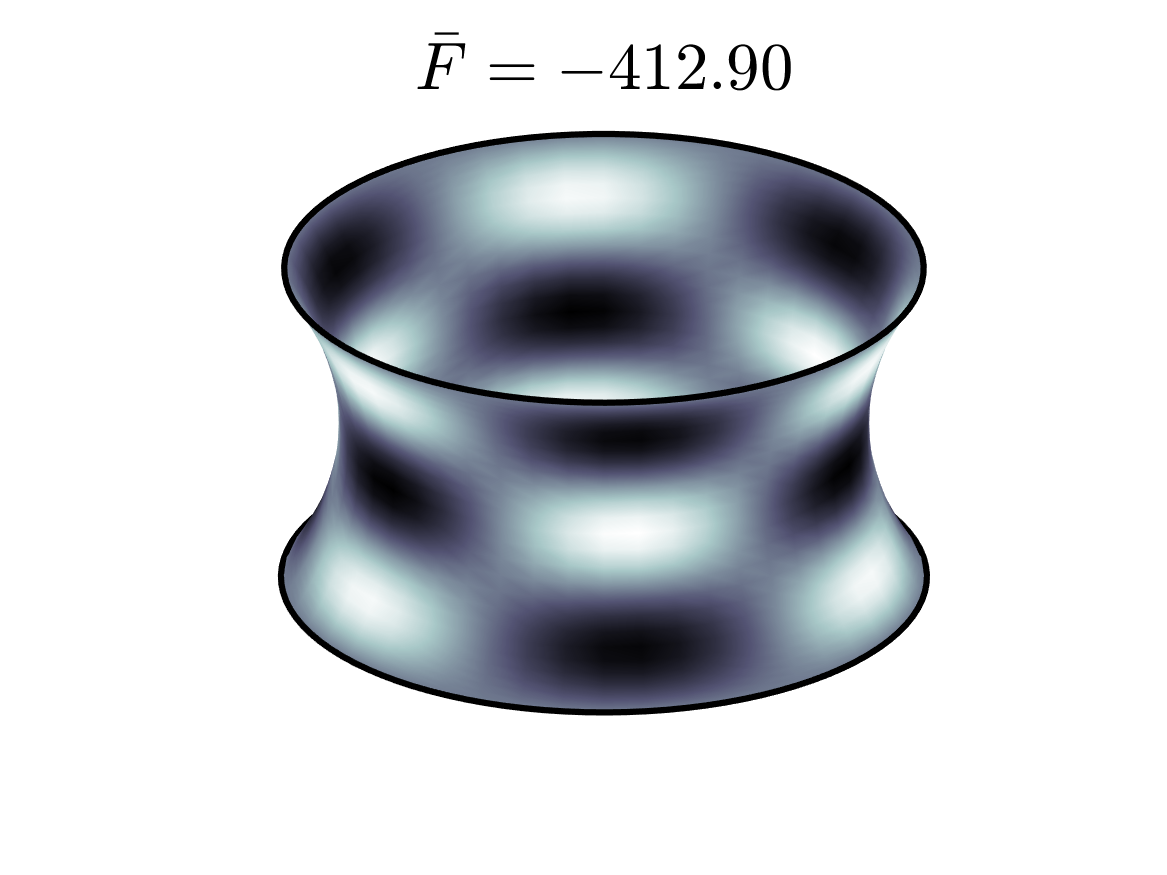}
    \end{subfigure} & \begin{subfigure}[b]{0.16\textwidth}
      \includegraphics[trim = 100pt 70pt 100pt 15pt, clip,width=\textwidth]{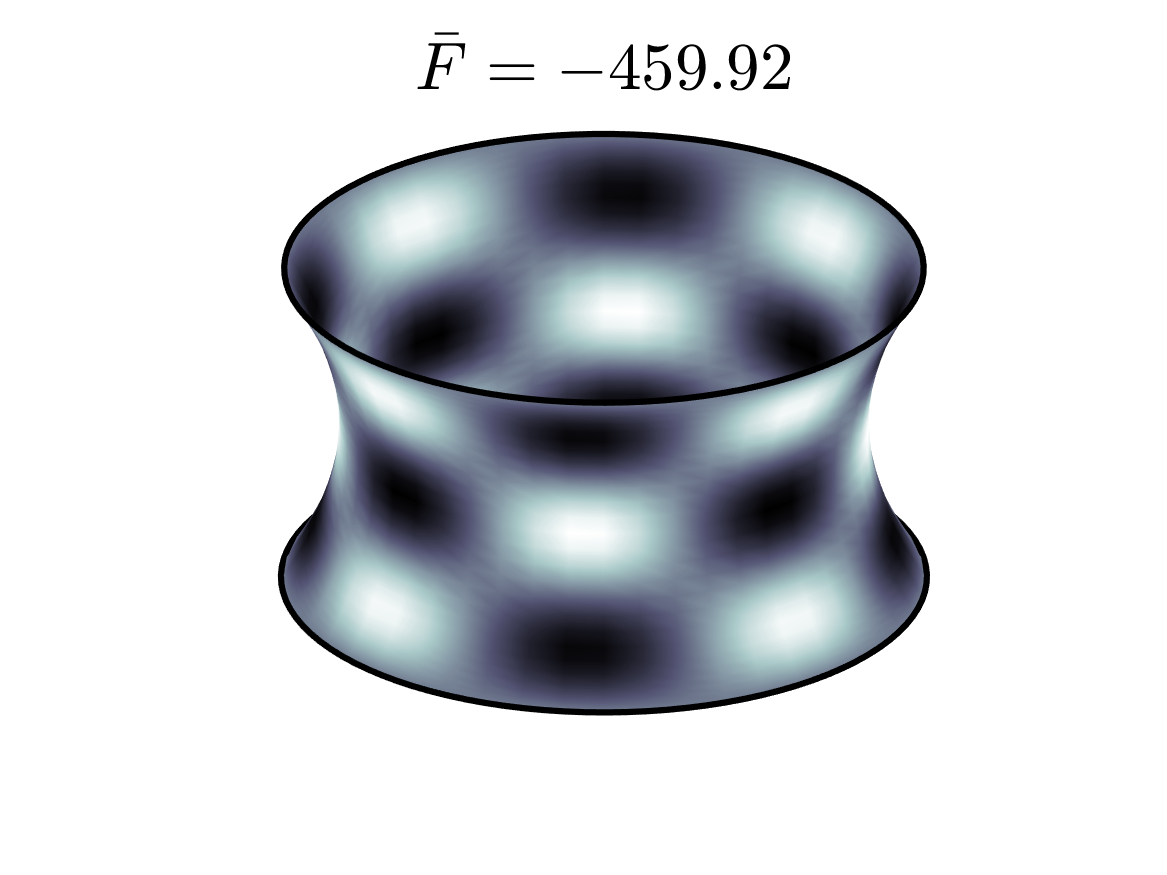}
    \end{subfigure}\\ 

    % --- ROW 9 ---
    &\rotatebox{90}{\textit{ \quad Branch 2}} &
    \begin{subfigure}[b]{0.16\textwidth}
      \includegraphics[trim = 100pt 84pt 100pt 15pt, clip,width=\textwidth]{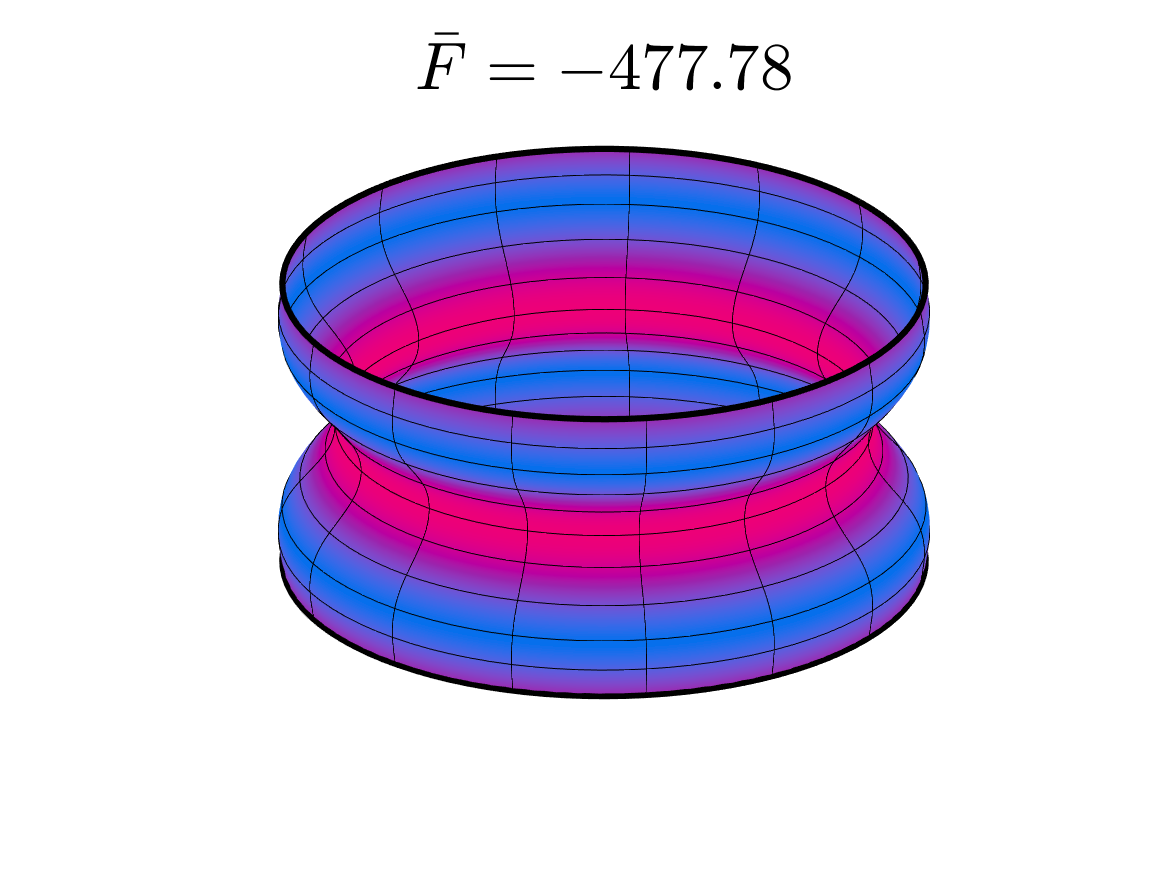}
    \end{subfigure} &
    \begin{subfigure}[b]{0.16\textwidth}
    \includegraphics[trim = 100pt 84pt 100pt 15pt, clip,width=\textwidth]{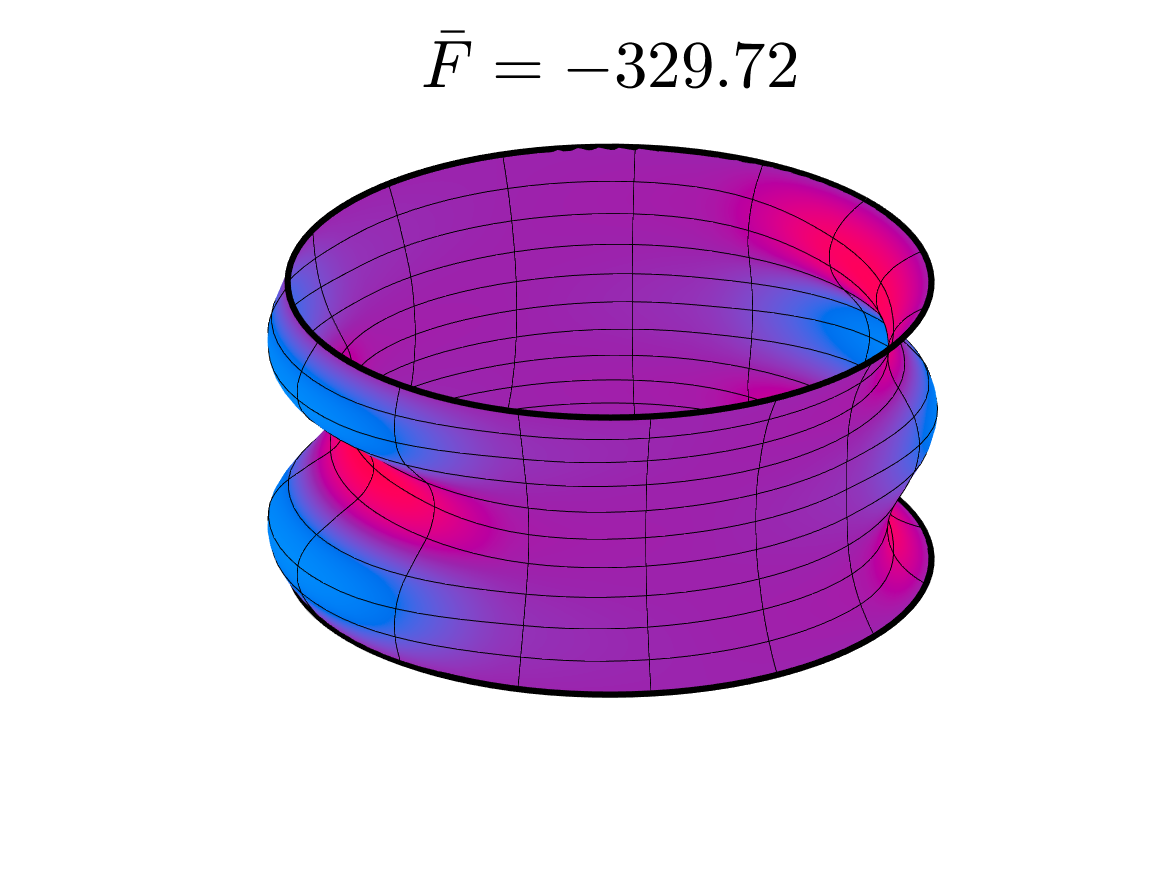}
    \end{subfigure} &
    \begin{subfigure}[b]{0.16\textwidth}
      \includegraphics[trim = 100pt 84pt 100pt 15pt, clip,width=\textwidth]{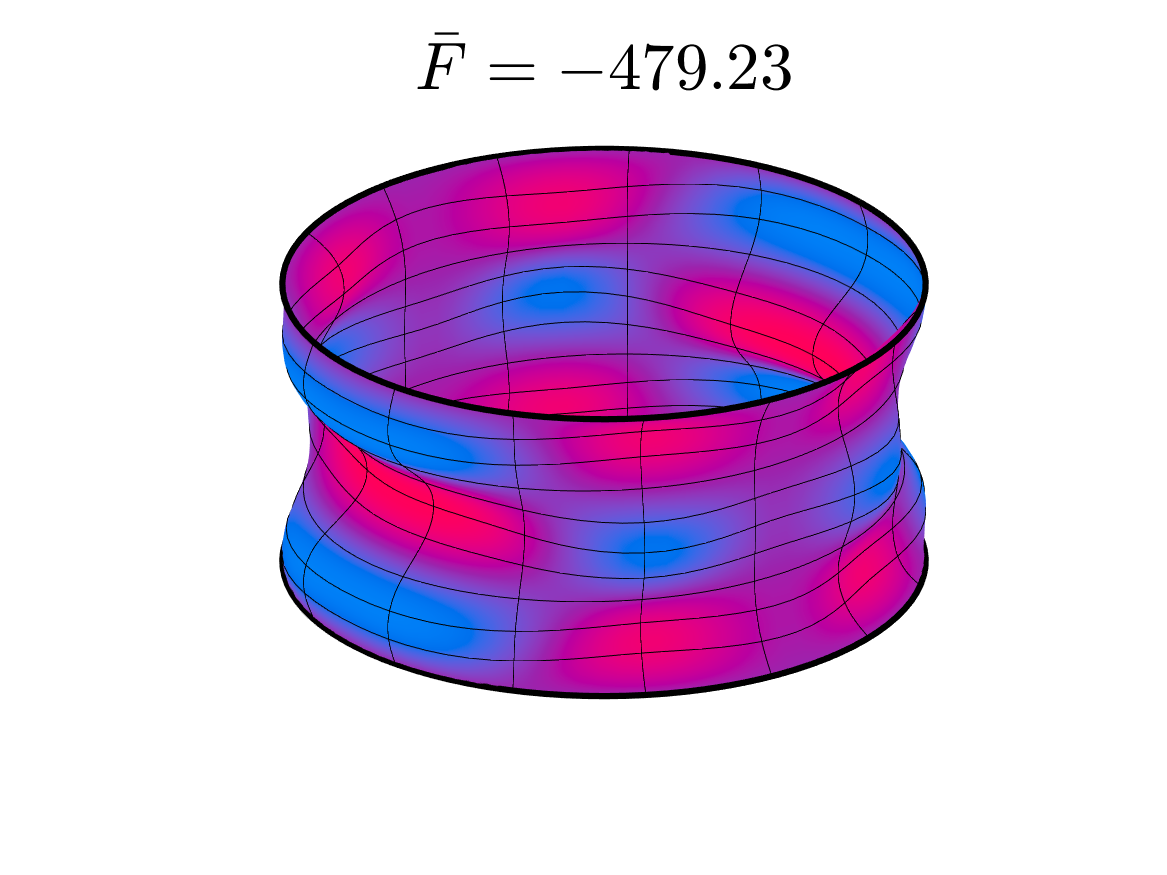}
    \end{subfigure} &
    \begin{subfigure}[b]{0.16\textwidth}
      \includegraphics[trim = 100pt 84pt 100pt 15pt, clip,width=\textwidth]{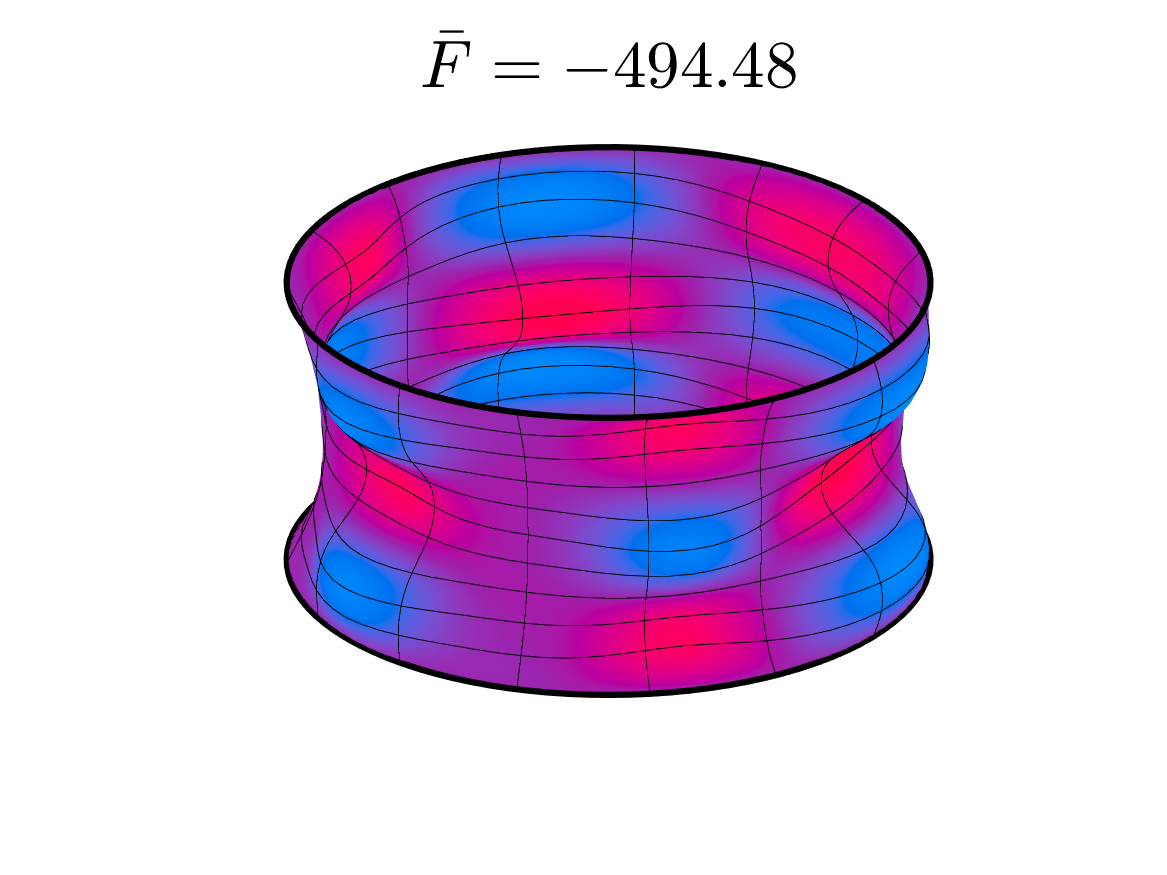}
    \end{subfigure} & \begin{subfigure}[b]{0.16\textwidth}
      \includegraphics[trim = 100pt 84pt 100pt 15pt, clip,width=\textwidth]{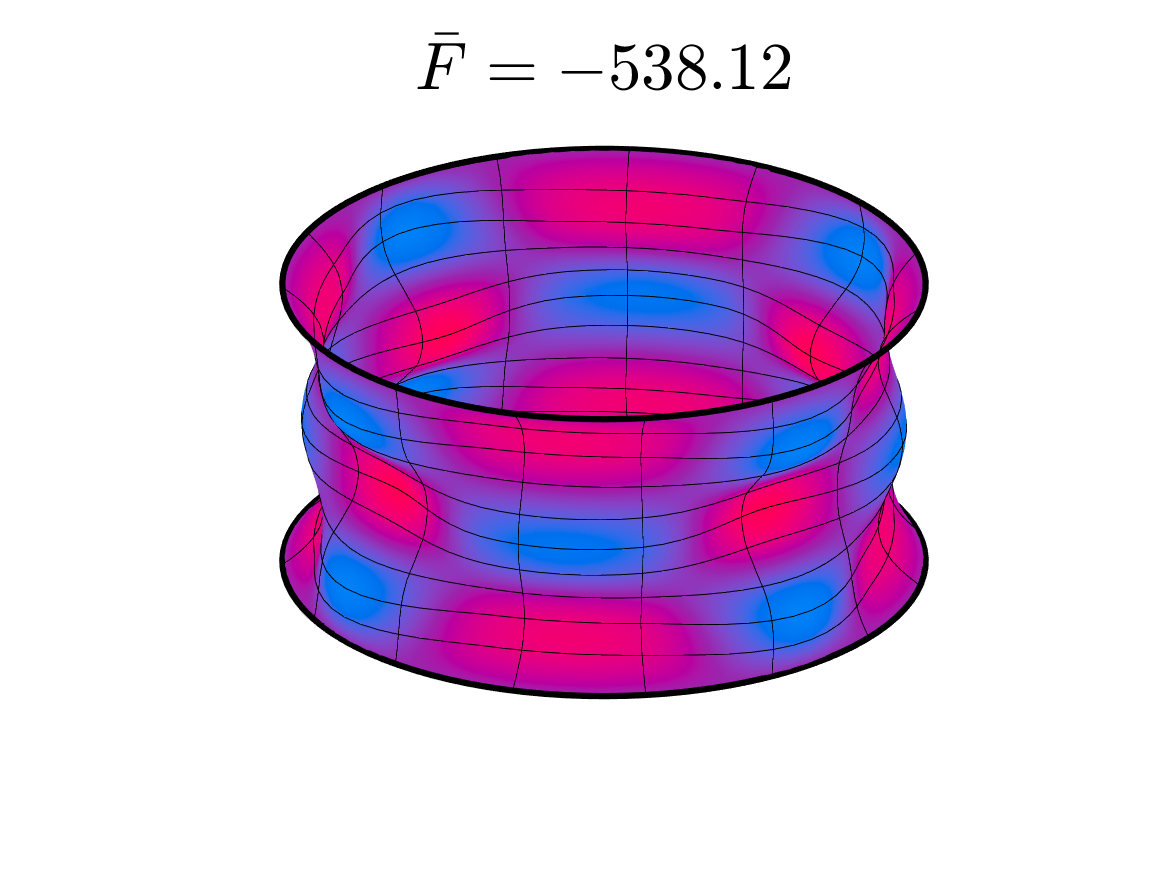}
    \end{subfigure}

    \end{tabular}
  \caption{The eigensurfaces ($h=h^*$), corresponding buckling modes ($h/h^*=0.9$), and dimensionless forces $\bar{F} = Fa/\kappa$, of an elastic fluid interface with circular rings ($\bar{A}=1$). }
  \label{fig:circular_modes}
\end{figure}

%\subsection{Numerical solution to the nonlinear problem}

We now turn to the fullly nonlinear problem, for which we use a pseudospectral method~\citep{trefethen_spectral_2000}. % to solve Eqns.~(\ref{ELeqn})--(\ref{notorqueBC}) in our geometry.
%The coordinate $z$ is discretized using Chebyshev nodes and $\phi$ is discretized uniformly. %Derivatives are approximated spectrally with fast Fourier transforms~\citep{trefethen_spectral_2000}. 
To conform with the traditional spectral formulation, we use a mixed Chebyshev--Fourier basis and treat both  $r$ and $H$ as independent variables and formulate the fourth-order PDE as a pair of second-order PDEs, Eqns.~(\ref{Heqn}) and (\ref{ELeqn}), with respective Dirichlet boundary conditions, Eqns.~(\ref{notorqueBC}a) and (\ref{notorqueBC}b). 
%We split the fourth-order PDE for $r$ into two second-order PDEs,  treating both $r$ and $H$ as independent variables. 
%If $r_{ij}$ and $H_{ij}$ represent their respective discretized versions ($i\in\{1,\hdots,N_{\zeta}\}$ and $j\in \{1,\hdots,N_\phi\}$ ), the first equation is $\mathcal{H}[r_{ij}] = H_{ij}$ 
%with Dirichlet boundary conditions $r_{1j} = (r^-)_j$ and $r_{N_{\zeta} j} = (r^+)_j$,where $\mathcal{H}$ is the mean curvature operator. %The associated boundary conditions are the essential boundary conditions enforcing continuity with the boundary rings, Eqn.~(\ref{dirichletBC}).The other PDE is a discretized version of Eqn.~(\ref{ELeqn}), 
    %     $\kappa (\mathcal{L}[H_{ij}] - 2\mathcal{K}[r_{ij}]H_{ij} + 2H_{ij}^3) - \mu H_{ij} = 0$,
    % with no-torque boundary conditions $H_{1j} = 
    %      H_{N_{\zeta}j} =0$, 
    % where $\mathcal{L}$ represents the Laplace--Beltrami operator and $\mathcal{K}$ represents the Gaussian curvature operator. %To enforce the boundary conditions, we use collocation at the boundary rings. %(i.e., the boundary conditions are enforced at $\zeta=\pm 1$ instead of the PDEs there). 
    We also consider the tension $\mu$ as an unknown, balancing this additional variable with an equation for
    %the area $\mathcal{A}[r_{ij}] = A_0$, where $\mathcal{A}$ represents 
    the surface area constraint. %calculated using the Clenshaw--Curtis rule (for $\zeta$) and the trapezoid rule (for $\phi$). 
    The  discretized system of nonlinear equations is solved using a trust region algorithm with a stopping condition of the dimensionless $\ell^\infty$ residual being less than $10^{-10}$. 
    %From numerical trials, we find that values of $N_{\zeta}$ and $N_\phi$ between 20 and 30 consistently result in an $\ell^\infty$ residual of less than $10^{-10}$. 
Figure~\ref{fig:circular_modes} shows solutions for $\bar{A}=1$ and $h/h^* = 0.9$ for the first few $n$ and $m$ modes, along with their corresponding eigensurfaces at $h=h^*$, found using the  solver. As $n$ and $m$ increase, the surfaces exhibit more oscillations, developing a  more pronounced checkerboard pattern. %The surfaces exhibit several noteworthy symmetries.  %When the rings possess planar symmetry, 
     Consistent with the results of the perturbation theory, each $(n,m)$ mode has two branches. The two branches consist of (sometimes subtly) distinct surfaces when $n$ is odd, while the branches for modes of even $n$ are mirror images of each other. The odd $n$ modes are symmetric about the midplane, that is, they are even functions of $z$, but the modes of even $n$ are not odd functions of $z$. 
     %As $n$ and $m$ increase, the mean curvature pattern increasingly resembles a checkerboard. %Because of the nonlinear nature of the problem, a solution born from $H^{(1/2)}$ and one from $-H^{(1/2)}$ will not be identical; thus, there are two distinct branches for each odd $n$ emanating from the same base state. 
   % The modes of even $n$ lack this symmetry: the two branches are mirror images of each other. 
    With increasing $m$ or $n$, the force $F$ generally becomes more negative (compressive); a notable exception to this trend is  $m=1$, for which the forces required are smaller than for all other $m$. We attribute this to the $m=1$ modes being essentially translations of cross-sections, involving comparatively little bending in the azimuthal direction and thus requiring less force, making the system more susceptible to a snap-through-type deformation~\citep{vasan_mechanical_2020}. %This is consistent with the fact that high-aspect-ratio structures are most susceptible to $m=1$ instabilities~\citep{goriely_nonlinear_2008}. 
    All of the shapes  in Fig.~\ref{fig:circular_modes} require $T=0$.

\section{Results: Noncircular Boundaries}

\begin{figure}
  \centering
  \begin{tabular}{c c| c c |c c c}
  
    % --- COLUMN HEADERS ---
    & {\textbf{$n=1,m=0$}}& \multicolumn{2}{c|}{\textbf{$n=1,m=1$}} & \multicolumn{2}{c}{\textbf{$n=2,m=3$}} &\\

    % --- ROW 1 ---
     \rotatebox{90}{\textit{ 
    \quad Branch 1}} &
    \begin{subfigure}[b]{0.16\textwidth}
      \includegraphics[trim = 100pt 70pt 100pt 15pt, clip,width=\textwidth]{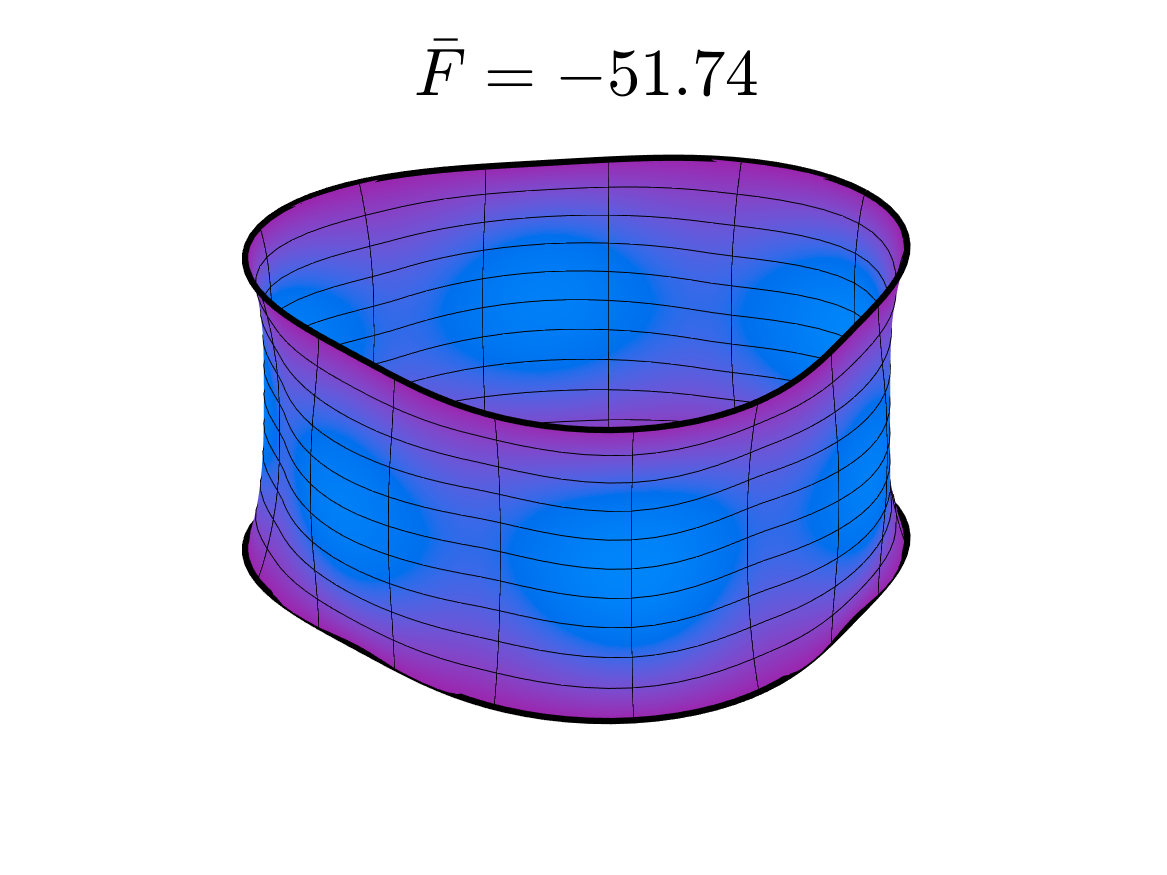}
    \end{subfigure} &
    \begin{subfigure}[b]{0.16\textwidth}
      \includegraphics[trim = 100pt 70pt 100pt 15pt, clip,width=\textwidth]{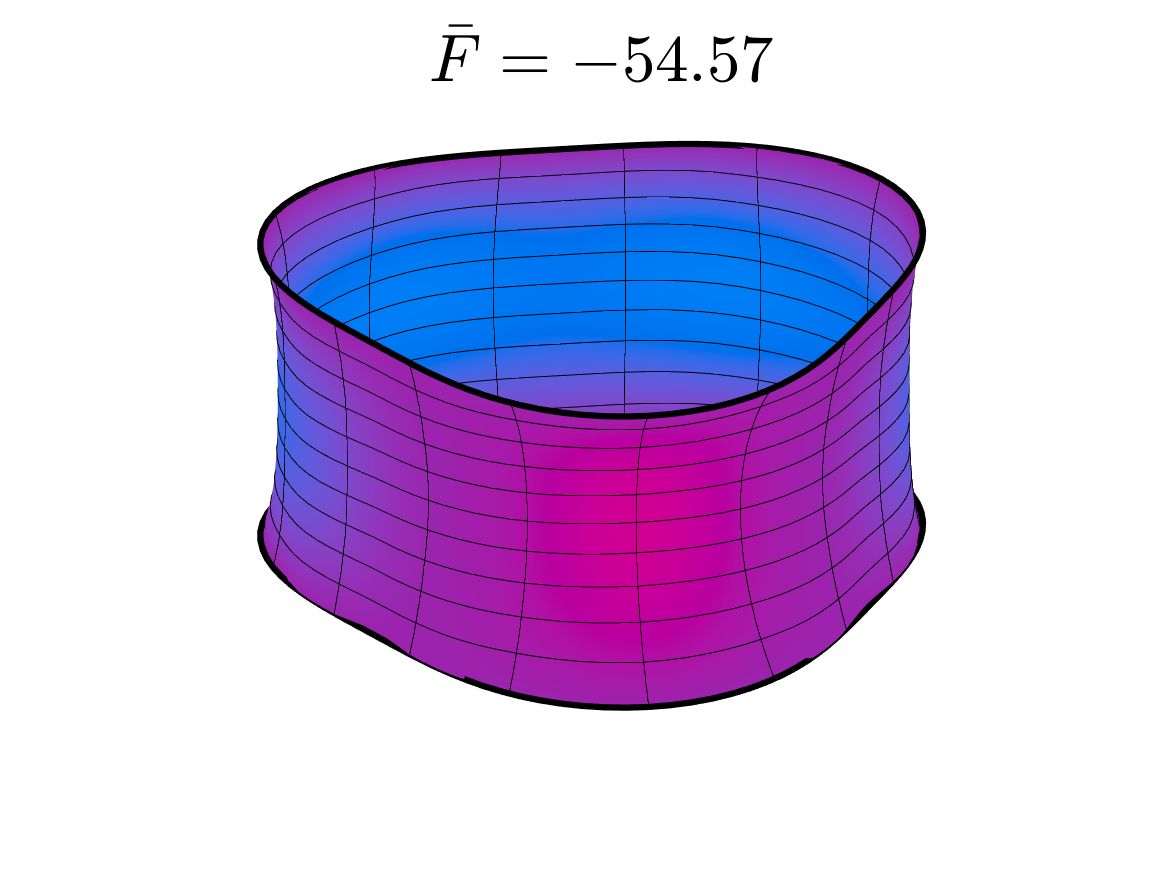}
    \end{subfigure} &
    \begin{subfigure}[b]{0.16\textwidth}
      \includegraphics[trim = 100pt 70pt 100pt 15pt, clip,width=\textwidth]{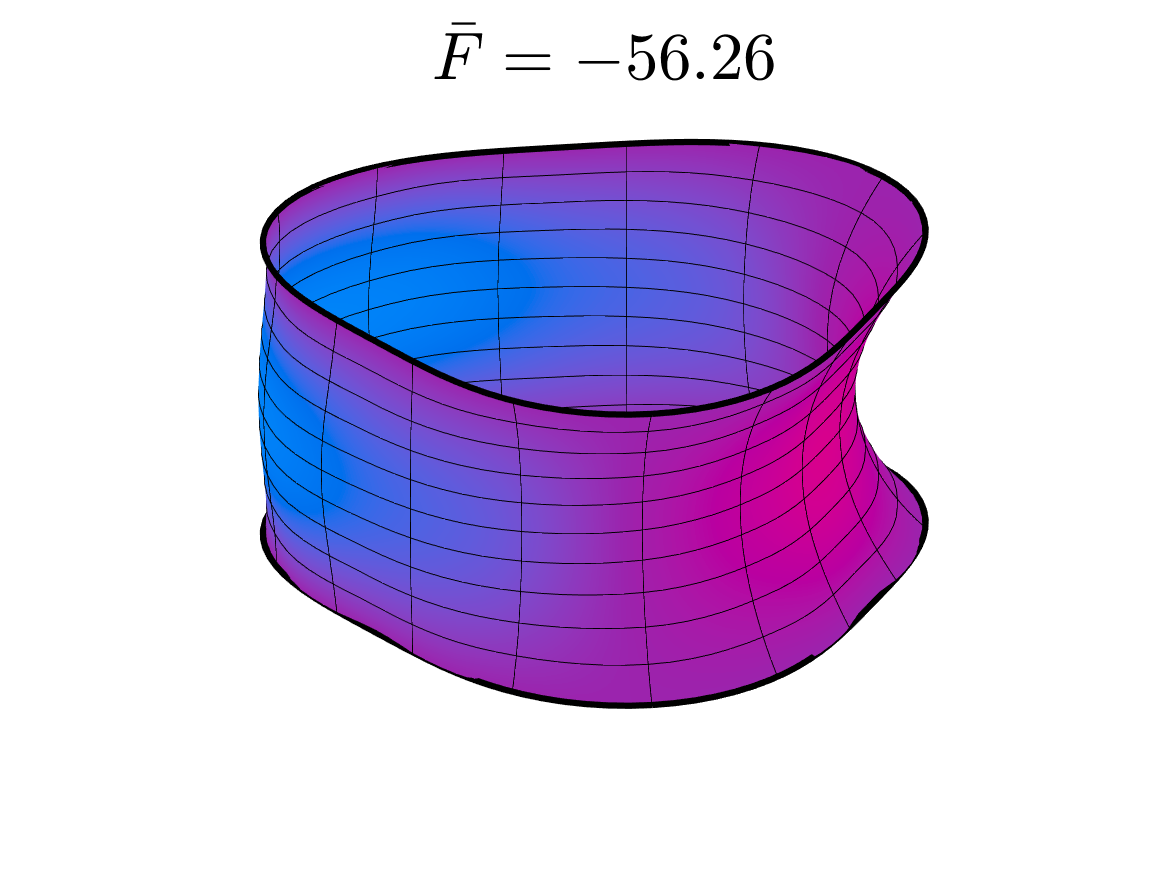}
    \end{subfigure} &
    \begin{subfigure}[b]{0.16\textwidth}
      \includegraphics[trim = 100pt 70pt 100pt 15pt, clip,width=\textwidth]{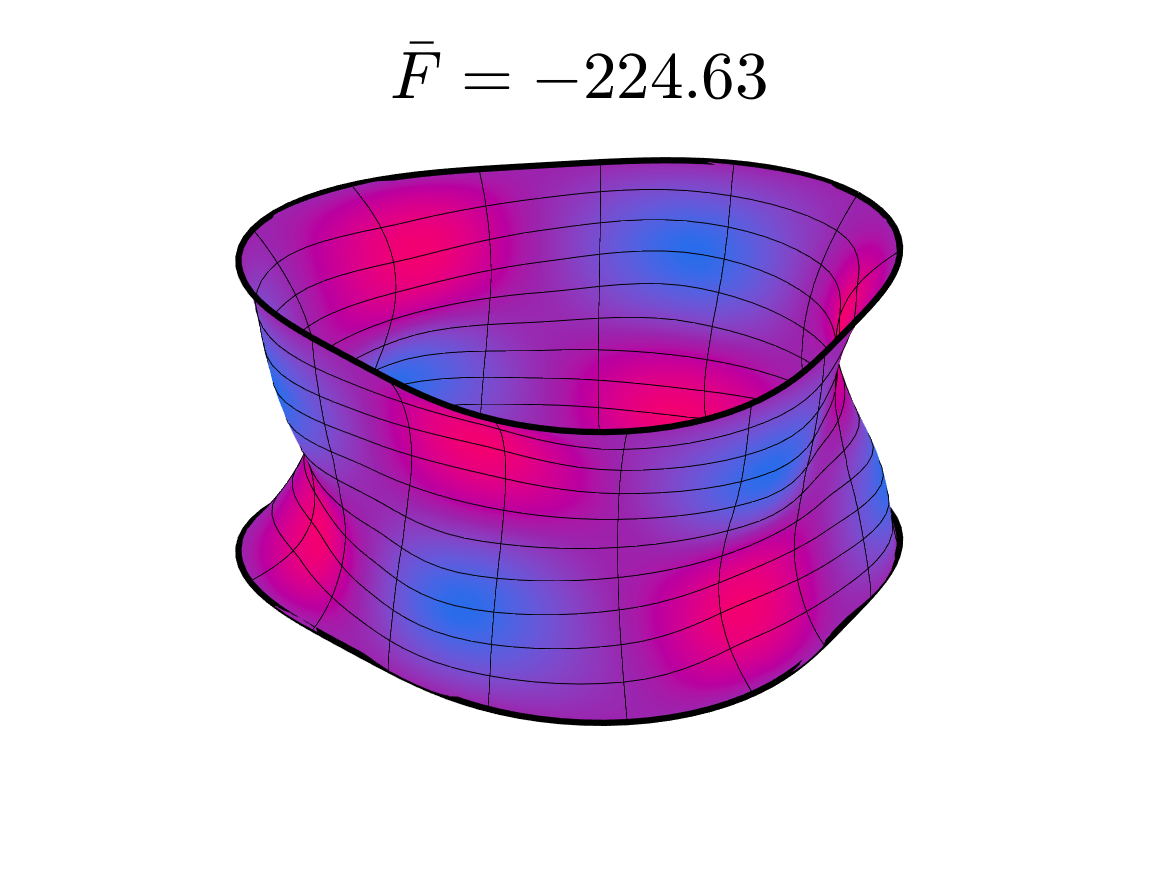}
    \end{subfigure} & \begin{subfigure}[b]{0.16\textwidth}
      \includegraphics[trim = 100pt 70pt 100pt 15pt, clip,width=\textwidth]{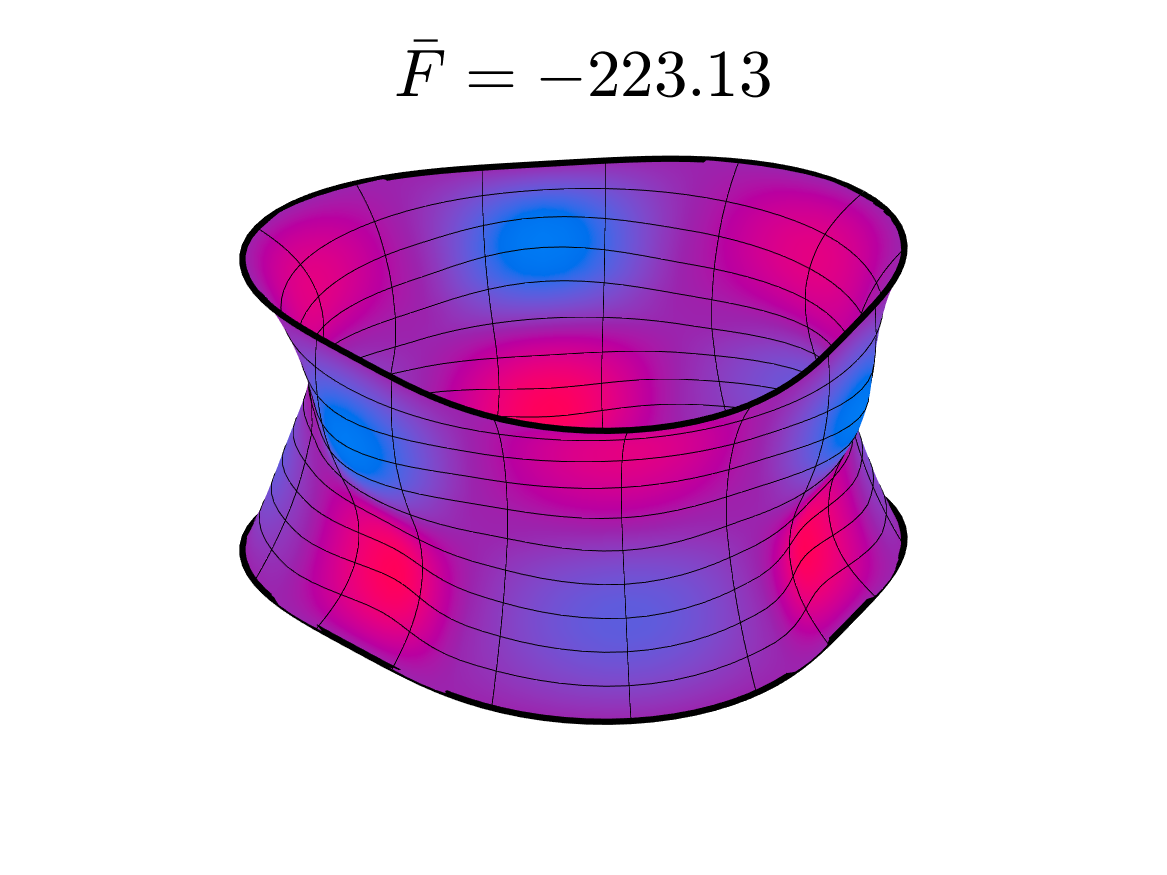}
    \end{subfigure} \\

    % --- ROW 2 ---
   \rotatebox{90}{\textit{ Eigensurfaces}} &
    \begin{subfigure}[b]{0.16\textwidth}
      \includegraphics[trim = 100pt 70pt 100pt 15pt, clip,width=\textwidth]{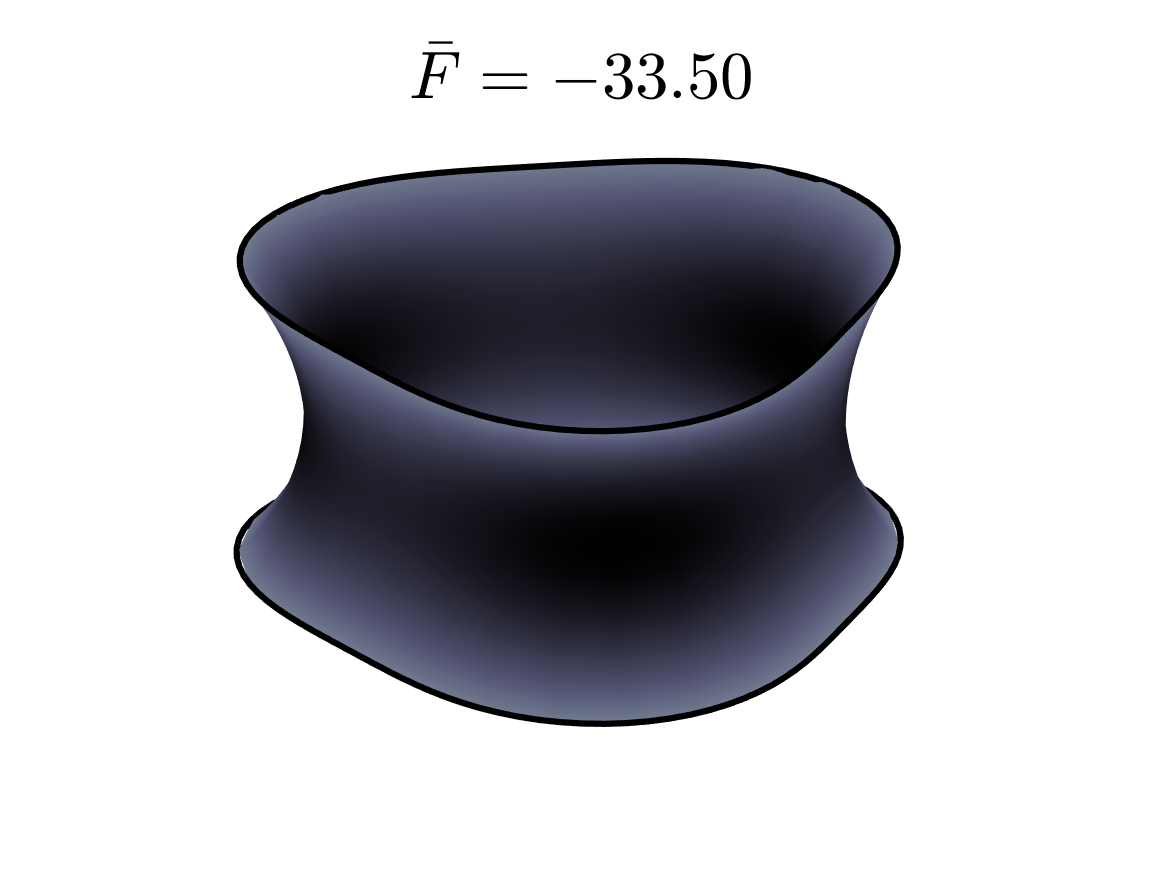}
    \end{subfigure} &
    \begin{subfigure}[b]{0.16\textwidth}
      \includegraphics[trim = 100pt 70pt 100pt 15pt, clip,width=\textwidth]{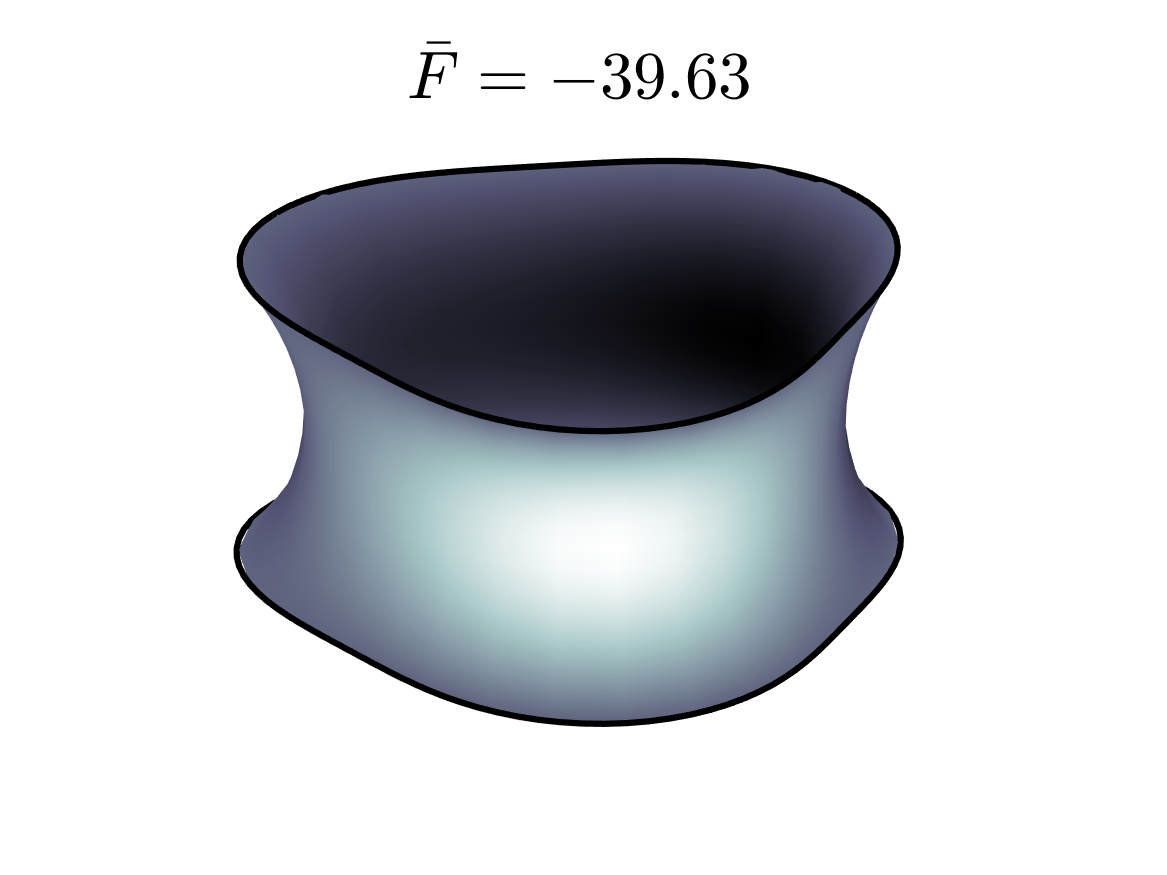}
    \end{subfigure} &
    \begin{subfigure}[b]{0.16\textwidth}
      \includegraphics[trim = 100pt 70pt 100pt 15pt, clip,width=\textwidth]{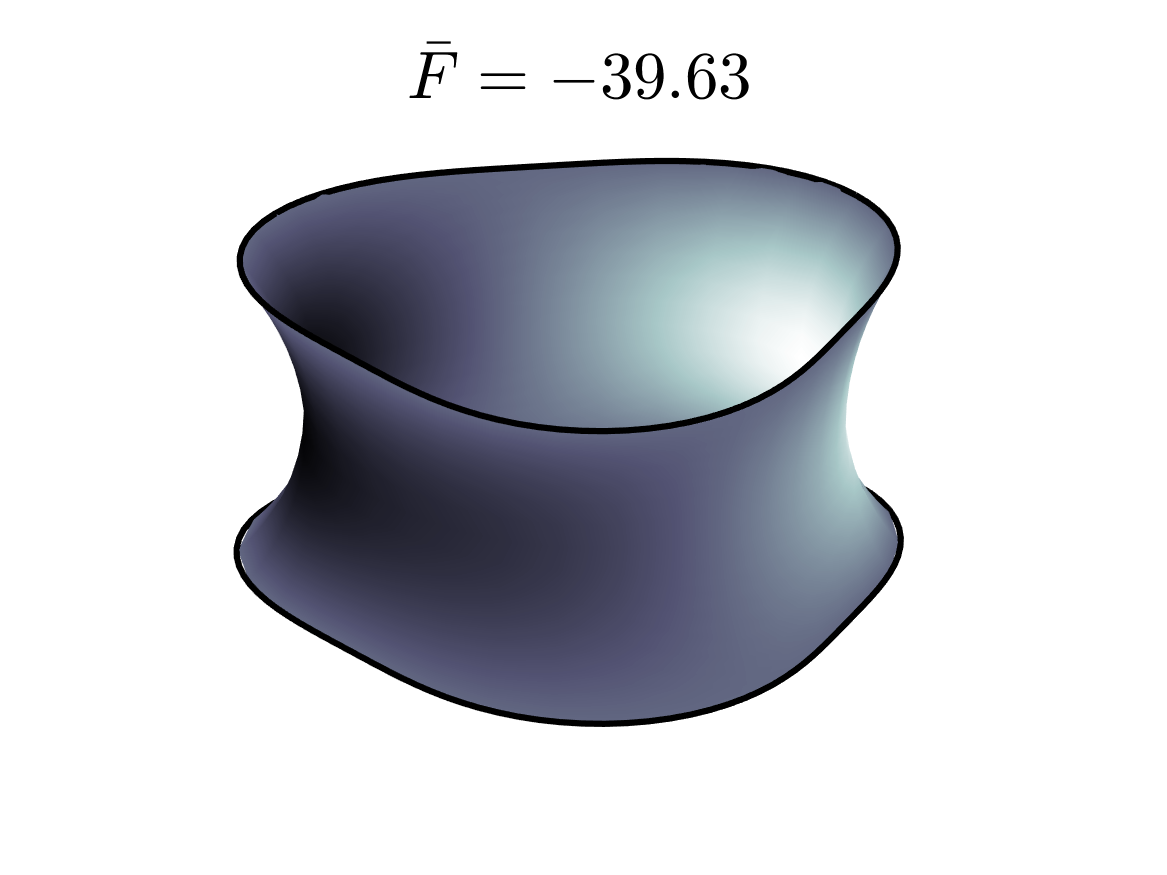}
    \end{subfigure} &
    \begin{subfigure}[b]{0.16\textwidth}
      \includegraphics[trim = 100pt 70pt 100pt 15pt, clip,width=\textwidth]{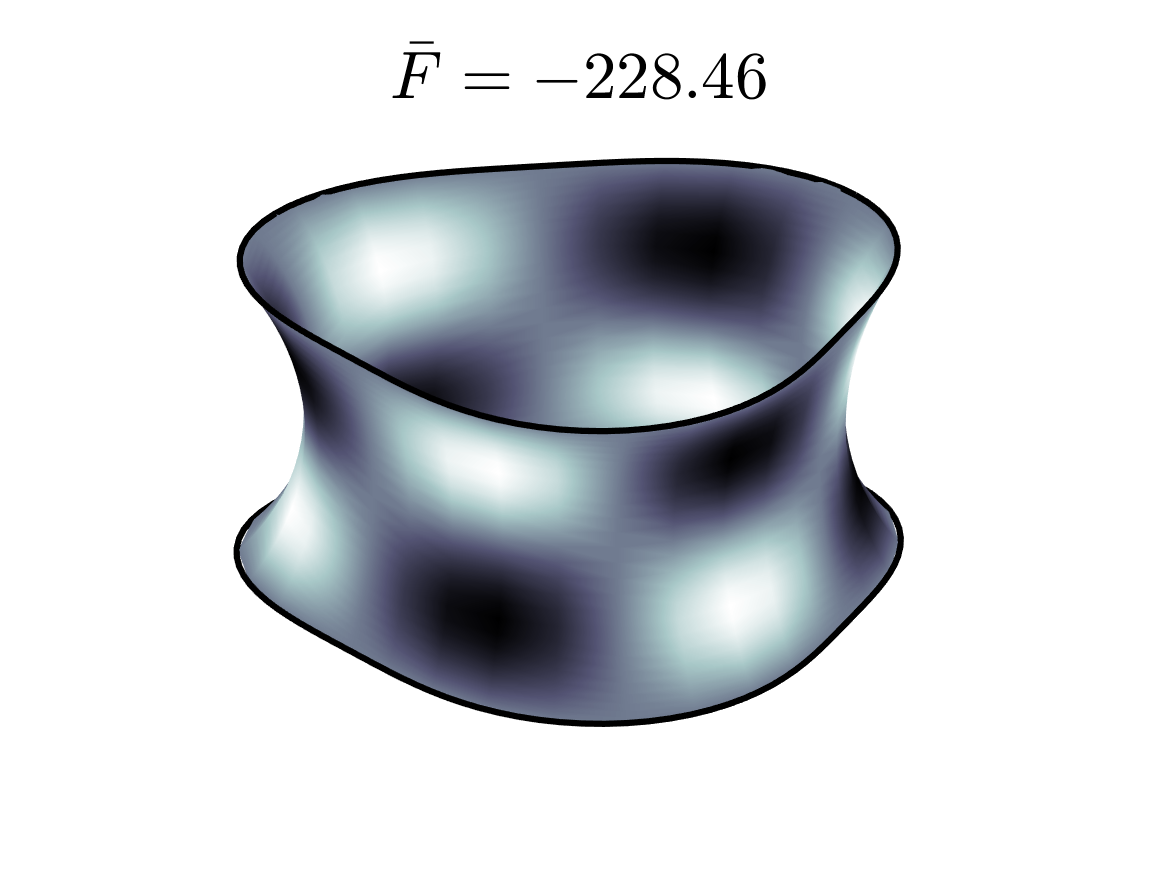}
    \end{subfigure} & \begin{subfigure}[b]{0.16\textwidth}
      \includegraphics[trim = 100pt 70pt 100pt 15pt, clip,width=\textwidth]{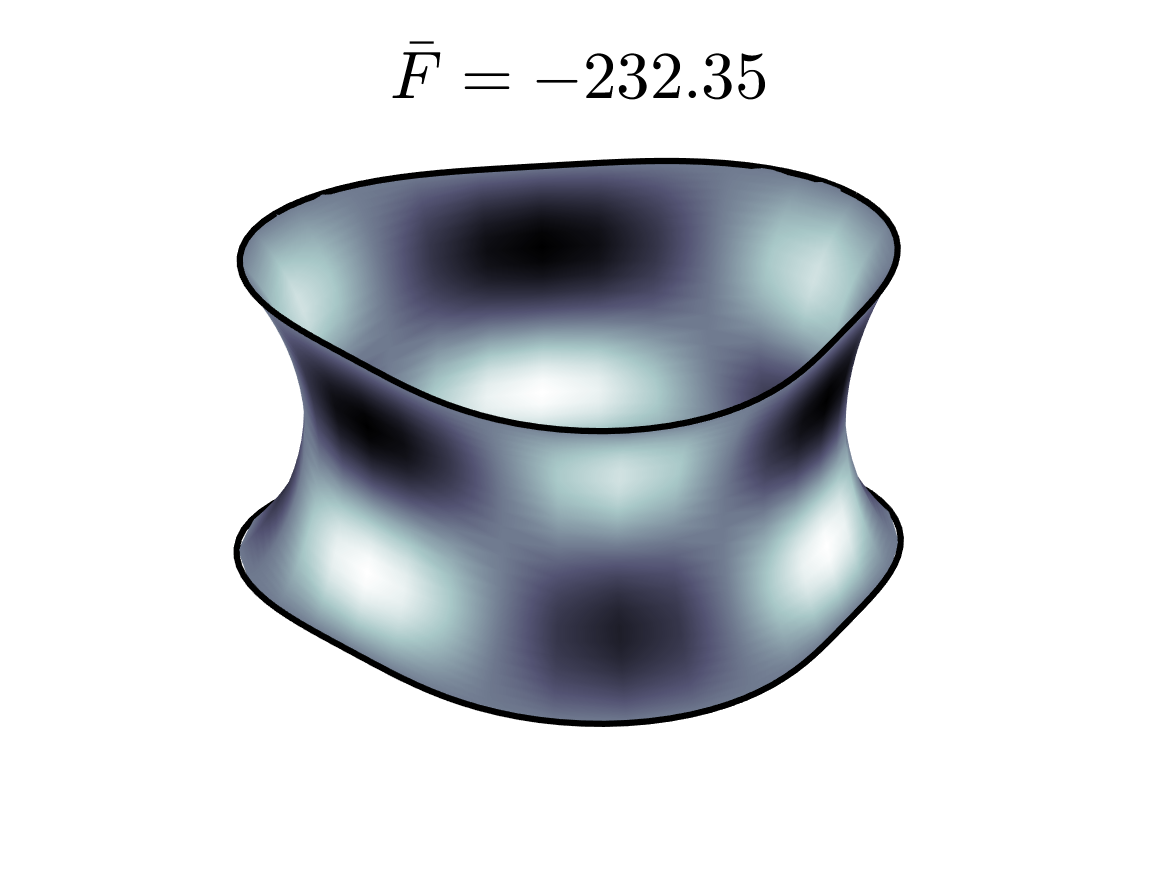}
    \end{subfigure}\\

    % --- ROW 3 ---
    \rotatebox{90}{\textit{ \quad Branch 2}} &
    \begin{subfigure}[b]{0.16\textwidth}
      \includegraphics[trim = 100pt 70pt 100pt 15pt, clip,width=\textwidth]{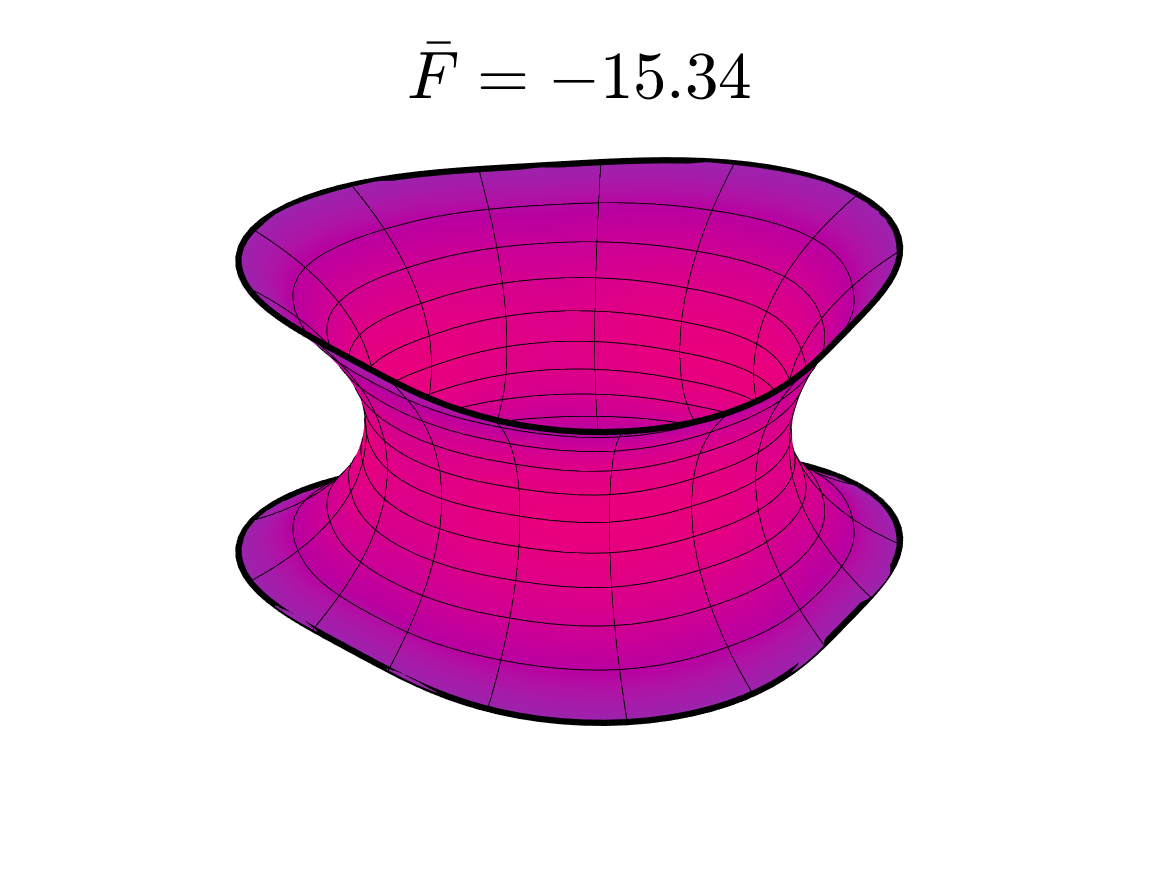}
    \end{subfigure} &
    \begin{subfigure}[b]{0.16\textwidth}
      \includegraphics[trim = 100pt 70pt 100pt 15pt, clip,width=\textwidth]{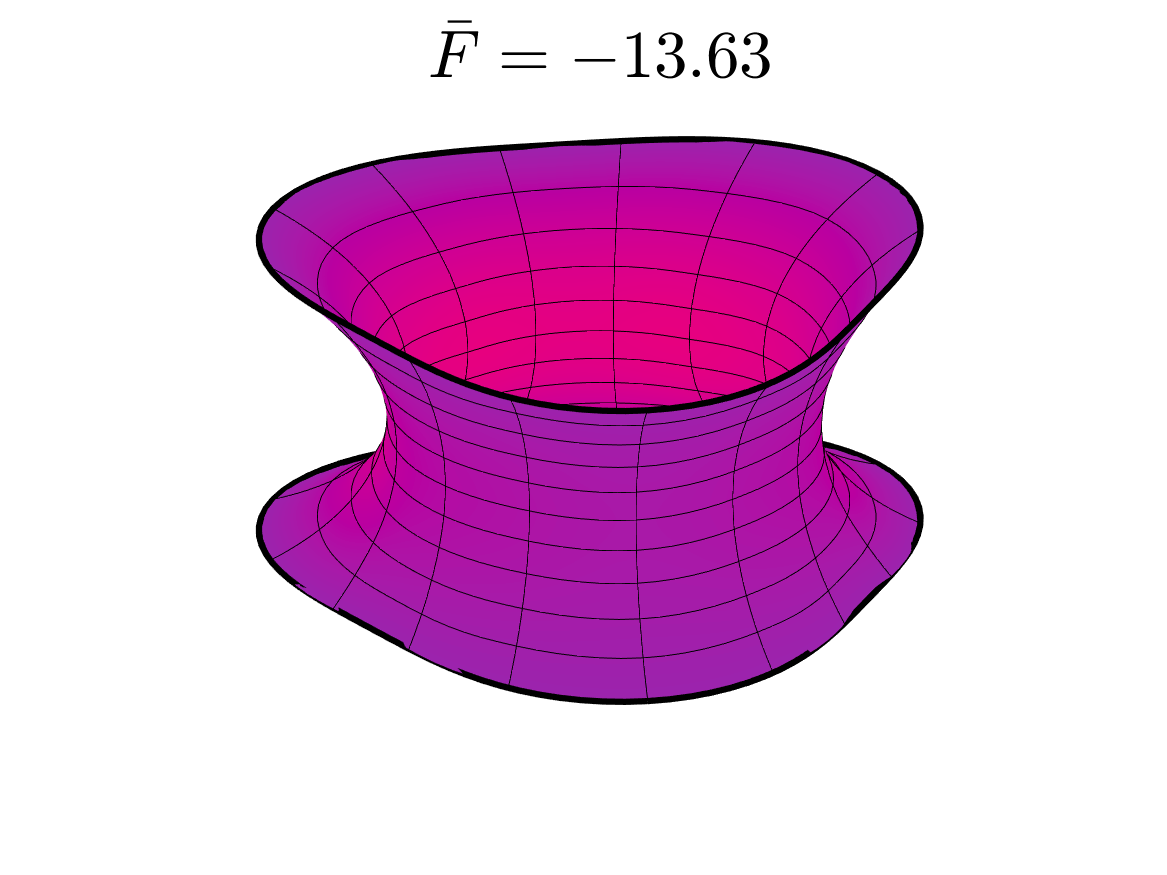}
    \end{subfigure} &
    \begin{subfigure}[b]{0.16\textwidth}
      \includegraphics[trim = 100pt 70pt 100pt 15pt, clip,width=\textwidth]{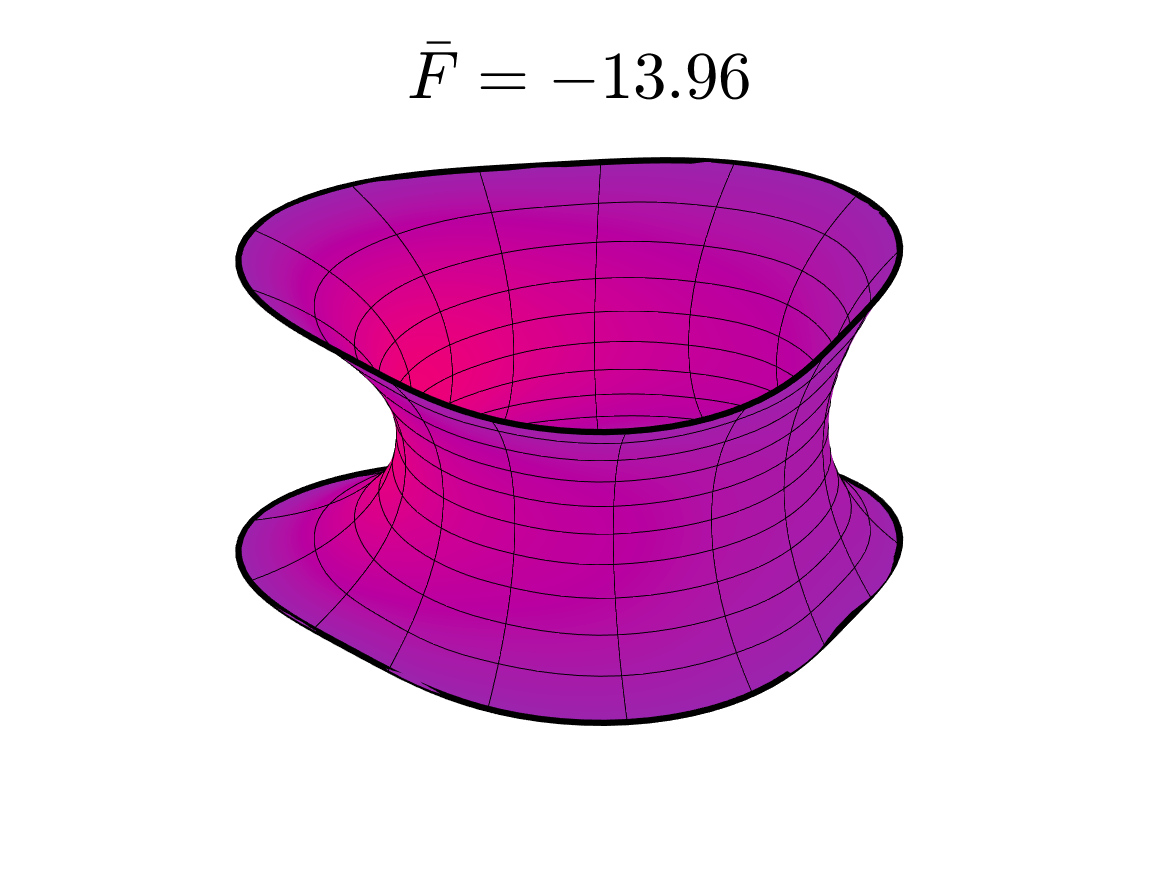}
    \end{subfigure} &
    \begin{subfigure}[b]{0.16\textwidth}
      \includegraphics[trim = 100pt 70pt 100pt 15pt, clip,width=\textwidth]{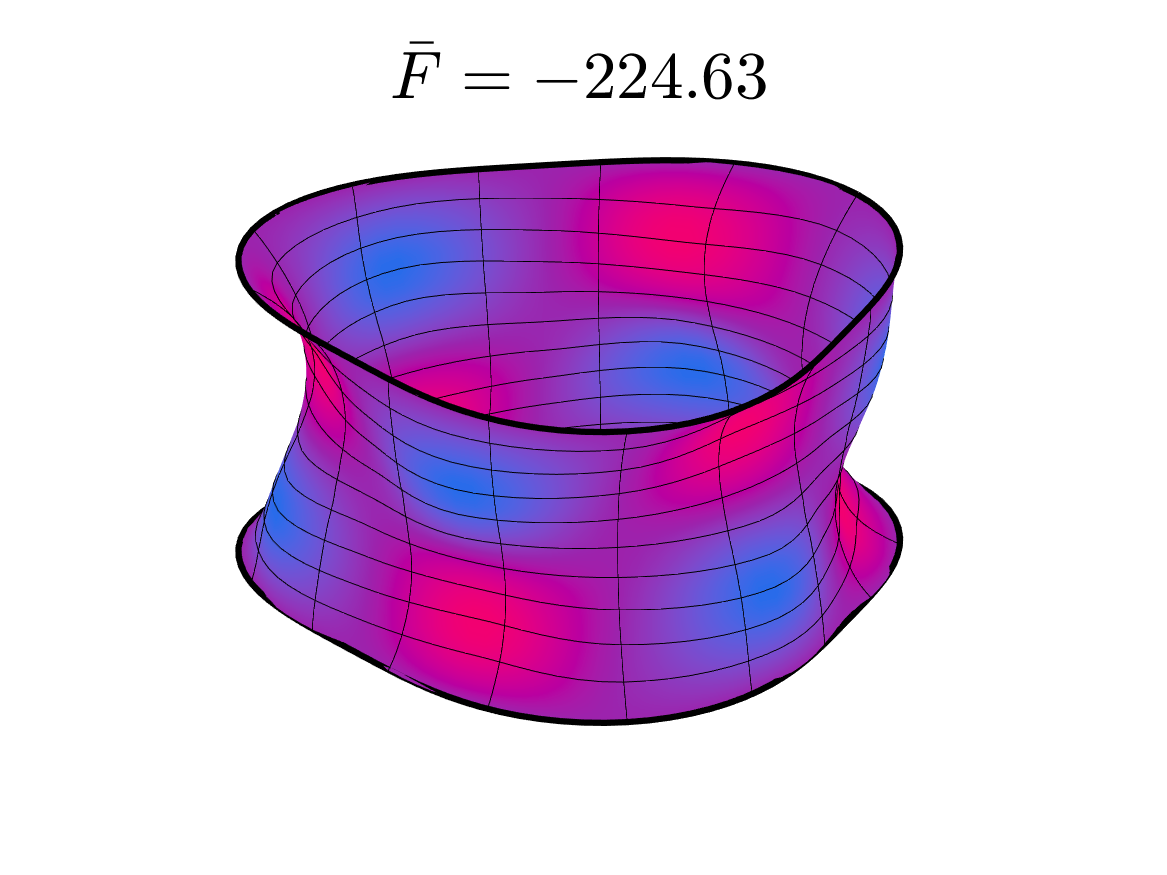}
    \end{subfigure} & \begin{subfigure}[b]{0.16\textwidth}
      \includegraphics[trim = 100pt 70pt 100pt 15pt, clip,width=\textwidth]{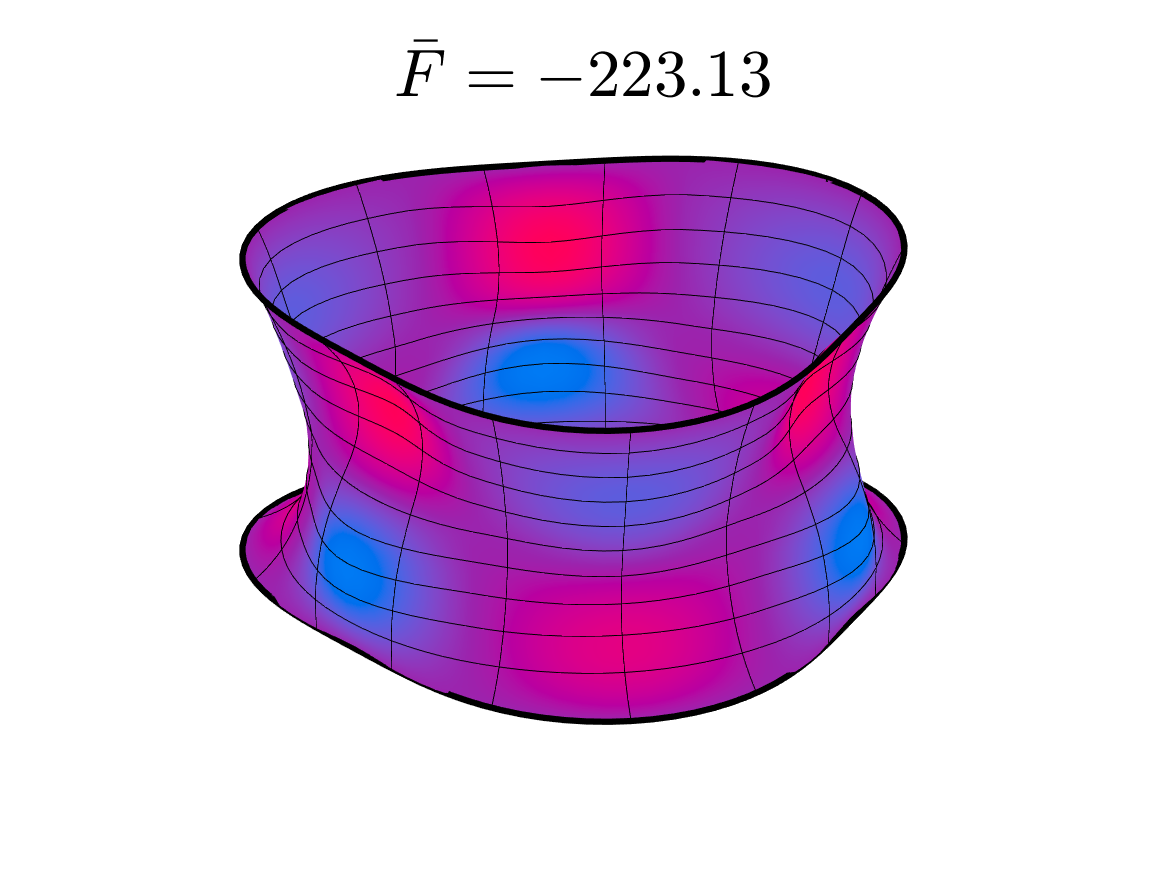}
    \end{subfigure} \\ 
    
    % --- ROW 4 ---
    &
    \begin{subfigure}[b]{0.16\textwidth}
      \includegraphics[trim = 10pt 16pt 10pt 325pt, clip, width=\textwidth]{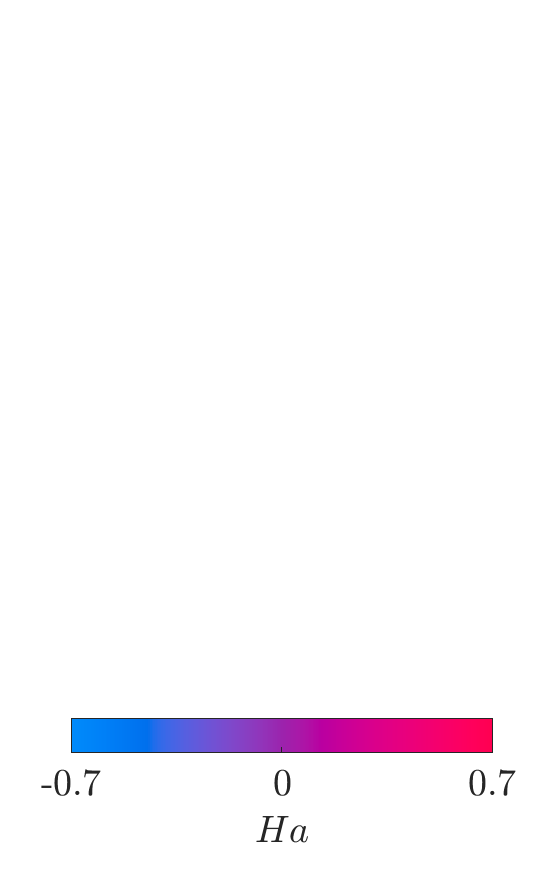}
    \end{subfigure} & \multicolumn{2}{c|}{ \begin{subfigure}[b]{0.16\textwidth}
      \includegraphics[trim = 10pt 16pt 10pt 325pt, clip, width=\textwidth]{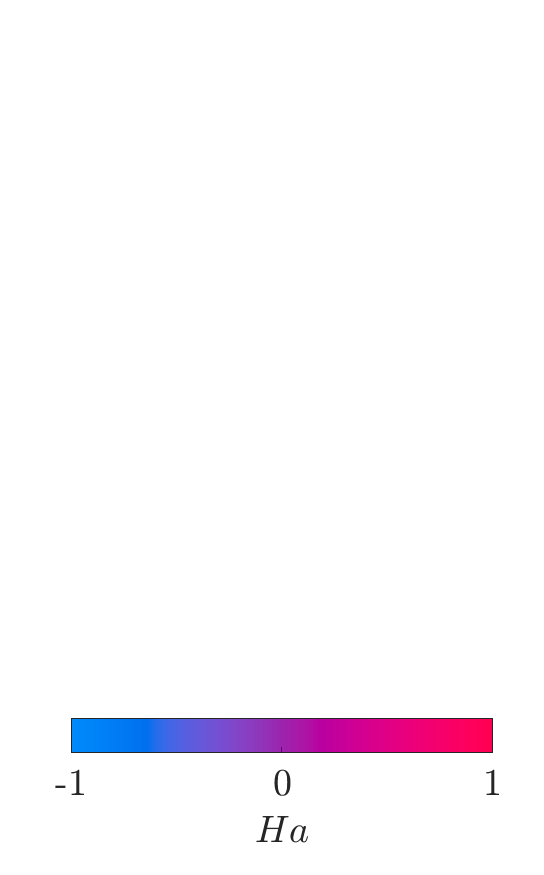}
      \end{subfigure}} & \multicolumn{2}{c}{ \begin{subfigure}[b]{0.16\textwidth}
      \includegraphics[trim = 10pt 16pt 10pt 325pt, clip, width=\textwidth]{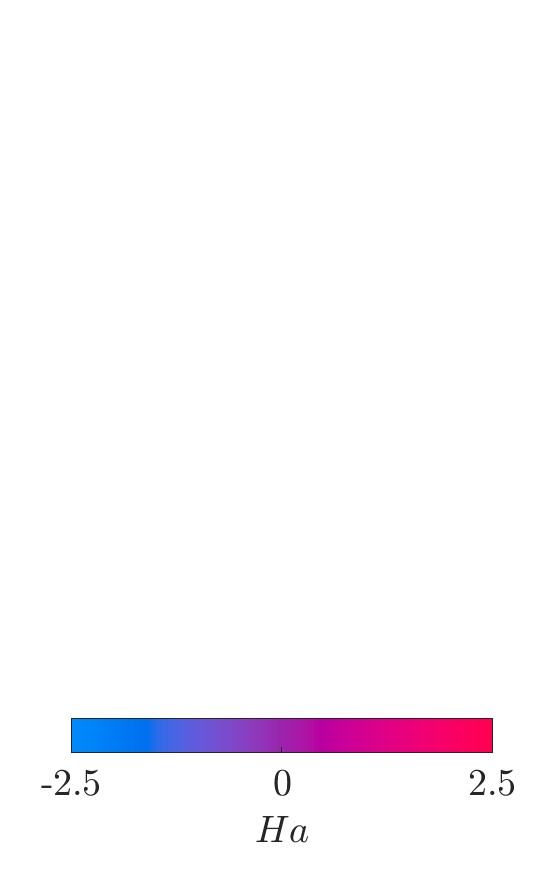}
      \end{subfigure}}

    \end{tabular}
  \caption{The eigensurfaces ($h=h^*$), corresponding buckling modes ($h/h^*= 0.9475...$), and dimensionless forces $\bar{F} = Fa/\kappa$, of an elastic fluid interface with  boundary rings of the form $r^\pm(\phi)/a = 1 + 0.1\cos (3\phi)$ ($\bar{A}=1.0341...$). Sine and cosine modes are distinct.}
  \label{fig:asym_modes}
\end{figure}

%In the general (nonaxisymmetric) case, we find that the modes can still be similarly catalogued in terms of $m$ and $n$.
We now explore how noncircular boundaries affect the surface shape and ring force. To this end, we consider boundaries of the form $r^\pm(\phi)/a = 1 + \delta \sum_k e^{\mathrm{i}k\phi}\hat{r}^\pm_k $,  $0<\delta\ll 1 $, and perturbatively calculate the minimal surface, taken to be of the form $r^{[0]}(z)+\delta r^{[1]}(\phi,z)+\hdots$, where $r^{[0]} = b\cosh (z/b)$ and $r^{[1]} = \sum_k e^{\mathrm{i}k\phi}\hat{r}_k(z)$. %We assume that the area is compatible with the boundary. 
 Expanding Eqn.~(\ref{Heqn}) to $O(\delta)$  gives a boundary value problem for the coefficients $r_k(z)$ (see also~\citep{walzel_perturbing_2022}):
\begin{equation}
    \frac{1}{2\cosh^3 \frac{z}{b}}\left[ \hat{r}''_k - \frac{2}{b}\tanh\frac{z}{b} \hat{r}_k' + \frac{1-k^2}{b^2}\hat{r}_k\right]=0, \quad \hat{r}_k(z=\pm h/2) = \hat{r}_k^\pm
    \label{perturbedbvp}
\end{equation}
We assume without loss that $\hat{r}^\pm_0=0$ so that the area constraint is automatically satisfied at $O(\delta)$. The general solution of Eqn.~(\ref{perturbedbvp}) can be written using hypergeometric functions; using these $\hat{r}_k(z)$, the $O(\delta)$ Laplace--Beltrami operator and Gaussian curvature are
{\setlength{\abovedisplayskip}{1pt}
\setlength{\belowdisplayskip}{0pt}
\begin{align}
    \Delta^{[1]}&=-\frac{1}{\cosh^3 \frac{z}{b}} \sum_k e^{\mathrm{i}k\phi}\left[\frac{2\mathrm{i}k \hat{r}_k}{b^3\cosh^2 \frac{z}{b}}\partial_\phi  + \tanh\frac{z}{b}\left(\frac{(1-k^2) \hat{r}_k}{b^2}-\frac{2\tanh \frac{z}{b} \hat{r}_k'}{b}+ \hat{r}_k''\right) \partial_z\right.\nonumber\\
    &\quad\left. + \frac{2\hat{r}_k}{b^3}\partial_\phi^2+ \frac{2\mathrm{i}k\hat{r}_k\tanh \frac{z}{b}}{b^2}\partial_z\partial_\phi + 2\tanh\frac{z}{b} \hat{r}_k' \partial_z^2\right] 
\end{align}}
\begin{equation}
    K^{[1]} = -\frac{1}{b\cosh^5 \frac{z}{b}} \sum_ke^{\mathrm{i}k\phi}\left[\hat{r}_k'' - \frac{4}{b}\tanh\frac{z}{b} \hat{r}_k' - \frac{1-k^2}{b^2}\hat{r}_k \right],
\end{equation}
from which we define $-\mathcal{J}^{[1]} = \Delta^{[1]}-2K^{[1]}$. Classical perturbation theory for degenerate self-adjoint operators gives formulas for the corrections to the eigenvalues~\citep{courant_methods_1989}: if $\{u_{1}^{[0]},u_{2}^{[0]}\}$ are two orthonormal eigenfunctions of $-\mathcal{J}^{[0]}$ for a given mode $(n,m)$, $m>0$, then the perturbed tensions $\mu_{nm}^{[1]}/\kappa$ are eigenvalues of the  $2\times 2$ matrix $M_{ij}= \langle-\mathcal{J}^{[1]}u_{i}^{[0]} , u_{j}^{[0]}\rangle$,
% \begin{equation}
%     \boldsymbol{M}=\left(\begin{array}{cc}
%     \langle-\mathcal{J}^{[1]}u_{nm}^{[0]} , u_{nm}^{[0]}\rangle & \langle -\mathcal{J}^{[1]}v_{nm}^{[0]} , u_{nm}^{[0]} \rangle\\
%     \langle-\mathcal{J}^{[1]}v_{nm}^{[0]} , u_{nm}^{[0]}\rangle & \langle-\mathcal{J}^{[1]}v_{nm}^{[0]} , v_{nm}^{[0]} \rangle\
%     \end{array}\right),
% \end{equation}
where angular brackets denote the inner product. Clearly, only eigenvalues for which $r_k \neq 0$ will receive a nonzero perturbation, lifting the degeneracy at this order. %;  for all other modes, $M_{ij}=0$. 
Proceeding to $O(\delta^2)$ reveals  that all eigenvalues for modes for which $m$ is a multiple of a $k$ with $r_k\neq0$ 
%(i.e., modes that respect the dihedral symmetries of the boundary) 
are perturbed.
%In the simple case of a single-mode perturbation of the boundary from a circle, this means that only $m$ values that are nonzero integer multiples of this mode will be perturbed at $O(\delta)$. 
The argument from Sec.~\ref{sec:linevp} continues to hold, and we again have two branches per eigenfunction; however, this time $\cos(m\phi)$ and $\sin(m\phi)$ are not the same because of the perturbed boundary. We thus expect  a total of \emph{four} branches per  $(n,m)$ pair in the $m\neq 0$ nonaxisymmetric case. Examples of these surfaces for $(1,0)$, $(1,1)$, and $(2,3)$  modes with a single-wavenumber boundary perturbation are shown in Fig.~\ref{fig:asym_modes}. We again see that some $m=1$ modes require less compression than $m=0$ .

%Depending on the dihedral symmetry of the boundaries, some of the eigenvalues will not be the same, as noncircular shapes break rotational symmetry, thus lifting the degeneracy of the eigenvalues in the catenoid case. An example of this for a simple single-mode boundary is shown in Fig...

%The perturbation argument from Sec.~\ref{sec:linevp} continues to apply, and compression of the rings again generates two solution branches per minimal eigensurface. Noncircular shapes break rotational symmetry, thus lifting the degeneracy of the eigenvalues in the catenoid case. An example of this for a simple single-mode boundary is shown in Fig...

\begin{figure}
    \centering
    (a)\includegraphics[trim = 20pt 0pt 40pt 20pt, clip,width=0.44\linewidth]{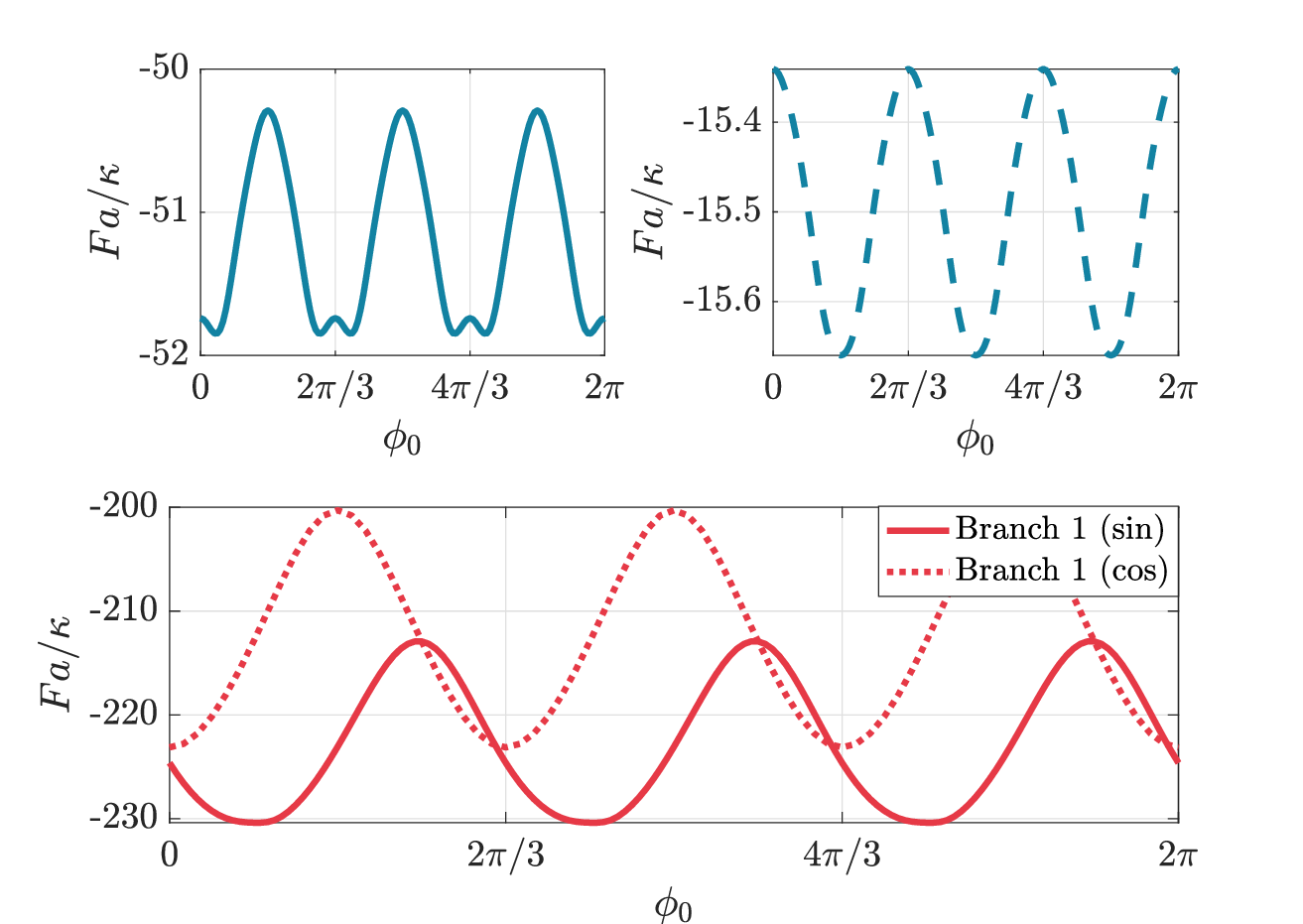}
    (b)\includegraphics[trim = 1pt 0pt 40pt 20pt, clip,width=0.42\linewidth]{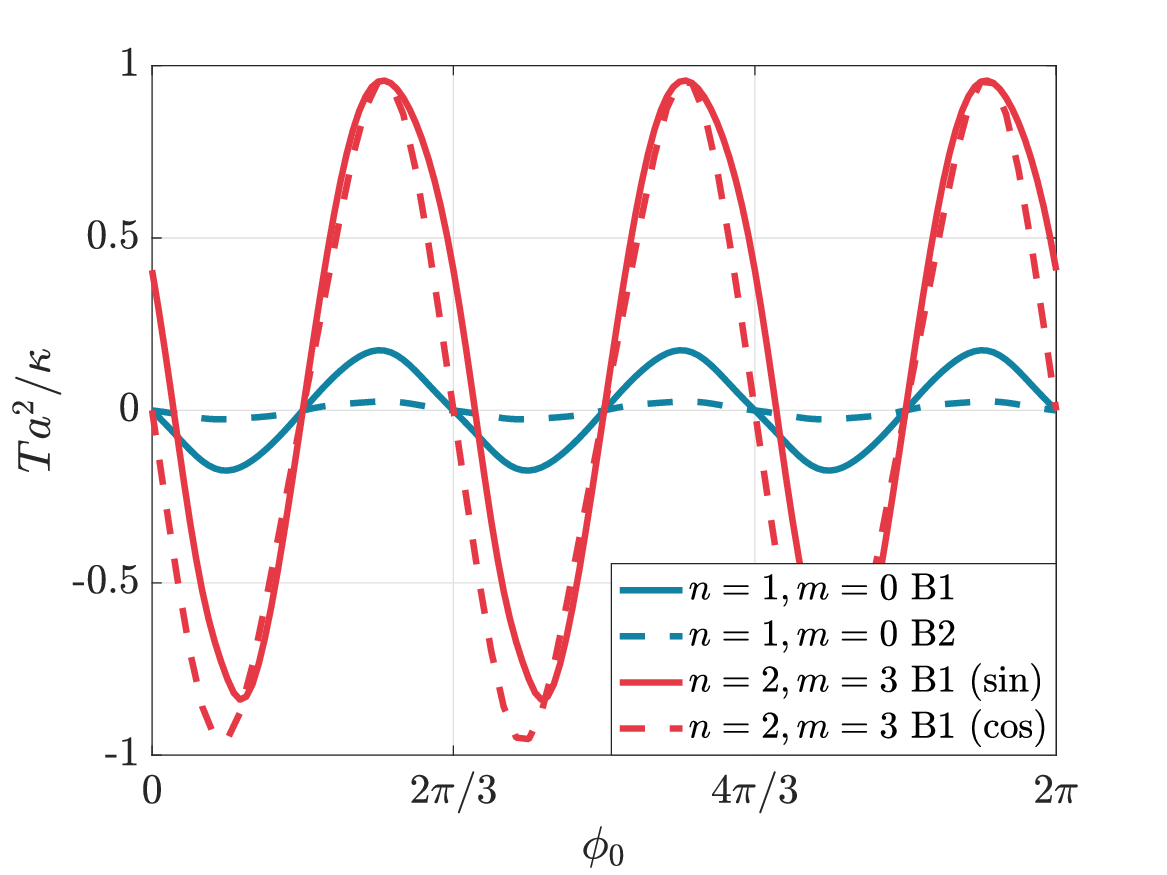}
    \caption{Force and torque as functions of relative ring twist $\phi_0$ for some of the surfaces in Fig.~\ref{fig:asym_modes}. (a)  Axial force for $(1,0)$ mode branch 1 [top left], $(1,0)$ mode branch 2 [top right], and the $(2,3)$ mode branches [bottom]. (b) Torque for the branches in (a). B means branch.}
    \label{fig:FTplots}
\end{figure}

\begin{figure}
    \centering
    (a)\includegraphics[trim = 40pt 0pt 40pt 0pt, clip,width=0.25\linewidth]{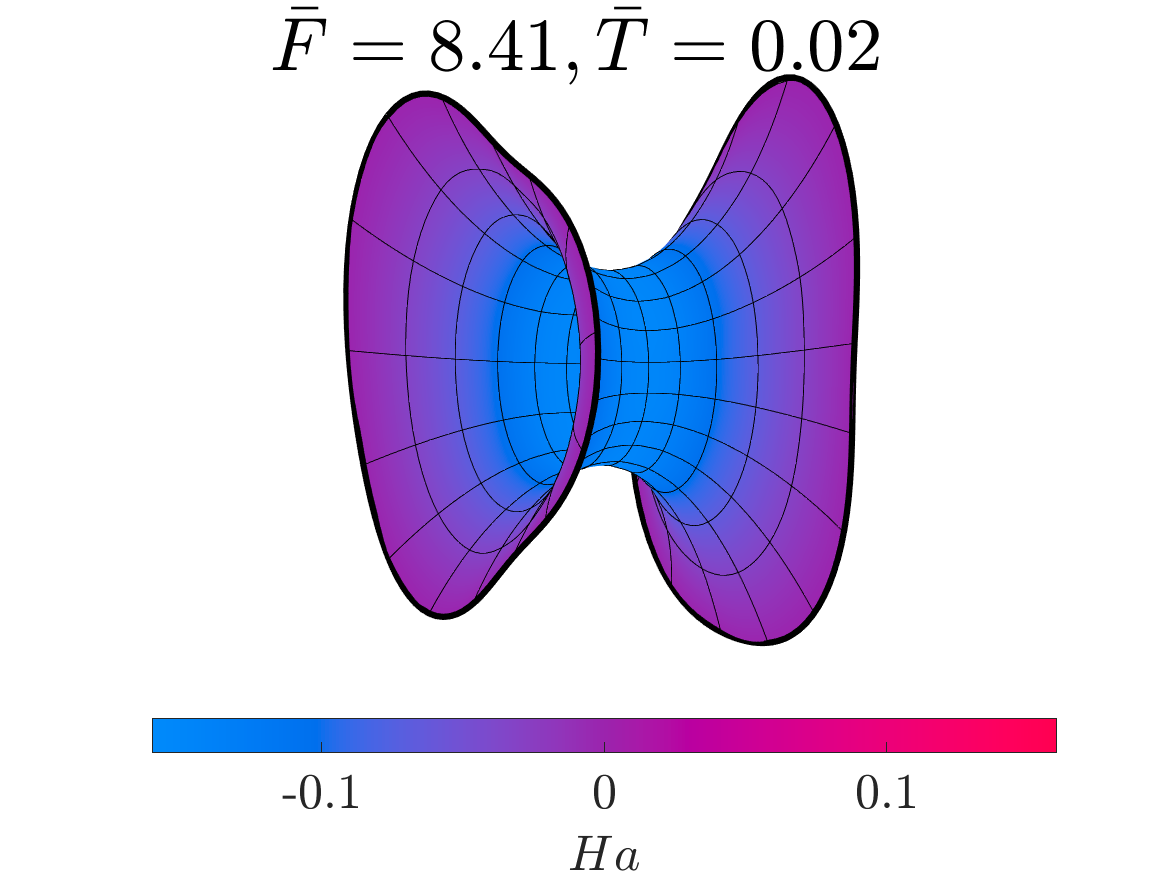}
    (b)\includegraphics[trim = 40pt 0pt 40pt 0pt, clip,width=0.25\linewidth]{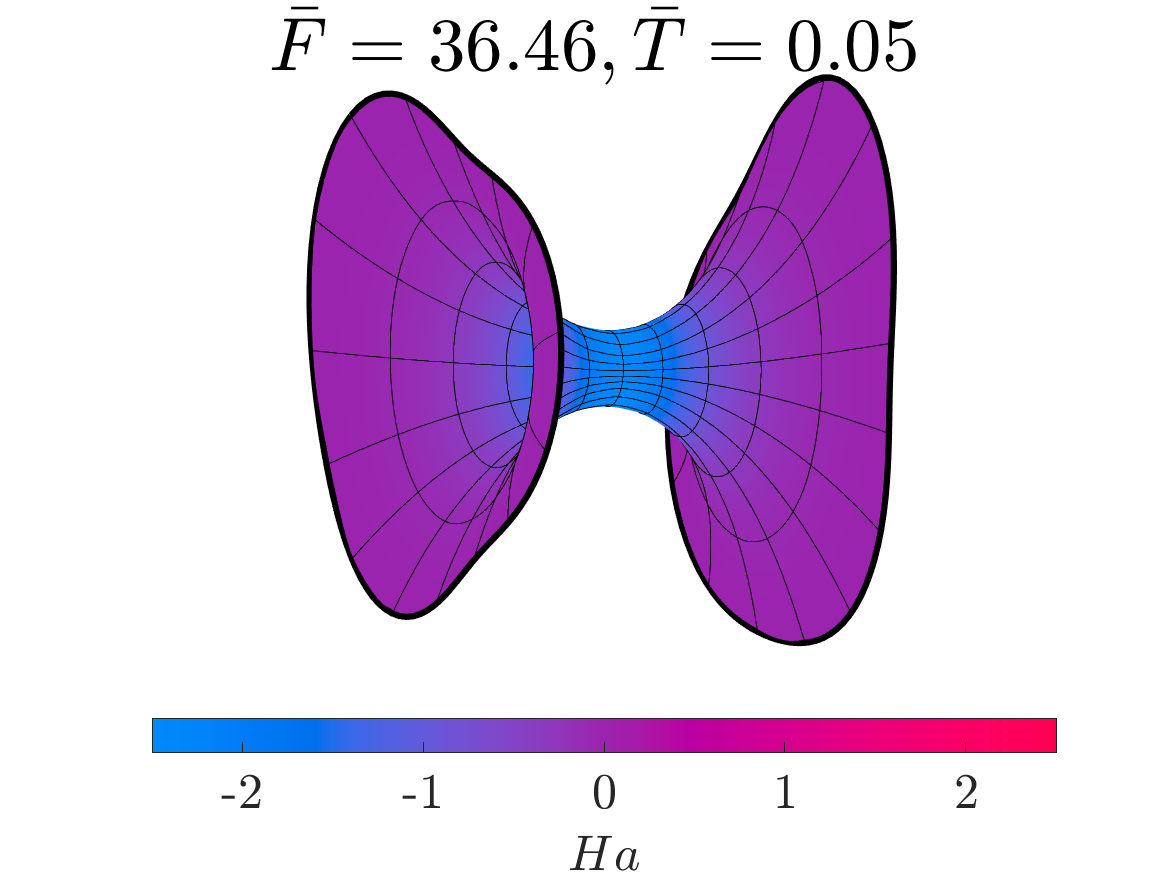}
    (c)\includegraphics[trim = 20pt 20pt 20pt 20pt, clip, width=0.23\linewidth]{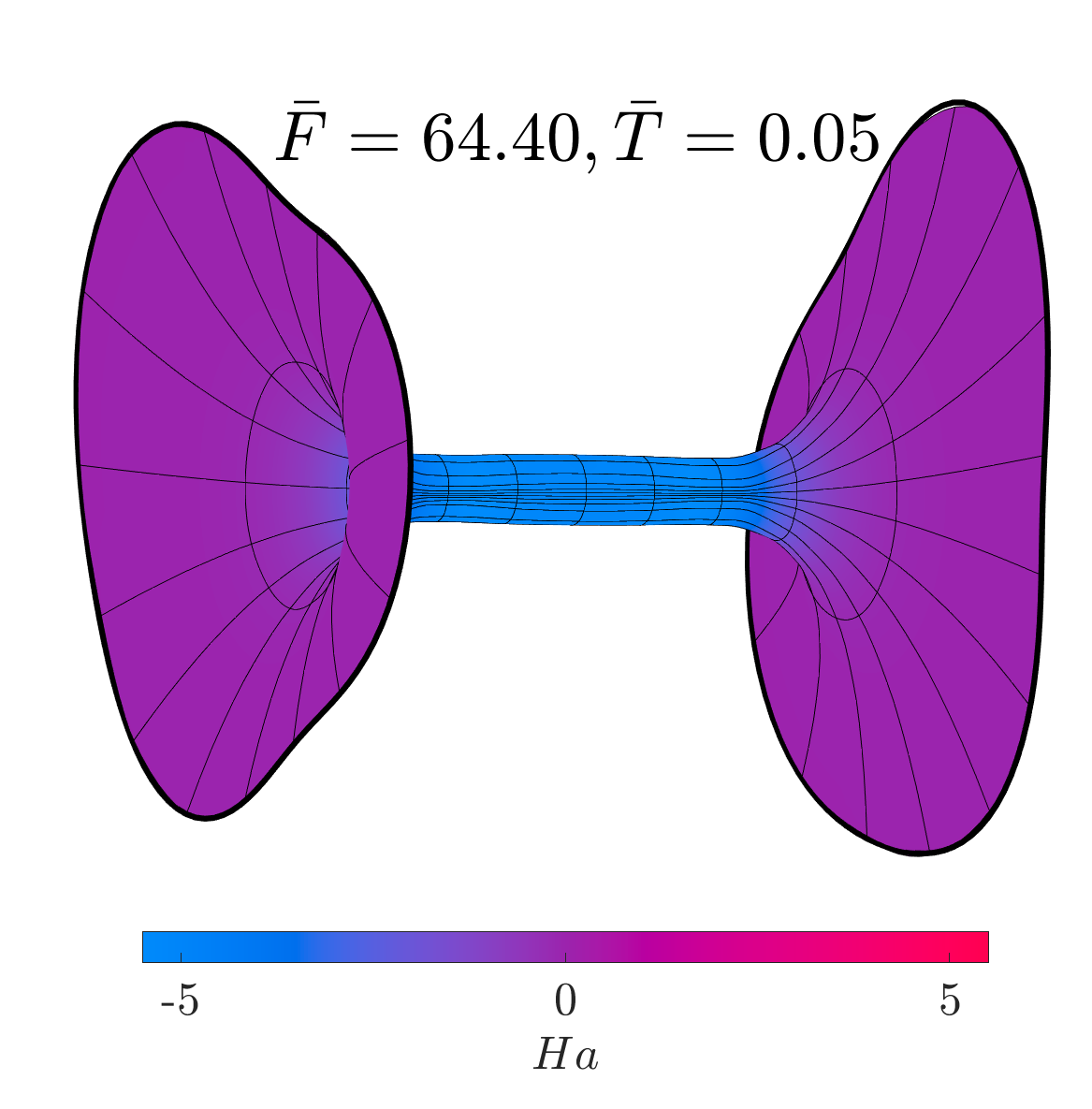}
    \caption{Transition of a $(1,0)$ mode into a tether with no maximal extension ($\bar{A}=1.1823\hdots$). Boundary rings are $r^-(\phi)/a = 1+0.1(\cos \phi + \sin 2\phi + \cos 3\phi)$ and $r^+(\phi) = r^-(\phi-\pi/3)$. (a) $h/a=1$, (b) $h/a=1.5$, (c) $h/a=3$.}
    \label{fig:tether}
\end{figure}

Noncircular boundary rings allow for the possibility of a torque, Eqn.~(\ref{Teqn}), due to rotational symmetry breaking. A boundary perturbation also introduces periodic $O(\delta)$ variations in the force. Force and torque as functions of relative twist $\phi_0$ for nonaxisymmetric boundary rings 
%of the form $r^+(\phi)/a = 1 + 0.1\cos (3(\phi-\phi_0))$ and $r^-(\phi)/a = 1 + 0.1\cos (3\phi)$ 
are plotted in Fig.~\ref{fig:FTplots}. We find that as $n$ increases, so does $T$. For $m>0$, sin and cos branches tend to require roughly the same magnitude of force and torque.  %If $n$ is even, then since the two branches are reflections of each other, the required torques are equal in magnitude but opposite in sign.
For completeness, we also explore the case of $\bar{A}>1$. Here, the $(1,0)$ branch of the thin minimal surface no longer has a maximal extension. For $h>h^*$, the surface becomes an elongated cylindrical tether that resembles a regularized Goldschmidt solution (Fig.~\ref{fig:tether}) connecting two halves of a minimal surface. It is natural that tethers only emerge from this  mode, as increasing $h$ requires a positive (extensile) force, suppressing the buckling instability. As in the axisymmetric case~\citep{powers_fluid-membrane_2002}, the cross sections of the tethers are nearly circular so that $\bar{T}=Ta^2/\kappa\approx 0$. The force $\bar{F}$, by contrast, is large, increasing linearly with extension for $h\gg a$.
%As the asymptotics of tethers have been thoroughly investigated in studies such as \citet{powers_fluid-membrane_2002}, we do not focus on these shapes here.


\section{Conclusion}

Roughly speaking, fixed-area elastic fluid interfaces can be understood as amalgamations of classical minimal surfaces and elastic shells. 
For sufficiently small surface area, such interfaces have a maximal extension $h^*$ and form a minimal surface as in the Plateau--Douglas problem. 
External confining forces, in conjunction with the area constraint, introduce the possibility of buckled solutions, analogous to how compressing the ends of a slender inextensible elastic rod gives rise to  buckling. Each of the infinitely many buckling modes characterized by integers $n$ and $m$ leads to multiple feasible branches as $h$ is compressed, the precise number of which depends on the boundary symmetry.
%As the rings are compressed away from $h^*$, there arise infinitely many buckling modes, each with two feasible branches, characterized by integers $n$ and $m$. %If the  area exceeds $\bar{A}_{\mathrm{r}}$, then tethers emerge as a solution with no maximal extension for $n=1$, $m=0$. %We also elucidated the effect of noncircular rings, whose dihedral symmetries can introduce more shapes. 
%Boundary perturbations can additionally influence the force and torque needed to hold these shapes in equilibrium and double the number of branches by breaking rotational symmetry. 
Interestingly, the $m=1$ modes can require less force than $m=0$, suggesting that the activation of these nonaxisymmetric modes could lower the energy barrier to shape transitions.

The work presented here broadens our understanding of feasible curved fluid configurations, providing a foundation for understanding biological processes involving asymmetry. Future work includes using a more general parameterization so that highly buckled shapes that are not functions of $z$ can be modeled. Such a parameterization could in principle also be used to treat nonplanar and deformable boundaries. A more general energy that includes a Gaussian curvature term also offers an interesting system to investigate, as Eqn.~(\ref{notorqueBC}) becomes inhomogeneous, prohibiting $H=0$. Singular perturbation theory will thus be needed.\\

\noindent
\textbf{Acknowledgements.} {We are grateful to James Hanna, Robert Pelcovits, and especially Thomas Powers for helpful discussions. Comments from Howard Stone and Ehud Yariv  helped inspire parts of the analysis in Sec.~\ref{sec:linevp}. LLJ acknowledges former employment at the National Institute of Standards and Technology.}\\

%\backsection[Funding]{XL acknowledges a Terminal Year Fellowship from the McCormick School of Engineering, Northwestern University.}

\noindent
\textbf{Declaration of interests.} {The authors report no conflict of interest.}\\

%\backsection[Data availability statement]{The data that support the findings of this study are openly available in [repository name] at http://doi.org/[doi], reference number [reference number]. See JFM's \href{https://www.cambridge.org/core/journals/journal-of-fluid-mechanics/information/journal-policies/research-transparency}{research transparency policy} for more information}

\noindent
\textbf{Author ORCIDs.} \\
{X. Liu, https://orcid.org/0009-0005-1459-6313; \\
N. A. Patankar, https://orcid.org/0000-0002-0665-3454; \\
L. L. Jia, https://orcid.org/0000-0003-1968-4767}



\section*{Supplementary material}

# Supplementary Material for "Shapes, forces, and torques of compressed elastic fluid interfaces: beyond axisymmetric configurations"

Xinyi Liu
Department of Engineering Sciences and Applied Mathematics, Northwestern University
2145 Sheridan Road, Evanston, Illinois 60208, USA
XinyiLiu2025@u.northwestern.edu

Neelesh A. Patankar
Department of Mechanical Engineering, Northwestern University
Department of Engineering Sciences and Applied Mathematics, Northwestern University
2145 Sheridan Road, Evanston, Illinois 60208, USA
n-patankar@northwestern.edu

Leroy L. Jia
Independent researcher
North Bethesda, MD 20852, USA
royjia117@gmail.com



## 1 Geometric quantities

Let the surface be given by

$$\boldsymbol{X}(\phi, z) = (r(\phi, z)\cos\phi, r(\phi, z)\sin\phi, z) \tag{1}$$

Tangent vectors:

$$\boldsymbol{X}_\phi = (r_\phi \cos\phi - r\sin\phi, r_\phi \sin\phi + r\cos\phi, 0) \tag{2}$$

$$\boldsymbol{X}_z = (r_z \cos\phi, r_z \sin\phi, 1) \tag{3}$$

Metric tensor $g_{ij} = \boldsymbol{X}_i \cdot \boldsymbol{X}_j$:

$$g_{ij} = \begin{bmatrix} r^2 + r_\phi^2 & r_\phi r_z \\ r_\phi r_z & r_z^2 + 1 \end{bmatrix} \tag{4}$$

Also, $g = \det(g_{ij}) = r^2(r_z^2 + 1) + r_\phi^2$.

Normal vector:

$$\boldsymbol{N} = \frac{(r_\phi \sin\phi + r\cos\phi, r\sin\phi - r_\phi\cos\phi, -rr_z)}{\sqrt{g}} \tag{5}$$

Second derivatives:

$$\boldsymbol{X}_{\phi\phi} = (r_{\phi\phi}\cos\phi - 2r_\phi \sin\phi - r\cos\phi, r_{\phi\phi}\sin\phi + 2r_\phi\cos\phi - r\sin\phi, 0) \tag{6}$$

$$\boldsymbol{X}_{\phi z} = (r_{z\phi}\cos\phi - r_z\sin\phi, r_{\phi z}\sin\phi + r_z\cos\phi, 0) \tag{7}$$

$$\boldsymbol{X}_{zz} = (r_{zz}\cos\phi, r_{zz}\sin\phi, 0) \tag{8}$$

Curvature tensor $b_{ij} = -\boldsymbol{N} \cdot \boldsymbol{X}_{ij}$:

$$b_{ij} = -\frac{1}{\sqrt{g}} \begin{bmatrix} r(r_{\phi\phi} - r) - 2r_\phi^2 & rr_{z\phi} - r_z r_\phi \\ rr_{z_\phi} - r_z r_\phi & rr_{zz} \end{bmatrix} \tag{9}$$

Also, $b = \det(b_{ij}) = (rr_{zz}[r(r_{\phi\phi} - r) - 2r_\phi^2] - (rr_{z\phi} - r_z r_\phi)^2)/g$.

Mean and Gaussian curvatures:

$$2H = g^{ij} b_{ij} = \frac{r(r_z^2+1)(r_{\phi\phi}-r) - 2r_\phi^2 - 2rr_z r_\phi r_{z\phi} + rr_{zz}(r^2+r_\phi^2)}{[r^2(r_z^2+1)+r_\phi^2]^{3/2}} \tag{10}$$

$$K = \frac{b}{g} = \frac{rr_{zz}[r(r_{\phi\phi}-r)-2r_\phi^2] - (rr_{z\phi}-r_z r_\phi)^2}{[r^2(r_z^2+1)+r_\phi^2]^2} \tag{11}$$

respectively.

# 2 Variational Formulation

## 2.1 Noether invariants

Let $\mathcal{E}(r, r_z, r_\phi, r_{zz}, r_{z\phi}, r_{\phi\phi})$ be defined by $E = \int \mathrm{d}\phi \mathrm{d}z\, \mathcal{E}$. Note that $\mathcal{E}$ includes the $\sqrt{g}$ factor and thus has dimensions of energy per unit length (or force). Since $\mathcal{E}$ does not depend on $z$ or $\phi$ explicitly, we have by the chain rule

$$\frac{\partial \mathcal{E}}{\partial z} = \frac{\partial \mathcal{E}}{\partial r} r_z + \frac{\partial \mathcal{E}}{\partial r_z} r_{zz} + \frac{\partial \mathcal{E}}{\partial r_\phi} r_{z\phi} + \frac{\partial \mathcal{E}}{\partial r_{zz}} r_{zzz} + \frac{\partial \mathcal{E}}{\partial r_{z\phi}} r_{z\phi z} + \frac{\partial \mathcal{E}}{\partial r_{\phi\phi}} r_{\phi\phi z} \tag{12}$$

$$\frac{\partial \mathcal{E}}{\partial \phi} = \frac{\partial \mathcal{E}}{\partial r} r_\phi + \frac{\partial \mathcal{E}}{\partial r_z} r_{z\phi} + \frac{\partial \mathcal{E}}{\partial r_\phi} r_{\phi\phi} + \frac{\partial \mathcal{E}}{\partial r_{zz}} r_{zz\phi} + \frac{\partial \mathcal{E}}{\partial r_{z\phi}} r_{z\phi\phi} + \frac{\partial \mathcal{E}}{\partial r_{\phi\phi}} r_{\phi\phi\phi} \tag{13}$$

Also, we have the Euler equation

$$\frac{\partial \mathcal{E}}{\partial r} - \left(\frac{\partial \mathcal{E}}{\partial r_z}\right)_z - \left(\frac{\partial \mathcal{E}}{\partial r_\phi}\right)_\phi + \left(\frac{\partial \mathcal{E}}{\partial r_{zz}}\right)_{zz} + \left(\frac{\partial \mathcal{E}}{\partial r_{z\phi}}\right)_{z\phi} + \left(\frac{\partial \mathcal{E}}{\partial r_{\phi\phi}}\right)_{\phi\phi} = 0 \tag{14}$$

which we use to eliminate $\partial\mathcal{E}/\partial r$. After some simplifying and rewriting of derivatives, we find

$$\mathcal{E}_z = \left[r_z \frac{\partial \mathcal{E}}{\partial r_z} + r_{zz}\frac{\partial \mathcal{E}}{\partial r_{zz}} - r_z\left(\frac{\partial \mathcal{E}}{\partial r_{zz}}\right)_z + r_{z\phi}\frac{\partial \mathcal{E}}{\partial r_{z\phi}}\right]_z + \left[r_z\frac{\partial \mathcal{E}}{\partial r_\phi} - r_z\left(\frac{\partial \mathcal{E}}{\partial r_{z\phi}}\right)_z + r_{z\phi}\frac{\partial \mathcal{E}}{\partial r_{\phi\phi}} - r_z\left(\frac{\partial \mathcal{E}}{\partial r_{\phi\phi}}\right)_\phi\right]_\phi \tag{15}$$

$$\mathcal{E}_\phi = \left[r_\phi \frac{\partial \mathcal{E}}{\partial r_z} + r_{z\phi}\frac{\partial \mathcal{E}}{\partial r_{zz}} - r_\phi\left(\frac{\partial \mathcal{E}}{\partial r_{zz}}\right)_z - r_\phi\left(\frac{\partial \mathcal{E}}{\partial r_{z\phi}}\right)_\phi\right]_z + \left[r_\phi\frac{\partial \mathcal{E}}{\partial r_\phi} - r_\phi\left(\frac{\partial \mathcal{E}}{\partial r_{z\phi}}\right)_z + r_{z\phi}\frac{\partial \mathcal{E}}{\partial r_{z\phi}} - r_\phi\left(\frac{\partial \mathcal{E}}{\partial r_{\phi\phi}}\right)_\phi\right]_\phi \tag{16}$$

In order to find the conserved quantities, we integrate both expressions in $\phi$ so that the $\phi$-derivative terms vanish. We thus conclude that

$$F := \oint \mathrm{d}\phi\ \left[\mathcal{E} - r_z\frac{\partial \mathcal{E}}{\partial r_z} - r_{zz}\frac{\partial \mathcal{E}}{\partial r_{zz}} + r_z\left(\frac{\partial \mathcal{E}}{\partial r_{zz}}\right)_z - r_{z\phi}\frac{\partial \mathcal{E}}{\partial r_{z\phi}}\right] \tag{17}$$

$$T := -\oint \mathrm{d}\phi\ \left[r_\phi\frac{\partial \mathcal{E}}{\partial r_z} + r_{z\phi}\frac{\partial \mathcal{E}}{\partial r_{zz}} - r_\phi\left(\frac{\partial \mathcal{E}}{\partial r_{zz}}\right)_z - r_\phi\left(\frac{\partial \mathcal{E}}{\partial r_{z\phi}}\right)_\phi\right] \tag{18}$$

are constant. Based on dimensions ($\mathcal{E}$ has dimensions of energy per length, i.e. force) and the expressions in the $r_\phi \to 0$ limit, we might expect $F$ to represent the axial force and $T$ to represent the torque (twisting moment) about the $z$-axis. We will show this below.

The full expressions for $F$ and $T$ are complicated, so we will calculate them numerically. To this end, we calculate

$$\frac{\partial \mathcal{E}}{\partial r_{zz}} = \sqrt{g}\left[\kappa(2H)\frac{\partial(2H)}{\partial r_{zz}}\right] = \kappa(2H)\frac{r(r^2+r_\phi^2)}{g} \tag{19}$$

$$\frac{\partial \mathcal{E}}{\partial r_{z\phi}} = \sqrt{g}\left[\kappa(2H)\frac{\partial(2H)}{\partial r_{z\phi}}\right] = -\kappa(2H)\frac{2rr_z r_\phi}{g} \tag{20}$$

$$\frac{\partial \mathcal{E}}{\partial r_z} = \left[\kappa(2H)\frac{\partial(2H)}{\partial r_z}\right]\sqrt{g} + \frac{r^2 r_z}{\sqrt{g}}\left[\frac{\kappa}{2}(2H)^2 + \mu\right] \tag{21}$$

$$\frac{\partial(2H)}{\partial r_z} = \frac{2rr_z(r_{\phi\phi}-r) - 2rr_\phi r_{z\phi}}{g^{3/2}} - \frac{3}{2}(2H)\frac{2r^2 r_z}{g} \tag{22}$$

$$\frac{\partial K}{\partial r_z} = \frac{2r_\phi(rr_{z\phi} - r_z r_\phi)}{g^2} - 2K\frac{2r^2 r_z}{g} \tag{23}$$

## 2.2 Principle of virtual work

For simplicity, we take the left ring to be fixed and move the right one. We will show that the conserved quantities above correspond to axial force and torque, using an argument that generalizes Jia et al. (2021). By the principle of virtual work

$$\delta E - \oint \mathrm{d}\ell \; f_1 \delta z_1 - \oint \mathrm{d}\ell \; \tau_1 \delta\phi_1 - \oint \mathrm{d}\ell \; m_1 \delta\theta_1 = 0 \tag{24}$$

Here, $\mathrm{d}\ell$ is the boundary line element, $f_1$ is the axial force per unit length on the boundary, $\tau_1$ is the twisting moment per unit length on the boundary, $m_1$ is the bending moment per unit length on the boundary, $\delta z_1$ is the virtual displacement of the ring, $\delta\phi_1$ is the virtual change in azimuthal angle by which the ring is rotated, $\delta\theta_1$ is the virtual change in the angle at the boundary (where $\tan\theta_1(\phi) = r_z(z_1, \phi)$), and the integrals are taken with respect to boundary arc length. In particular,

$$\mathrm{d}\ell = \mathrm{d}\phi \, \sqrt{\bar{r}^2 + \bar{r}_\phi^2} \tag{25}$$

where $\bar{r}(\phi)$ is the curve representing the boundary ring.

Next, we calculate the first variation. Let $\mathcal{E}$ be the energy density as above.

$$\begin{aligned}
\delta E &= \int_0^{2\pi} \mathrm{d}\phi \int_{-z_1}^{z_1} \mathrm{d}z \; \left[\mathcal{E}(r+\delta r, \ldots, r_{\phi\phi} + \delta r_{\phi\phi}) - \mathcal{E}(r, \ldots, r_{\phi\phi})\right] + \oint \mathrm{d}\phi \; \delta z_1 \mathcal{E} \\
&= \int_0^{2\pi} \mathrm{d}\phi \int_{-z_1}^{z_1} \mathrm{d}z \; \left[\delta r \frac{\partial \mathcal{E}}{\partial r} + \ldots + \delta r_{\phi\phi} \frac{\partial \mathcal{E}}{\partial r_{\phi\phi}}\right] + \oint \mathrm{d}\phi \; \delta z_1 \mathcal{E} \\
&= \int_0^{2\pi} \mathrm{d}\phi \int_{-z_1}^{z_1} \mathrm{d}z \; \delta r \frac{\delta \mathcal{E}}{\delta r} + \oint \mathrm{d}\phi \left\{ \delta z_1 \mathcal{E} + \delta r \left[ \frac{\partial \mathcal{E}}{\partial r_z} - \left(\frac{\partial \mathcal{E}}{\partial r_{zz}}\right)_z - \left(\frac{\partial \mathcal{E}}{\partial r_{z\phi}}\right)_\phi \right] + \delta r_z \frac{\partial \mathcal{E}}{\partial r_{zz}} \right\}
\end{aligned} \tag{26}$$

where

$$\frac{\delta \mathcal{E}}{\delta r} = \frac{\partial \mathcal{E}}{\partial r} - \left(\frac{\partial \mathcal{E}}{\partial r_z}\right)_z - \left(\frac{\partial \mathcal{E}}{\partial r_\phi}\right)_\phi + \left(\frac{\partial \mathcal{E}}{\partial r_{zz}}\right)_{zz} + \left(\frac{\partial \mathcal{E}}{\partial r_{z\phi}}\right)_{z\phi} + \left(\frac{\partial \mathcal{E}}{\partial r_{\phi\phi}}\right)_{\phi\phi} \tag{27}$$

is assumed to be zero in the Euler–Lagrange equation. Next, we need to write $\delta r$ and $\delta r_z$ in terms of other variations. Let $\delta r_1$ represent the difference in $r$ for the point at the boundary (i.e., the point whose Lagrangian label is $t = 0$).

$$\delta r_1 = (r + \delta r)(z_1 + \delta z_1, \phi_1 + \delta\phi_1) - r(z_1, \phi_1) = r_z(z_1, \phi_1)\delta z_1 + r_\phi(z_1, \phi_1)\delta\phi_1 + \delta r(z_1, \phi_1) + \ldots \tag{28}$$

$$\implies \delta r(z_1, \phi_1) = \delta r_1 - r_z(z_1, \phi_1)\delta z_1 - r_\phi(z_1, \phi_1)\delta\phi_1 \tag{29}$$

What is the right formula for the bending moment when the axis and force aren't orthogonal?

Let $\boldsymbol{e}_1$ be the tangent vector of the boundary curve and $\boldsymbol{e}_3$ be the surface normal. Then, $\boldsymbol{e}_2 = \boldsymbol{e}_3 \times \boldsymbol{e}_1$ is tangent to the surface. The angle $\theta_1$ is the angle between $\boldsymbol{e}_2$ and the unit vector in the $z$-direction. We find that

$$\cos\theta_1 = \boldsymbol{e}_2 \cdot \boldsymbol{e}_z|_{z=z_1,\phi=\phi_1} = \left.\frac{\sqrt{r^2 + r_\phi^2}}{\sqrt{r^2(1 + r_z^2) + r_\phi^2}}\right|_{z=z_1,\phi=\phi_1} \tag{30}$$

Alternatively, we could work with $\tan\theta_1$ to be consistent with Jia et al. (2021). In this case, we will want to vary

$$\tan\theta_1 = \left.\frac{r r_z}{\sqrt{r^2 + r_\phi^2}}\right|_{z=z_1,\phi=\phi_1} \tag{31}$$

We perturb $\phi_1 \to \phi_1 + \delta\phi_1$, $z \to z + \delta z_1$, $r \to r + \delta r$. We also use the expression for $\delta r$ that we derived above to eliminate it in favor of $\delta r_1$, $\delta z_1$, and $\delta\phi_1$. The computation is lengthy, but the result is

$$\begin{aligned}
\delta\theta_1 \sec^2\theta_1 = \delta\theta_1 \frac{g}{r^2 + r_\phi^2} &= \frac{r\delta z_1}{\sqrt{r^2 + r_\phi^2}} \left( r_{zz} - \frac{r_z r_\phi r_{z\phi}}{r^2 + r_\phi^2} \right) + \frac{r\delta\phi_1}{\sqrt{r^2 + r_\phi^2}} \left( r_{z\phi} - \frac{r_z r_\phi r_{\phi\phi}}{r^2 + r_\phi^2} \right) \\
&\quad + \frac{\delta r_1}{\sqrt{r^2 + r_\phi^2}} \left( r_z - \frac{r^2 r_z}{r^2 + r_\phi^2} \right) - \frac{\delta r_\phi}{\sqrt{r^2 + r_\phi^2}} \frac{r r_z r_\phi}{r^2 + r_\phi^2} + \frac{r\delta r_z}{\sqrt{r^2 + r_\phi^2}}
\end{aligned} \tag{32}$$

We want to solve for $\delta r_z$, so we rearrange to get

$$\delta r_z = \delta\theta_1 \frac{g}{r\sqrt{r^2+r_\phi^2}} - \delta z_1 \left( r_{zz} - \frac{r_z r_\phi r_{z\phi}}{r^2+r_\phi^2} \right) - \delta\phi_1 \left( r_{z\phi} - \frac{r_z r_\phi r_{\phi\phi}}{r^2+r_\phi^2} \right) - \delta r_1 \frac{r_z r_\phi^2}{r(r^2+r_\phi^2)} + \delta r_\phi \frac{r_z r_\phi}{r^2+r_\phi^2} \tag{33}$$

We use the fact that $\delta r_\phi = (\delta r_1)_\phi - r_{z\phi}\delta z_1 - r_{\phi\phi}\delta\phi_1$ (for our setup, $\delta z_1$ and $\delta\phi_1$ are constants). Plugging this into the boundary integral yields

$$\oint \mathrm{d}\phi \; \delta r_z \frac{\partial \mathcal{E}}{\partial r_{zz}} = \oint \mathrm{d}\phi \; \left[ \delta\theta_1 \frac{g}{r\sqrt{r^2+r_\phi^2}} - \delta z_1 r_{zz} - \delta\phi_1 r_{z\phi} \right] \frac{\partial \mathcal{E}}{\partial r_{zz}} - \oint \mathrm{d}\phi \; \frac{\delta r_1}{r} \left( \frac{r r_z r_\phi}{r^2+r_\phi^2} \frac{\partial \mathcal{E}}{\partial r_{zz}} \right)_\phi \tag{34}$$

The rest of the terms are seen to cancel out. This equation implies that if $\bar{r}(\phi)$ and $\delta r_1$ are not constant, then the radial force will need to be modified by an extra term. However, in our setup, the rings are rigid so $\delta r_1 = 0$.

Adding in the external force, torque, and moment via the principle of virtual work, we find

$$\oint \mathrm{d}\phi \left\{ \mathcal{E} - r_z \left[ \frac{\partial \mathcal{E}}{\partial r_z} - \left( \frac{\partial \mathcal{E}}{\partial r_{zz}} \right)_z - \left( \frac{\partial \mathcal{E}}{\partial r_{z\phi}} \right)_\phi \right] - r_{zz} \frac{\partial \mathcal{E}}{\partial r_{zz}} - f_1 \sqrt{r^2+r_\phi^2} \right\} \delta z_1 = 0 \tag{35}$$

$$\oint \mathrm{d}\phi \left\{ -r_\phi \left[ \frac{\partial \mathcal{E}}{\partial r_z} - \left( \frac{\partial \mathcal{E}}{\partial r_{zz}} \right)_z - \left( \frac{\partial \mathcal{E}}{\partial r_{z\phi}} \right)_\phi \right] - r_{z\phi} \frac{\partial \mathcal{E}}{\partial r_{zz}} - \tau_1 \sqrt{r^2+r_\phi^2} \right\} \delta\phi_1 = 0 \tag{36}$$

$$\oint \mathrm{d}\phi \; \left\{ \frac{g}{r\sqrt{r^2+r_\phi^2}} \frac{\partial \mathcal{E}}{\partial r_{zz}} - m_1 \sqrt{r^2+r_\phi^2} \right\} \delta\theta_1 = 0 \tag{37}$$

for arbitrary deformations at the boundary. Setting the integrands to zero yields the expressions given in the previous section after integration by parts is performed. In particular, one can calculate that

$$\frac{\partial \mathcal{E}}{\partial r_{zz}} = \frac{r(r^2+r_\phi^2)}{g} [\kappa(2H)] \tag{38}$$

for our energy. This yields the expression for the bending moment (per unit length) that we are familiar with: $m_1 = \kappa(2H)$.


\bibliographystyle{plain}
\bibliography{Perturbed-Membranes}

\end{document}